\documentclass[aps,prd,nofootinbib,twocolumn,superscriptaddress,preprintnumbers,balancelastpage,longbibliography]{revtex4-1}

\pdfoutput=1
\usepackage{subfigure}
\usepackage{graphicx}
\usepackage{epstopdf}
\usepackage{mathrsfs}
\usepackage{amssymb}
\usepackage{verbatim}
\usepackage{color}
\usepackage{multirow}
\usepackage{amsmath}
\usepackage{hyperref}
\usepackage{physics}

\hypersetup{pdfstartview=FitV,colorlinks=true,linkcolor=blue,citecolor=red,filecolor=black,urlcolor=blue}

\def\trh{T_{\rm RH}}
\def\tmax{T_{\rm max}}
\def\tcross{T_{\cross}}
\def\across{a_{\cross}}

\newcommand{\aend}{a_{\rm end}}
\newcommand{\arh}{a_{\rm RH}}
\newcommand{\amax}{a_{\rm max}}
\newcommand{\rhorh}{\rho_{\rm RH}}
\newcommand{\rhoend}{\rho_{\rm end}}

\newcommand{\rhomax}{\rho_{\rm max}}
\newcommand{\beq}{\begin{equation}}
\newcommand{\eeq}{\end{equation}}
\newcommand{\bea}{\begin{eqnarray}}
\newcommand{\eea}{\end{eqnarray}}

\begin{document}

\preprint{UMN--TH--4408/24, FTPI--MINN--24/27}

\title{Aspects of Gravitational Portals and Freeze-in during Reheating }

\author{Stephen E. Henrich}
    \email[Correspondence email address: ]{henri455@umn.edu}
    \affiliation{William I. Fine Theoretical Physics Institute, School of Physics and Astronomy, University of Minnesota, Minneapolis, Minnesota 55455, USA}

\author{Yann Mambrini}
     \email[Correspondence email address: ]{mambrini@ijclab.in2p3.fr}
\affiliation{Universit\'e Paris-Saclay, CNRS/IN2P3, IJCLab, 91405 Orsay, France}

\author{Keith A. Olive}
    \email[Correspondence email address: ]{olive@umn.edu}
    \affiliation{William I. Fine Theoretical Physics Institute, School of Physics and Astronomy, University of Minnesota, Minneapolis, Minnesota 55455, USA}


\date{\today} 

\begin{abstract}
We conduct a systematic investigation of freeze-in during reheating while taking care to include both direct and indirect production of dark matter (DM) via gravitational portals and inflaton decay. Direct production of DM can occur via gravitational scattering of the inflaton, while indirect production occurs through scattering in the Standard Model radiation bath. We consider two main contributions to the radiation bath during reheating. The first, which may dominate at the onset of the reheating process, is produced via gravitational scattering of the inflaton. The second (and more standard contribution) comes from inflaton decay. We consider a broad class of DM production rates parameterized as $R_{\chi} \propto T^{n+6}/\Lambda^{n+2}$, and inflaton potentials with a power-law form $V(\phi) \propto \phi^{k}$ about the minimum. We find the relic density produced by freeze-in for each contribution to the Standard Model bath for arbitrary $k$ and $n$, and compare these with the DM density produced gravitationally by inflaton scattering. We find that freeze-in production from the gravitationally-produced radiation bath can exceed that of the conventional decay bath and account for the observed relic density provided that $m_{\chi} > \trh$, with additional $k$- and $n$-dependent constraints. For each freeze-in interaction considered, we also find $m_{\chi}$- and $\trh$-dependent limits on the BSM scale, $\Lambda$,  for which gravitational production will exceed ordinary freeze-in production.

\end{abstract}

\keywords{dark matter, re-heating, inflation, gravity}

\maketitle

\section{Introduction} \label{sec:outline}

Reheating after the period of accelerated expansion is an essential component of any model of inflation \cite{reviews}. For many purposes, it may be sufficient to consider reheating as an instantaneous process. In this case, the accelerated expansion ends, a period of matter domination due to inflaton oscillations (about a quadratic minimum) continues until 
inflaton decay occurs when the inflaton decay rate, $\Gamma_\phi \sim H$, where $\phi$ is the inflaton and $H$ is the Hubble parameter. For example, the abundance of gravitinos produced at the end of inflation can be approximated by \cite{nos,ehnos,kl}
$Y_{3/2} \sim \Gamma_{3/2}/H \sim \trh/M_P$, where $\Gamma_{3/2}$ is the gravitino production rate, $\trh$ is the reheating temperature, and $M_P \simeq 2.4 \times 10^{18}$~GeV is the reduced Planck mass. The final abundance is relatively insensitive to the details of reheating and depends only on $\trh$. The same is true for many dark matter candidates produced by the freeze-in mechanism \cite{fimp,Bernal:2017kxu}.

However the radiation bath which dominates the energy density of the Universe at $T < \trh$,  does not just appear instantaneously. When the inflaton begins to oscillate, 
decays begin to occur and the Universe very quickly heats up to a maximum temperature, $\tmax$ \cite{Scherrer:1984fd,Giudice:2000ex,GKMO1,GKMO2}. 
Then, depending on the particle production rate, and the reheating temperature, the abundance may be sensitive to the details of the reheating process. In the example of the weak scale gravitino, its production cross-section is largely independent of temperature. 
In contrast, the production cross-section in models of high scale supersymmetry scales as $T^6$ \cite{Benakli:2017whb,grav2,grav3},
and as a result the abundance is sensitive to $\tmax$ \cite{Garcia:2017tuj}. This is sometimes referred to as ultra-violet freeze-in. 

The gravitational production of particles, or production through a gravitational portal may also occur during the process of reheating \cite{ema,Garny:2015sjg,Tang:2016vch,Bernal:2018qlk,Ema:2019yrd,Chianese:2020yjo,Ahmed:2020fhc,Kolb:2020fwh,Redi:2020ffc,Ling:2021zlj,MO,Bernal:2021kaj,Barman:2021ugy,Haque:2021mab,cmov,Clery:2022wib,Co:2022bgh,Garcia:2022vwm,Kaneta:2022gug,Mambrini:2022uol,Basso:2022tpd,Barman:2022qgt,Haque:2023yra,Kolb:2023dzp,Garcia:2023awt,Kaneta:2023kfv,Garcia:2023qab,Garcia:2023dyf,kkmov,Garcia:2023obw,Kolb:2023ydq,Clery:2024dlk,Racco:2024aac,Dorsch:2024nan,Choi:2024bdn,Verner:2024agh,Jenks:2024fiu}. Since this is a pure gravitational phenomenon, it is most important when the energy in the process is greatest. In this case, that would correspond to the period immediately after inflation ends, when $T\simeq \tmax$.  Indeed, in addition to the radiation bath produced by decays, there is also a bath produced gravitationally directly from the inflaton condensate \cite{Haque:2021mab,cmov}. Depending on $\trh$, this bath may achieve a higher temperature than the $\tmax$ obtained by decays. There are then (at least) three sources for particle production that can contribute to the relic density of dark matter: 1) freeze-in from the thermal bath
produced by inflaton decays; 2) the often neglected freeze-in from the gravitationally produced thermal bath; and 3) direct gravitational production from the inflaton condensate. We plan to compare these sources of particle production.

In typical non-instantaneous models of reheating, the inflaton
energy density is transferred to the Standard Model radiation bath via a Yukawa-
like coupling where the inflaton decay rate is given by~\footnote{For a recent review of inflaton decay in Starobinsky-like models, see \cite{Ema:2024sit}. } 
\begin{equation} 
\Gamma_{\phi} = \frac{y^2}{8 \pi} m_{\phi}\, .
\end{equation} 
Inflaton decay products then thermalize, and we will refer to this thermal bath of density $\rho_R^y$ as the  {\it decay bath}. We assume here that the thermalization is instantaneous compared to other time-scales, (see  \cite{therm,Harigaya:2013vwa,therm2,therm3,Mukaida:2015ria,Garcia:2018wtq,Drees:2021lbm,Passaglia:2021upk,Drees:2022vvn,Mukaida:2022bbo} for discussions of the thermalization process). In most cases of interest, the production of the decay bath leads to reheating and for a potential with a quadratic minimum, the reheating temperature is
\beq
\trh \propto (\Gamma_\phi M_P)^{\frac12} \propto y ({m_\phi}{M_P})^{\frac12} \, .
\eeq

In addition to inflaton decay, there is an unavoidable gravitational coupling between the inflaton and the Standard Model (SM). Specifically, gravitational scattering of the inflaton ($\phi \phi \rightarrow h_{\mu \nu} \rightarrow \text{SM fields}$, where $h_{\mu\nu}$ represents the graviton\footnote{This corresponds to an expansion around the flat Minkowski spacetime metric 
$g_{\mu \nu}=\eta_{\mu \nu}+\frac{2h_{\mu \nu}}{M_P}$.}) will generate another Standard Model radiation bath of density $\rho_R^h$, which we will refer to as the {\it gravitational bath}. It is produced soon after inflation ends, when the energy density stored in the inflaton condensate is still very high.  As we will see, once this bath is produced, its energy density will scale as
\begin{equation} 
\rho^{h}_{R} = \rhomax^{h} \left( \frac{\amax}{a} \right)^{4}\,, 
\label{rhoh} 
\end{equation} 
since its production rate is too weak to counterbalance its Hubble dilution.
While this is the typical scaling for radiation in an expanding universe,  it  
differs from the initial scaling of the decay bath 
\begin{equation}
\rho^{y}_{R} = \rhomax^{y} \left( \frac{\amax}{a} \right)^{\frac{3}{2}},
\end{equation} 
when the inflaton oscillates about a quadratic minimum, and \cite{GKMO2}
\beq
\rho_R^y=
\begin{cases}
\rhomax^y\left(\frac{\amax}{a}\right)^{\frac{6k-6}{k+2}}~~~~k<7\,,
\\
\rhomax^y\left(\frac{\amax}{a}\right)^4~~~~~~~~k>7
\,,
\end{cases}
\eeq
for a general potential $V(\phi)\propto \phi^k$ about its minimum.\footnote{Note that for large $k\ge 6$, the evolutionary behavior of $\rho^y_R$ is altered at late times due to the effective masses of the decay products. We do not include these effects here. See \cite{GKMO2} for more details.} The decay bath dilutes more slowly as radiation is continually introduced until the inflaton decay rate exceeds the Hubble rate when subsequently $\rho_R^y \propto a^{-4}$.

On the other hand, the gravitational production mechanism leads to a lower bound for the maximal temperature of order $10^{12}$ GeV \cite{cmov} 
and is larger than the maximum temperature of the decay bath when $y \lesssim 2 \times 10^{-6}$ (for the case of a quadratic potential).
Here, we consider the possible consequences for particle production when the gravitational bath dominates over the more conventional decay bath which eventually leads to reheating when it dominates over the energy density of the inflaton condensate\footnote{We note that in the context of minimal (Einstein) gravity with a quadratic inflaton potential, the gravitational bath does not lead to sufficient reheating \cite{cmov,Haque:2021mab,Clery:2022wib, Barman:2022qgt} and inflaton decays are required for reheating.}.

In what follows, we will first briefly review the production of the gravitational bath and compare it to the more standard decay bath in Section \ref{sec:gravbath}. We will then proceed in Section \ref{sec:FI} to set up the Boltzmann equations for the freeze-in production of dark matter (DM) and provide solutions for the production through the gravitational portal. 
The corresponding production from the thermal baths are
computed in Section \ref{sec:FIscat} for the case where $m_\chi < \trh$ and the production rates are characterized by an effective interaction with scale $\Lambda > \tmax^h$.
Since both $T^h_{\rm max}$ and $T^y_{\rm max}$ are in principle greater than the reheating temperature $\trh$, it is possible to produce particles with masses $m_\chi > \trh$. This will also be considered in Section \ref{sec:mgtrtrh}. Interesting results are found when
$\Lambda < \tmax^h$ and this case is set up in Section \ref{sec:FIscat3}.

Having established the formalism for each of the above cases, our results will be presented in Section \ref{sec:develop}.
In Section \ref{sec:developk2}, we concentrate on the results for a quadratic minimum, namely $k=2$. In Section \ref{sec:developkneq2},
we show some results for other values of $k$.
Our conclusions will be given in Section \ref{sec:conclusions}
where we also comment on the possibility of attaining thermal equilibrium in which case freeze-out is the dominant form of DM production. Finally we note other contexts for which the gravitational bath may be important to be discussed in more detail in future work. 

\section{Approach} \label{sec:appr}

\subsection{Production of the thermal baths}
\label{sec:gravbath}

For definiteness, we will consider a class of inflationary models known as T-models~\cite{Kallosh:2013hoa} as a specific example, 
\begin{equation}
    V(\phi) \; = \;\lambda M_P^{4}\left|\sqrt{6} \tanh \left(\frac{\phi}{\sqrt{6} M_P}\right)\right|^{k} \, .
\label{Vatt}
\end{equation}
For $k=2$, this potential is similar to the Starobinsky potential \cite{Staro}. 
When expanded about the minimum of the potential,
\begin{equation}
    \label{Eq:potmin}
    V(\phi)= \lambda M_P^4 \left(\frac{\phi}{M_{P}}\right)^k, \quad \phi \ll M_{P} \, .
\end{equation}
For $k=2$, we have a quadratic minimum and the inflaton mass is $m^2_\phi = 2 \lambda M_P^2$. We consider only even values of $k$.
Inflation ends when accelerated expansion stops. This occurs when $\dot \phi_{\rm end}^2 = V(\phi_{\rm end})$ where $\phi_{\rm end}$ is the inflaton field value when inflation ends and is approximately \cite{GKMO2}
\beq
\phi_{\rm end} \;\simeq\;\sqrt{\frac{3}{8}}\, M_P \ln\left[ \frac{1}{2} + \frac{k}{3}\left(k+\sqrt{k^2+3}\right) \right]\,.
\eeq
The inflation slow roll parameters are largely insensitive to $k$, however the normalization of the CMB spectrum does depend on $k$ with
\beq
\lambda \simeq \frac{18 \pi^2 A_s}{6^{k/2} N_*^2} \, ,
\eeq
where $A_s = 2.1\times 10^{-9}$ as determined by Planck \cite{Planck} and $N_*$ is the number of e-folds from the exit of the horizon scale with wave number $k_* = 0.05$~Mpc$^{-1}$, at $\phi=\phi_*$, to $\phi_{\rm end}$. Note that $N_*$ also depends on the reheating temperature and hence on $y$ \cite{Liddle:2003as,Martin:2010kz,egnov}.
We will ignore the subleading dependence on $y$ and fix $\lambda$ (and $\rho_{\rm end} = \rho_\phi(\phi_{\rm end}) = \frac32 V(\phi_{\rm end})$) for each value of $k$ considered. Following \cite{cmov}, we set
$y=10^{-7}$ as a base value, and for $k=2$, we have $\lambda=2.5 \times 10^{-11}$  and 
$\rho_{\rm end}^{1/4} = 5.2 \times 10^{15}$ GeV. For $k=4$, $\lambda=3.3\times 10^{-12}$ and $\rho_{\rm end}^{1/4}=4.8\times 10^{15}\,{\rm GeV}$  whereas for $k=6$, 
$\lambda=4.6\times 10^{-13}$ and $\rho_{\rm end}^{1/4}=4.6\times 10^{15}\,{\rm GeV}$. For more on the determination of these parameters, see \cite{GKMO2}. 

The gravitational bath is produced from the inflaton condensate 
transfer of energy through the exchange of a graviton. 
The energy density of this bath is found by solving the Boltzmann equation
\beq
\frac{d \rho^h_R}{dt} + 4 H \rho^h_R = N \frac{\rho_\phi^2 \omega}{16 \pi M_P^4}
\Sigma^h_k \, ,
\label{Eq:rhorsigma}
\eeq
where $N=4$ is the number of real scalars in the Standard Model.
Indeed, it is easy to see from conformal invariance arguments, that
the exchange of a graviton is unable to produce massless fermions or massless gauge bosons.
$\omega$ is the inflaton frequency of oscillation about its minimum \cite{GKMO2}
\beq
\label{eq:angfrequency}
\omega=m_\phi \sqrt{\frac{\pi k}{2(k-1)}}
\frac{\Gamma(\frac{1}{2}+\frac{1}{k})}{\Gamma(\frac{1}{k})},
\eeq
with $m_\phi(\phi) = \frac{\partial^2 V(\phi)}{\partial \phi^2}$.
For $k=2$, the frequency $\omega = m_\phi$ is independent of the field value. 
The Hubble parameter is given by
\beq
H^2 = \left( \frac{\dot{a}}{a} \right)^2 =  \frac{1}{3M_P^2}(\rho_\phi + \rho_R^h + \rho_R^y)\ ,
\eeq
and is heavily dominated by the inflaton density prior to reheating.
Finally, the factor
\beq
\Sigma^h_k=\sum_{n=2}^\infty n |{\cal P}_n^k|^2 \, ,
\label{Eq:sumhk}
\eeq
where the ${\cal P}^k_n$ are coefficients of the Fourier expansion of $V(\phi)$ in terms of $\rho_\phi$,
\beq
V(\phi)=V(\phi_0) \sum_{-\infty}^{+\infty}{\cal P}^k_ne^{-in\omega t}
=\rho_\phi\sum_{-\infty}^{+\infty}{\cal P}^k_ne^{-in\omega t}\,,
\label{Vexp}
\eeq
where $\omega$ is given by Eq.~(\ref{eq:angfrequency})
and $\phi_0$ is the envelope of $\phi(t)$.
From ${\cal P}^2_2=\frac{1}{4}$, one deduces 
$\Sigma^h_2 = \frac{1}{8}$.  For more details, see \cite{GKMO2,cmov}. 

The solution to Eq.~(\ref{Eq:rhorsigma}) is \cite{cmov}
\begin{eqnarray}
\rho^h_R&=&N\frac{\sqrt{3}M_P^4\gamma_k\Sigma^h_k}{16 \pi}
\left(\frac{\rhoend}{M_P^4}\right)^{\frac{2k-1}{k}}
\frac{k+2}{8k-14} \nonumber \\
&& \times \left[\left(\frac{\aend}{a}\right)^4
-\left(\frac{\aend}{a}\right)^{\frac{12k-6}{k+2}}\right] 
\, ,
\label{Eq:rhorsigma_sol}
\end{eqnarray}
where $\gamma_k = (\omega/m_\phi) \lambda^\frac1k \sqrt{k(k-1)}$.
We see that for $a \gg \aend$, $\rho_R^h$ scales as $a^{-4}$ as noted earlier. 
This radiation
density peaks at 
\beq
\frac{a^h_{\rm max}}{\aend} = \left(\frac{2k+4}{6k-3}\right)^{\frac{k+2}{8k-14}}\,,
\label{ahmax}
\eeq
which implies a maximum temperature,
\bea
\alpha \left(T^h_{\rm max}\right)^4 =\rho^h_{\rm max} & = & \frac{\sqrt{3}M_P^4\gamma_k\Sigma^h_k}{4 \pi}
\left(\frac{\rhoend}{M_P^4}\right)^{\frac{2k-1}{k}} \nonumber \\
&& \times \frac{k+2}{12k-6} \left(\frac{2k+4}{6k-3}\right)^{\frac{2k+4}{4k-7}} \, ,
\label{Eq:rhomaxbis}
\eea
where $\alpha = g \pi^2/30$ with $g = 427/4$ in the Standard Model.
For $k=2$, we have
\beq
\left.T_{\rm max}^h\right|_{k=2}\simeq 1.0 \times 10^{12} 
~{\rm GeV} \, ,
\label{Eq:tmax}
\eeq
where we used values of $\lambda$ and $\rhoend$ appropriate for $k=2$.

\begin{figure}[!ht]
\centering
\vskip .2in
\includegraphics[width=3.in]{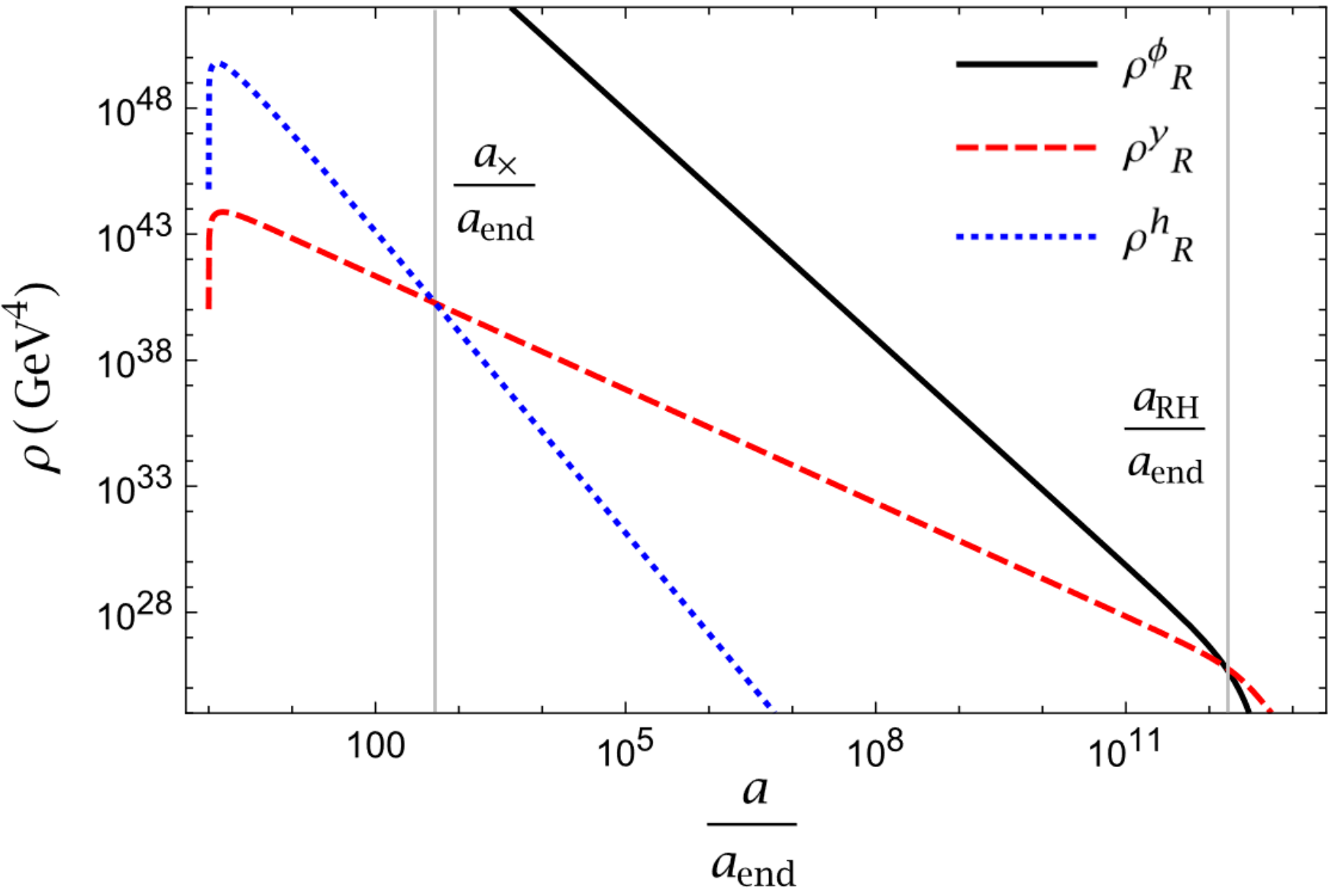}
\caption{\em \small Evolution of the inflaton density, $\rho_\phi$ (black-solid), the radiation density produced by inflaton decay, $\rho_R^y$ (red-dashed) and the radiation density produced by gravitational inflaton scattering, $\rho_R^h$ (blue-dotted) as a function of $a/a_{\rm end}$ for a Yukawa-like coupling $y=2 \times 10^{-9}$ with $\rhoend^\frac14 = 5.2 \times 10^{15}$~GeV (corresponding to $\trh \sim 10^6$~GeV) and $k=2$. The two vertical lines show the positions of $\across$ and $\arh$. 
}
\label{Fig:rhoR_num}
\end{figure}

The evolution of $\rho_R^h$ in Eq.~(\ref{Eq:rhorsigma_sol}) is shown in Fig.~\ref{Fig:rhoR_num} by the blue dotted curve where it is compared with the solution for the radiation bath produced by inflaton decay (red dashed curve) \cite{cmov},
\begin{eqnarray}
\rho^y_R&=& \frac{\sqrt{3}M_P^4\gamma_k^3 y^2 \Sigma^y_k}{8 \pi}
\left(\frac{\rhoend}{M_P^4}\right)^{\frac{k-1}{k}} \lambda^{-\frac2k}
\frac{k+2}{7-k} \nonumber \\
&& \times \left[
\left(\frac{\aend}{a}\right)^{\frac{6k-6}{k+2}} - \left(\frac{\aend}{a}\right)^4 \right]\, ,
\label{Eq:rhorsigma_sol2}
\end{eqnarray}
with\footnote{Note that the sum here begins at $n=1$ and not $n=2$ as in the case given in Eq.(\ref{Eq:sumhk}), because the first mode contributing to the decay process corresponds to the frequency $\omega$, whereas it is $2\omega$ for scattering.} 
\beq
\Sigma_k^y = \sum_{n=1}^\infty n^3 |\tilde{\cal P}_n^k|^2 \, ,
\label{Eq:sumyk}
\eeq
and where $\tilde {\cal P}_n^k$ are coefficients of the Fourier expansion of 
$\phi\sim \phi_0 \tilde {\cal P}$ which can account for the decay and are distinct from the coefficients of $V(\phi)$ present in Eq.~(\ref{Vexp}). 
For $k=2$, $\tilde {\cal P}_1^2=\frac12$, and $\Sigma_2^y = \frac14$.

The density
$\rho_R^y$ has a maximum at 
\beq
\frac{a^y_{\rm max}}{\aend} = 
\left(\frac{2k+4}{3k-3}\right)^{\frac{k+2}{14-2k}}
\label{aymax}
\eeq
and at the maximum
\begin{eqnarray}
\alpha \left(T^y_{\rm max}\right)^4 = \rho_{\rm max}^y & = & \frac{y^2 \gamma_k^3 \sqrt{3}}{16 \pi}
\lambda^{-\frac{2}{k}}
M_P^4 \left(\frac{\rhoend}{M_P^4}\right)^{1-\frac{1}{k}}
\nonumber \\
& & \times \left(\frac{3k-3}{2k+4}\right)^{\frac{3k-3}{7-k}}  \Sigma_k^y \, .
\end{eqnarray}  
For $k=2$,
\beq
\left. T^y_{\rm max}\right|_{k=2} \simeq 7.4\times 10^{14} \sqrt{y}~{\rm GeV}\,.
\label{Eq:tymax}
\eeq
Comparing Eqs.~(\ref{Eq:tmax}) and (\ref{Eq:tymax}), we see that for sufficiently small $y$, $T^y_{\rm max} < T^h_{\rm max}$.

\begin{figure}[!ht]
\centering
\vskip .2in
\includegraphics[width=3.in]{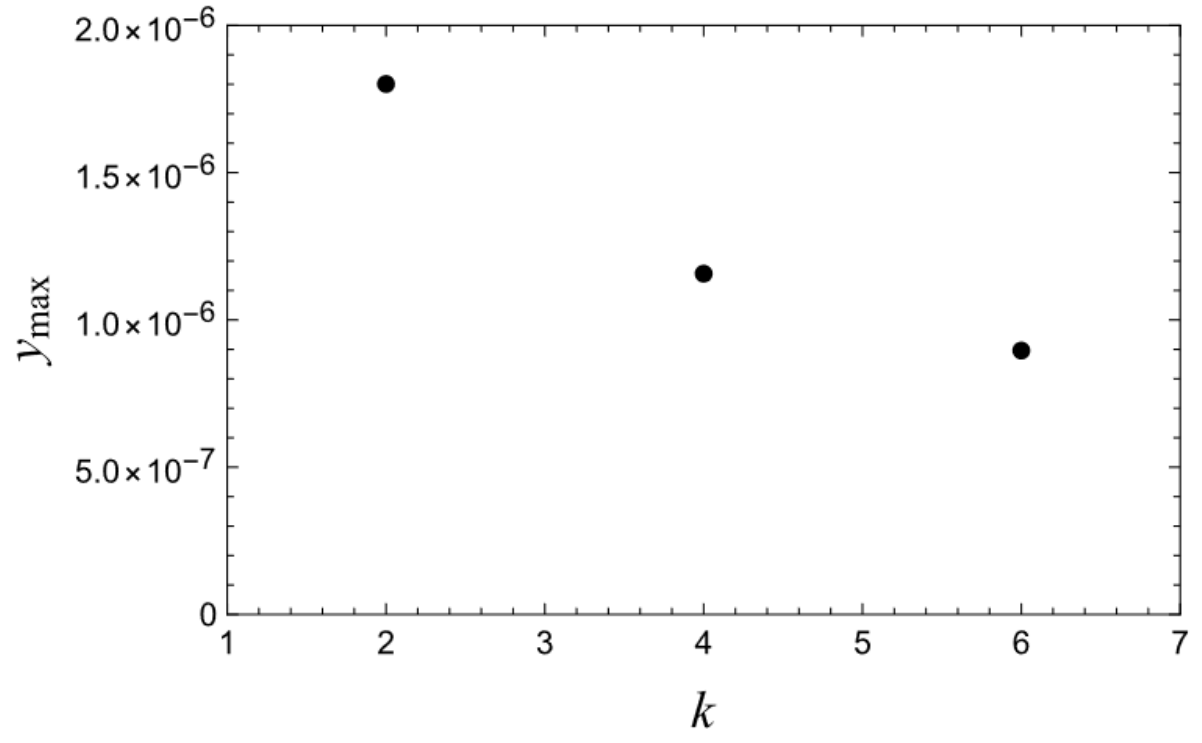}
\caption{\em \small The value of the decay coupling $y_{\rm max}$, such that the peak value of the the energy densities of the two radiation baths are equal, for three values of $k = 2, 4, 6$ and a potential $V(\phi)\propto \phi^k$.} 
\label{Fig:ymax}
\end{figure}

\begin{figure}[!ht]
\centering
\vskip .2in
\includegraphics[width=3.in]{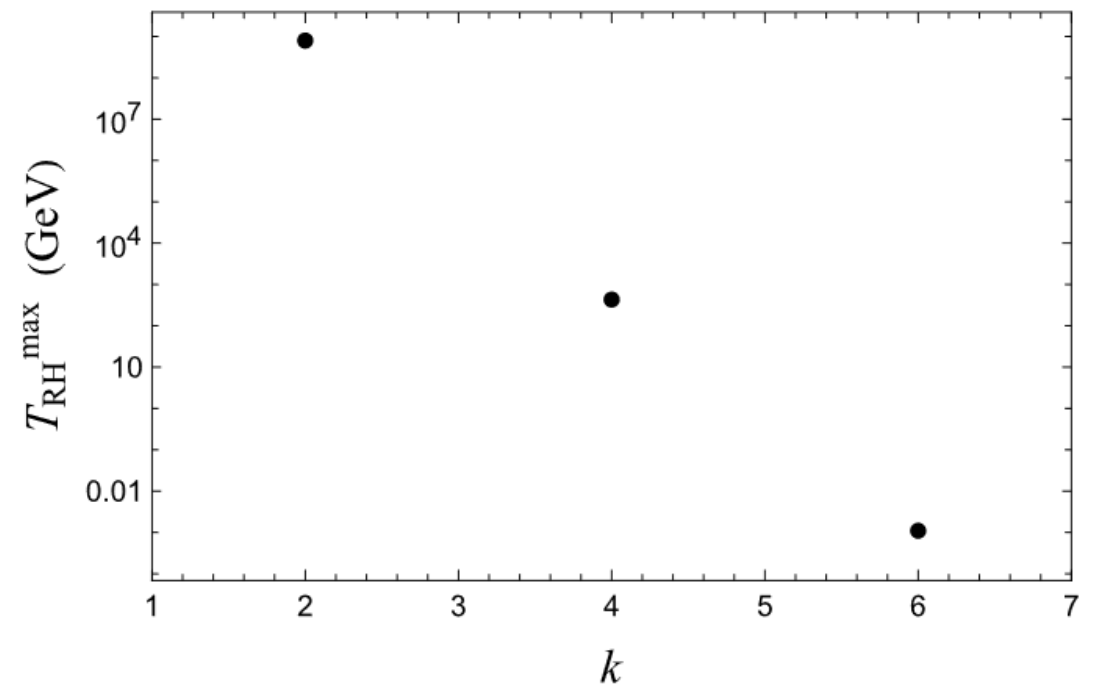}
\caption{\em \small The value of $\trh$, such that the peak value of the the energy densities of the two radiation baths are equal.  }
\label{Fig:TRHmax}
\end{figure}

We show in Fig.~\ref{Fig:ymax} the value of $y=y_{\rm max}$ such that the peak value of the the energy densities of the two radiation baths are equal for different values of $k$. This value of $y_{\rm max}$ corresponds to the maximum reheating temperature shown in Fig.~\ref{Fig:TRHmax}.
We see a strong dependence of $T_{\rm RH}^{\rm max}$ on $k$, 
and for any value of reheating temperature {\it below} 
$\trh^{\rm max}$ the thermal bath is dominated by the gravitational bath rather than the decay bath early in the reheating process.  For example,
when the reheating temperature is below $\sim 10^9$ GeV, a quadratic
potential generates a gravitational bath which dominates the overall radiation 
content. However, one needs to have a reheating temperature below $\sim 1$~MeV for $k=6$ for the gravitational bath to dominate. This comes from the fact that for
higher $k$, the dilution of $\phi$ is greater, which renders the scattering amplitude $\propto \phi^2$ much less efficient than the decay amplitude $\propto \phi$. The lack of efficiency needs to be compensated by a weaker decay coupling if there is a period where the radiation density is dominated by the gravitational bath. In any case, 
for $\trh < \trh^{\rm max}$ and $k<7$, the gravitational bath dominates only temporarily as seen in Fig.~\ref{Fig:rhoR_num}. 
Since $\rho_R^h\propto a^{-4}$, and $\rho_R^y\propto a^{-\frac{6k-6}{k+2}}$ eventually the decay bath begins to dominate. 
On the other hand as discussed in more detail below, for $k>7$, as both bath redshift as $a^{-4}$, reheating is completely determined by $\rho_R^h$ if $T_{\rm max}^y < T_{\rm max}^h$.

\begin{figure}[!ht]
\centering
\vskip .2in
\includegraphics[width=3.in]{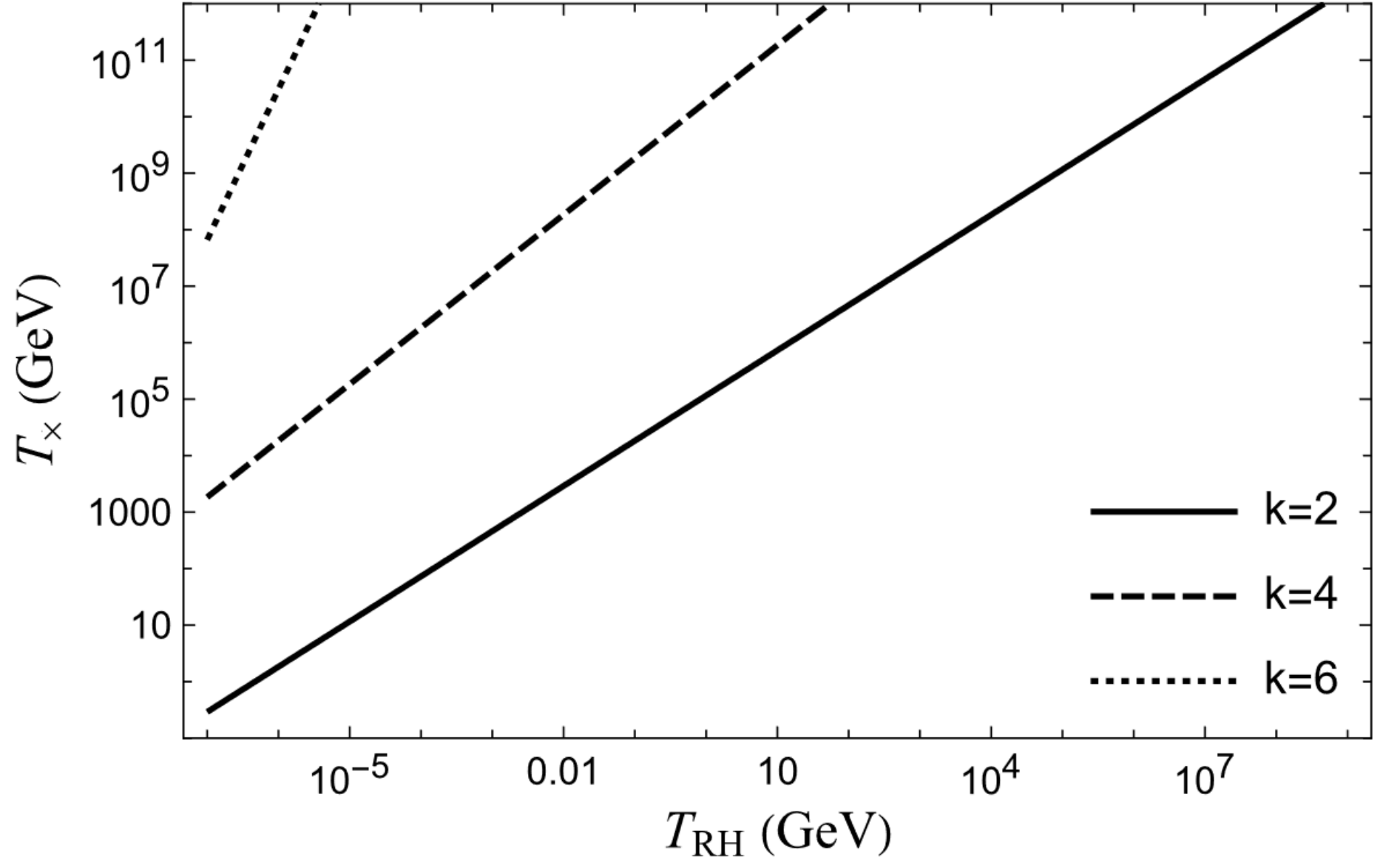}
\caption{\em \small $T_{\times}$ as a function of $T_{RH}$ for $k=2,4,6$. $T_{\times}$ will never exceed $10^{12}$ GeV, since this is the peak value of $T^{h}_{max}$.
}
\label{Fig:Tcross}
\end{figure}

An important point in the evolution of $\rho_R^h$ and $\rho_R^y$ 
is the "crossing" scale factor $\across$ defined when
$\rho^h(\across)=\rho^y(\across)$. 
This point is represented in Fig.~\ref{Fig:rhoR_num} at the intersection of the blue-dotted and red-dashed lines.  For $k\leq 6$, from 
\beq
\rho_R^h = \rho^h_{\rm max}\left(\frac{\amax}{a}\right)^4\,, ~~ {\rm and} ~~ \rho_R^y = \rhorh\left(\frac{\arh}{a}\right)^{\frac{6k-6}{k+2}}\,,
\eeq
one deduces 
\beq
\frac{\across}{\aend}=\left(\frac{\rho_{\rm max}^h}{\rhoend^{1-\frac{1}{k}}\rhorh^\frac{1}{k}}\right)^{\frac{k+2}{14-2k}}\,.
\label{Eq:across}
\eeq
This corresponds to a density
\begin{equation}
\alpha T_{\cross}^4 = \rho_{\cross} = (\rho^{h}_{\rm max})^{\frac{3-3k}{7-k}}\left(\rho_{\rm end}^{1-\frac{1}{k}}\rho_{\rm RH}^{\frac{1}{k}}\right)^{\frac{2k+4}{7-k}}
\label{Eq:rhocross}
\end{equation}
which gives 
\beq
\tcross \simeq
1.2 \times 10^5 \left(\frac{\trh}{1~\rm{GeV}}\right)^\frac{4}{5}~\rm{GeV}\,,
\label{Eq:tcross}
\eeq
for $k=2$.
The temperature $\tcross$ at which $\rho^h(\across) =\rho^y(\across)$ is shown in 
Fig.~\ref{Fig:Tcross} as a function of the reheating temperature, for three values of $k=2$, 4 and 6.  We clearly see 
that for $k=6$, the gravitational bath never dominates for any allowable $\trh > 4$~MeV. For $k=4$, the gravitational bath can dominate only if $\trh \lesssim 100$ GeV.
Note that, as expected, there is no solution to Eq.(\ref{Eq:across})
for $k\ge7$, because $\rho_R^h$
and $\rho_R^y$ both redshift $\propto a^{-4}$. 

The inflaton density for $\aend \ll a \ll \arh$ is
\beq
\rho_\phi(a) = \rho_{\rm end} \left(\frac{a}{a_{\rm end}} \right)^{-\frac{6k}{k+2}}  \, ,
\label{Eq:rhophi}
\eeq
and is also shown in Fig.~\ref{Fig:rhoR_num} by the black solid line.
It redshifts faster than the gravitational bath with $\rho_R^h \sim a^{-4}$ for $k>4$, allowing for the possibility of pure gravitational reheating.
As one can see for $k=2$, $\rho_R^y \sim a^{-3/2}$ whereas $\rho_\phi \sim a^{-3}$ (until $\Gamma_\phi \sim H$), allowing for  reheating through the decay bath. 
As an illustration, in Fig.~\ref{Fig:rhoR_num}, we have chosen, $y = 2 \times 10^{-9}$ corresponding to a reheating temperature $\trh \simeq 10^6$~GeV at $a = \arh$, defined by $\rho_\phi(\arh) = \rho^y_R(\arh)=\rhorh$.

    \subsection{Freeze-In Production through the Gravitational Portal}
    \label{sec:FI}

The production of dark matter through freeze-in during reheating has (at least) three sources. Since all matter is coupled to gravity, the scattering of the inflaton condensate though the gravitational portal \cite{MO,cmov} 
will produce all types of matter including dark matter. In addition, scattering from either of the two radiation baths ($\rho_R^h$ and $\rho_R^y$) 
will also lead to the production of dark matter if the dark matter is coupled to Standard Model particles. Even though the gravitational production
through inflaton scattering is suppressed by a factor $M_P^{-4}$, the density of energy stored in the condensate at the end of inflation, $\rhoend$, is sufficient
to compensate for this suppression. 
This is because the scattering rates within the thermal bath are proportional to $(\rho_R^h)^2$ and 
$(\rho_R^y)^2 \ll \rho_\phi^2$.
In the following,
we compute the relic density of dark matter (both scalar and fermionic) arising from freeze-in for each of the three aforementioned sources.

  If there is an interaction between a dark matter particle $\chi$
  and the Standard Model whose rate, $\Gamma$ exceeds the expansion rate, $H$, then $\chi$ enters into thermal equilibrium with the Standard Model bath and its relic density is given
 by the WIMP paradigm \cite{Arcadi:2017kky,Arcadi:2024ukq}.  
 We will comment on this possibility in Section \ref{sec:conclusions}. However, in many cases, the interaction is too weak for thermalization to occur. 
 In this case, particle production occurs through the so-called freeze-in mechanism \cite{fimp,Bernal:2017kxu}. 
Examples of this include interactions with heavy mediators like a $Z'$ from SO(10)-motivated scenarios \cite{Mambrini:2013iaa}, or moduli field models \cite{Chowdhury:2018tzw}.
 The number density of DM particles $\chi$ in the early universe depends on the hierarchy between 
 its production rate $R_{\chi}$ (from either inflaton scattering, or scattering within the thermal bath) and the Hubble parameter $H$.
 In this case, the DM number density $n_{\chi}$ obeys the classical Boltzmann equation \begin{equation}
    \frac{dn_{\chi}}{dt} + 3 H n_{\chi} = R^i_{\chi} \, ,
    \end{equation} 
    where $R^i_{\chi}$ is the production rate per unit time per unit volume for either the gravitational portal ($i=\phi^k$) or 
    one of the two radiation baths ($i=h,~y$)
    for which the rates are obviously equal as function of the temperature (but not as function of the scale factor $a$ of course). 
     It is often more convenient to write this equation in terms of the dynamical variable $a$ rather than the time $t$ as follows:
    \begin{equation}
    \frac{dn_{\chi}}{da} + 3 \frac{n_{\chi}}{a} = \frac{R^i_{\chi}(a)}{H a}.
    \end{equation} 
    We may then define the co-moving number density $Y_{\chi} = n a^3$, which gives
    \begin{equation} 
    \frac{dY_{\chi}}{da} = \frac{a^{2} R^i_{\chi}(a)}{H}.
    \label{dY/da} 
    \end{equation} 
    Furthermore, if we assume that $H$ is dominated by the inflaton
    during the entire DM production process, 
    \beq
    H = \sqrt{\frac{\rhorh}{3 M_P^2}} \left(\frac{\arh}{a}\right)^{\frac{3k}{k+2}}\,,
    \eeq
and Eq.(\ref{dY/da}) becomes
    \beq
\frac{dY_X}{da}=\frac{\sqrt{3}M_P}{\sqrt{\rhorh}}a^2\left(\frac{a}{a_{\rm RH}}\right)^{\frac{3k}{k+2}}R^i_\chi(a) \, .
\label{Eq:boltzmann4}
\eeq

The gravitational portal production is largely model independent and depends primarily on the inflaton energy density at the end of inflation. 
This comes from the fact that the (gravitational) production rate decreases rapidly with $a$, and is only efficient at the very beginning of the reheating process. 
Moreover, 
the rate for the production of scalar or fermionic particles differs in an important way.
    The scalar production rate is 
    \beq
\label{Eq:ratephi0}
R^{\phi^k}_0=\frac{2 \times \rho_\phi^2}{16 \pi M_P^4} \Sigma_0^k \, ,
\eeq
where the factor of two accounts for the fact we produce two dark matter particles per scattering, with
\begin{equation}
   \Sigma_0^k = \sum_{n = 1}^{\infty}  |{\cal P}^k_n|^2\left[1+\frac{2m^2_\chi}{E_n^2}\right]^2 
\sqrt{1-\frac{4m_\chi^2}{E_n^2}} \, ,
\label{Sigma0k}
\end{equation}
where $E_n = n \omega$ is the energy of the $n$-th inflaton oscillation mode and $m_\chi$ is the mass of the produced dark matter.
For fermionic dark matter, the rate is
\beq
\label{eq:rateferm}
R^{\phi^k}_{1/2}=\frac{2 \times \rho_\phi^2}{4 \pi M_P^4}
\frac{m_\chi^2}{m_\phi^2}
\Sigma_{1/2}^k \, ,
\eeq
where the factor of two accounts for the  sum over the particle and antiparticle final states, with\footnote{In this work we consider dark matter candidates that are either
real scalars or a Dirac fermion.}
\begin{equation}
    \label{eq:ratefermion2}
    \Sigma_{1/2}^k = \sum_{n=1}^{+\infty} |{\cal P}_n^k|^2
    \frac{m_\phi^2}{E_n^2}
\left[1-\frac{4 m_\chi^2}{E_n^2}\right]^{3/2} \, .
\end{equation}
In Eqs.~(\ref{Sigma0k}) and (\ref{eq:ratefermion2}) the Fourier coefficients are given by Eq.~(\ref{Vexp}).
For $k=2$, $\Sigma_0^k = \frac{1}{16}$ and $\Sigma_{1/2}^k=\frac{1}{64}$. 
Note that the gravitational production of fermions is helicity suppressed and carries a factor of $(m_\chi/m_\phi)^2$. While the 
production rate is highly peaked at $a \approx \aend$, 
the final relic density will depend on the subsequent evolution of the Universe and as such will depend on the reheating temperature. 

It is now straightforward to obtain the relic density of $\chi$ 
at $\arh$ by integrating Eq.~(\ref{Eq:boltzmann4}) between $\aend$
and $\arh$. We consider first
$R_\chi^i=R_0^{\phi^k}$ or $R_{1/2}^{\phi^k}$,
noting that the rates depend on $a$ through $\rho_\phi$.\footnote{For the production of fermions when $k>2$, there is an additional dependence on the scale factor through $m_\phi$ which is $\phi$-dependent and hence $a$-dependent.} 
The resulting density evaluated at $\arh \gg \aend$ for scalars is 
\beq
n_0^\phi(a_{\rm RH})\simeq
\frac{\sqrt{3}\rhorh^{3/2}}{8 \pi M_P^3}
\frac{k+2}{6k-6}
\left(\frac{\rhoend}{\rhorh}\right)^{1-\frac{1}{k}}\Sigma^k_0\,.
\label{n0phi}
\eeq
The fraction of the present critical density is then 
easily obtained from $n(\arh)$ \cite{mybook}
\beq
\frac{\Omega_\chi h^2}{0.12} = 4.9 \times 10^7~
\frac{n_\chi(\arh)}{\trh^3} \frac{m_\chi}{{\rm GeV}}\,,
\label{Omega}
\eeq
so that for scalars and for $k=2$,
\bea
\frac{\Omega_0^\phi h^2}{0.12} &\simeq& 1.1 \left(\frac{\trh}{10^{10}~{\rm GeV}}\right) 
\left( \frac{m_\chi}{10^7 ~{\rm GeV}}\right) \,.
\label{Eq:omega0k2}
\eea
Similarly, for fermions, 
\beq
n_{1/2}^\phi(a_{\rm RH})\simeq 
\frac{m_\chi^2\sqrt{3}(k+2)\rhorh^{\frac{1}{2}+\frac{2}{k}}}{12 \pi k(k-1)\lambda^{\frac{2}{k}}M_P^{1+\frac{8}{k}}}
\left(\frac{\rhoend}{\rhorh}\right)^{\frac{1}{k}}
\Sigma_{\frac{1}{2}}^k\,,
\label{Eq:nhalf}
\eeq
where we used
\beq
m_\phi^2=V''(\phi_0)=k(k-1)\lambda^{\frac{2}{k}}M_P^2
\left(\frac{\rho_\phi}{M_P^4}\right)^{1-\frac{2}{k}}.
\eeq
Then, the fraction of critical density for fermions for $k=2$ is
\bea
\frac{\Omega_{1/2}^\phi h^2}{ 0.12} &=&
3.5\times 10^{-14} \left(\frac{\trh}{10^{10}~{\rm GeV}}\right)
\left(\frac{m_\chi}{ 10^{7} ~{\rm GeV}}\right)^3   \, .
\label{Eq:omegaphihalfk2}
\eea
General expressions for the relic density of bosons and fermions in terms of $k$ are given in the Appendix in Eqs.~(\ref{Eq:omega0}) and (\ref{Eq:omegaphihalf}).

\subsection{Freeze-in Production through Thermal Scattering for $m_\chi < \trh$ and $\Lambda > \tmax$}
\label{sec:FIscat}

 Next, we consider DM production processes
 from scattering within the radiation bath.
Of course the thermal production rate from either the gravitational 
bath, $\rho_R^h$,
or the decay bath, $\rho_R^y$, as function of the temperature $T$ 
are the same and we 
 parametrize these rates as 
 \begin{equation}
 R_{\chi}^{h,y} \propto \frac{T^{n+6}}{\Lambda^{2+n}},
 \label{R(T)} 
 \end{equation} 
 where $\Lambda$ is some high energy scale ($\Lambda \geq \tmax$) that suppresses the interaction. 
 
Some comments on this parametrization are in order. If there is a contact interaction between scalars in the thermal bath and a scalar dark matter candidate $\chi$, with coupling $\kappa/4$,
the thermally averaged rate is found to be $\frac{\kappa^2}{2304 \pi} T^4 \simeq 10^{-4} \kappa^2 T^4$ \cite{cmov}
assuming only a SM Higgs in the thermal bath.
This corresponds to $n = -2$ and we can parameterize the rate by the thermal coupling
$\sigma^2 = \frac{\kappa^2}{2304 \pi}$, with 
$R_\chi^{h,y}(T)=\sigma^2 T^4$. Renormalizablity probably requires $\sigma \lesssim 0.1$. A contact interaction between SM scalars and DM fermions is necessarily an effective interaction corresponding to $n=0$. 
Other possible (renormalizable) interactions 
to consider are SM scalars producing DM scalars mediated by vector exchange (SVS). 
If the mass scale of the mediator is related to $\Lambda$, then for $T \gg \Lambda$, we obtain a rate with $n=-2$ (as in the contact interaction), while for $T \ll \Lambda$, 
$n=2$. Noting the change in $n$ will be particularly important if values of $\Lambda$
lie between $T_{\rm max}$ and $\trh$. This will be discussed in subsection \ref{sec:FIscat3}.
Scalar to scalar interactions mediated by scalars (SSS) produce rates with $n=-6$ for $T \gg \Lambda$ (in this case the rate is proportional to the mass scale associated with the scalar vertex to the 4th power), and with $n=-2$ for $T\ll \Lambda$. 
Other interactions with fermionic DM (SVF or FVF or FSF) also lead to $n=-2$ for $T \gg \Lambda$ and $n=2$ when $T \ll \Lambda$. 
Scalars producing fermions via scalar exchange (SSF) lead to $n=-4$ for $T \gg \Lambda$ and $n=0$ for $T \ll \Lambda$.
Of course other non-renormalizable interactions can be envisioned which lead to 
larger values of $n$. All factors coming from thermal averaging are assumed to be included in the scale $\Lambda$.

Since the production rates of the two baths
are defined in terms of temperature, as a function of the scale factor  (used to integrate the Boltzmann equation (\ref{Eq:boltzmann4})), the rates for the 
gravitational and decay baths
 differ. 
For the gravitationally produced SM radiation bath, we know that  $\rho^{h}_{R} = \alpha T_{h}^{4}$, we can use (\ref{Eq:rhorsigma_sol}) to find $T_{h} \propto a^{-1}$ for $a \gg \aend$. In contrast, for the decay bath, $\rho^{y}_{R} = \alpha T_{y}^{4}$, and from Eq.~(\ref{Eq:rhorsigma_sol2}) $T_{y} \propto a^{-\frac{3(k-1)}{2(k+2)}}$ for $k<7$ and $T_{y} \propto a^{-1}$ for $k>7$.
For large values of $k > 7$, if $y<y_{\rm max}$ such that $\rho^y_{\rm max} < \rho^h_{\max}$ the Universe will not reheat from inflaton decays
(because both baths redshift as $a^{-4}$), however, reheating does occur from scattering when $\rho_R^h = \rho_\phi$. However, unless $k\ge 10$, the reheating temperature is too low, and is conflict with BBN
and even for higher $k$ (which implies a larger $\trh$), 
there is a conflict from BBN from the production of gravitational waves \cite{Barman:2022qgt}.
This constraint applies because modes of gravitational waves entering the horizon during 
a period dominated by a stiff equation of state ($k>4$) before $\trh$,
can easily exceed the energy density of the radiation at the BBN epoch \cite{Yeh:2022heq}.
For this reason, we will restrict our attention to $k = 2, 4$, and 6.\footnote{Higher values are allowed if $y$ is sufficiently large so that reheating occurs via inflaton decays (and avoids the effects of fragmentation \cite{Garcia:2023dyf}), but in this case the effects of the gravitational bath are irrelevant.}

Combining Eqs.(\ref{Eq:boltzmann4}) and (\ref{R(T)}), the Boltzmann equation for the production of dark matter for each radiation bath is then
\begin{align} 
\frac{dY_{\chi}}{da} =&  
 \frac{\sqrt{3} M_P}{\sqrt{\rhorh}} \frac{T^{n+6}(a)}{\Lambda^{n+2}} a^{2}\left( \frac{a}{\arh}\right)^{\frac{3k}{k+2}} \, ,
\end{align} 
which must be 
integrated from $\aend$ to $\arh$. 
To obtain $n_\chi$ from each bath we replace $T(a)$ from either Eq.~(\ref{Eq:rhorsigma_sol}) for $n^h_\chi$ or from Eq.~(\ref{Eq:rhorsigma_sol2}) for $n^y_\chi$. Clearly a full analytical solution is not viable. However, if we make the approximation $T(a) = T^h_{\rm max} (a_{\rm max}/a)$, we have 
    \begin{equation} 
    \begin{split} 
    n_{\chi}^{h}(\arh)= & \left[ \frac{\sqrt{3} M_{P} \left(T^h_{\rm max}\right)^{n+6}}{\sqrt{\rhorh} ~\Lambda^{2+n} }\right]\left( \frac{k+2}{nk+2n+6}\right) \\ & \times \left(\frac{a_{\rm max}}{\arh} \right)^{\frac{6(k+1)}{k+2}} \left[  1 -\left( \frac{a_{\rm max}}{\arh}\right)^{\frac{nk+2n+6}{k+2}} \right] . 
    \end{split} 
    \label{gravNum} 
    \end{equation} 
    Since  $a_{\rm max} \ll \arh$, the first term in the final brackets will dominate for $n > -\frac{6}{k+2} = n_c^h$, while for $n < n_c^h$, the second term in the final brackets will dominate. In this case, the number  density redshifts independently of $k$ as it is $\propto (a_{\rm max} / \arh)^{n+6}$. In particular, for $k=2$, the transition occurs at  $n^h_c =  -\frac{3}{2}$.
    In all of the examples discussed above with rates determined by a mediator with mass $\mathcal{O}(\Lambda)$ for both fermionic and scalar dark matter, $n < n^h_c$ when $\Lambda \gg T^h_{\rm max}$.

    The relic density derived from Eq.~(\ref{gravNum})
    for $k=2$
    is 
    \beq
    \frac{\Omega_{\chi}^h h^2}{ 0.12} \simeq \frac{0.8 \times 10^{6+12n}}{(3+2n) \Lambda^{n+2}} \trh m_\chi \, ,
    \label{Ohngtnck2}
    \eeq
    for $n > n_c^h$. Units for dimensionful parameters are all in GeV. 
    The more general expression for arbitrary $k$ is given in the Appendix in Eq.~(\ref{Ohngtnc}).
    For $n < n_c^h$ the relic density is dominated at low temperatures where the decay bath is dominant (particularly if $\trh$ is larger than the BBN limit). However when $m_\chi > \trh$
    the gravitational bath can be significant and this case is treated in the next subsection.

     The number density from the decay bath can be determined in a similar way, giving us (for $n \neq \frac{10-2k}{k-1}$) 
     \begin{equation} 
     \begin{split} 
     n_{\chi}^{y}(&a_{\rm RH}) =  \left[\frac{ \sqrt{3} M_{P}\left(T^y_{\rm max}\right)^{n+6}}{\sqrt{\rhorh}~\Lambda^{2+n} }\right] \left(\frac{2k+4}{3n-3nk+30-6k}\right)\\
 & \times  \left(\frac{a_{\rm max}}{a_{\rm RH}}\right)^\frac{(k-1)(3n+18)}{2k+4}
 \left[1-\left(\frac{a_{\rm max}}{\arh}\right)^\frac{3n-3nk-6k+30}{2k+4}\right]
     \label{nychiarh}
     \end{split} 
     \end{equation}
 for $k<7$, 
 and in this case the 1st term in the final brackets dominates when $n < n^y_c = \frac{10-2k}{k-1}$. For $n=n^y_c$, the integration must be redone and a logarithmic dependence on $T_{\rm max}/\trh$ will appear. 
 For larger $k$, $\rho^y_R \propto a^{-4}$, and
$n_\chi^y$ is given by Eq.~(\ref{gravNum}) with $T^h\to T^y$.
Note that in both Eqs.~(\ref{gravNum}) and (\ref{nychiarh}), the fraction $(a_{\rm max}/\arh)$ can be replaced using 
\beq
\frac{a_{\rm max}}{\arh} =\left(\frac{a_{\rm max}}{a_{\rm end}}\right)\left(\frac{\alpha \trh^4}{\rhoend}\right)^\frac{k+2}{6k} \, ,
\label{aearh}
\eeq
where the ratio $\frac{a_{\rm max}}{a_{\rm end}}$ is distinct for the two baths and they are given by Eqs. (\ref{ahmax}) and (\ref{aymax}).

The present relic density can then be obtained using Eq.~(\ref{Omega}). The general expressions for the relic density are given in the Appendix in Eqs.~(\ref{ohnltncy})-(\ref{ohngtncy}) for $n<n_c^y$, $n=n_c^y$, and $n>n_c^y$. For $k=2$, and $n<n_c^y$, Eq.~(\ref{ohnltncy}) is
\beq
\frac{\Omega_{\chi}^y h^2}{ 0.12} \simeq \frac{9.3 \times 10^{25}}{6-n}\frac{\trh^{n+1} m_\chi}{\Lambda^{n+2}} \, .
\label{ohnltncyk2}
\eeq

However, we need to be careful to account for any  significant DM production that may occur at temperatures below $\trh$. To determine whether 
production after reheating is significant compared to the production during reheating, we simply need to integrate the Boltzmann equation from 
$\arh$ to $a_{m}$ where $a_{m}$ is the scale factor at which the temperature of the radiation bath is $T = m_{\chi}$. Note that this will 
only be necessary when $m_{\chi}<\trh$. Upon integration, and using the fact that the Hubble parameter is now dominated by the radiation density, we find for $n< \frac{10-2k}{k-1}$ 
\begin{align} 
\label{nam}
n&^{y}_{\chi}(a_{m}) = \frac{\sqrt{3} M_{P}
\left(T^y_{\rm max}\right)^{n+6}}{\sqrt{\rho_{\rm RH}}\Lambda^{2+n} } \left(\frac{a_{\rm max}}{a_{\rm RH}}\right)^{\frac{(k-1)(3n+18)}{2k+4}}\left(\frac{a_{\rm RH}}{a_{m}}\right)^{3} 
\nonumber 
\\\times & \left[\left(\frac{2k+4}{3n-3nk+30-6k}\right) 
 +\left(\frac{1}{1+n}\right)\left(1-\left(\frac{a_{\rm RH}}{a_{m}}\right)^{n+1}\right)\right]  \\
 =& 
 \frac{\sqrt{3} M_{P}\trh^{n+1} m_{\chi}^{3}}{\sqrt{\alpha} \Lambda^{n+2}}\left[\left(\frac{2k+4}{3n-3nk+30-6k}\right) \right. \nonumber \\ 
& \left. +\left(\frac{1}{1+n}\right)\left(1-\left(\frac{m_{\chi}}{\trh}\right)^{n+1}\right)\right] \nonumber
\end{align} 
The first term in large brackets corresponds to production prior to reheating, while the second term corresponds to production after reheating. Thus, we find that when $n < -1$ there is significant production after $\trh$ and we must perform the integration all the way down to $a_{m}$. However, for $n > -1$, the production after $\trh$ is negligible compared to production during reheating.\footnote{To obtain the second equality, we used $\tmax = \alpha^\frac{1-k}{4k} \trh^\frac1k \rhoend^\frac{k-1}{4k} \left(\frac{(3k-3)}{(2k+4}\right)^\frac{3k-3}{4(7-k)}$.}
This can be understood from the fact that for larger $n$, 
the production at the beginning of process (at $\trh$ for $a>\arh$)
dominates over the late time production, at $a_m$.
Therefore, the number density at $\arh$ displayed in Eq.~(\ref{nychiarh}) is sufficient to determine the relic density when $n =0,2$, or $6$. We will use $a = a_{m}$ as the lower limit of integration for $n=-2$ and $m_{\chi} < \trh$.
Note that for $n > \frac{10-2k}{k-1}$ the result is given by Eq. \ref{nychiarh}, with a density $n^y_\chi\propto 
\left(\frac{\aend}{\arh}\right)^{\frac{6k+6}{k+2}}$, and is independent of $n$, because it is produced in the UV.\footnote{We will refer to energy scales at or near $\tmax$ as
the UV, and energy scales at or below $\trh$ as the IR.}

The relic density in this case is given in Eq.~(\ref{pastrh}) and for $k=2$ is
\begin{align}
\frac{\Omega^{y}_{\chi}h^{2}}{0.12} &=
3.5 \times 10^{25} \frac{\trh^{n+1} m_{\chi}}{ \Lambda^{n+2}} \nonumber \\ 
\times &\left[\left(\frac{8}{18-3n}\right)  +\left(\frac{1}{1+n}\right)\left(1-\left(\frac{m_{\chi}}{\trh}\right)^{n+1}\right)\right] 
\label{pastrhk2}
\end{align}

\subsection{Freeze-In Production for $m_{\chi} > \trh$, and $\Lambda >\tmax^h$}
\label{sec:mgtrtrh}

All of the expressions above for the DM number density produced by the gravitational bath and the decay bath hold for $\trh>m_{\chi}$. In the 
case where the dark matter mass is greater than the reheating temperature, 
the DM number density at $\trh$ may or may not change depending on $n$ and $k$. Interestingly, the DM number density at reheating for the 
gravitational bath does not change when $m_{\chi}>\trh$ provided $n > n_c^h$, or $n>-2$ for $k\geq 2$. 
 This corresponds to the case when the lower bound of integration, $m_\chi$, does not affect the total number of particles 
produced, as one can see from Eq.~(\ref{gravNum}), replacing $\arh$
by $a_m$ in the final set of brackets.
However, we have seen that there are examples of 
interactions where effectively,  $n=-2, -4$, and $-6$, and in these cases the number density if affected when $m_{\chi} > \trh$, 
because the lower limit of integration dominates Eq.~(\ref{gravNum}). It is these cases to which we now turn our attention.

To compute the DM number density at $\trh$ for masses larger than the reheating temperature, we proceed similarly as above. However, instead of 
integrating the Boltzmann equations from $\aend$ to $\arh$, we integrate from $\aend$ to $a_{m}$,  
and then compute the subsequent dilution from $a_{m}$ to $\arh$,
which depends on the value of $k$. On the other hand, direct 
production of dark matter from the inflaton condensate is unaffected as long as $m_\chi < m_\phi$. At still larger masses, the production of dark 
matter is kinematically forbidden. 

We write the resulting number density for the production from the gravitational bath
in terms of the scale factors, keeping only the dominant term when 
$n < n_c^h$, or $nk+2n+6 <0$,
\begin{align} 
n^{h}_{\chi}(a_{RH})&=-\left[\frac{\sqrt{3}M_{P}\left(T^h_{\rm max}\right)^{n+6}}{\sqrt{\rhorh}\Lambda^{2+n}}\right]  \left(\frac{k+2}{nk+2n+6}\right) \nonumber \\ 
& \times \left(\frac{a_{\rm max}}{\arh} \right)^{\frac{6(k+1)}{k+2}} \left(\frac{a_{\rm max}}{a_{m}}\right)^{\frac{nk+2n+6}{k+2}}
\label{nhbigm}\, .
\end{align}
The dependence on the scale factors can be removed using
Eq.~(\ref{aearh}) and $\frac{a_{\rm max}}{a_{m}}=\frac{m_{\chi}}{\tmax^{h}}$.
 As we already remarked, for $n > n_c^h$, the result given in Eq.~(\ref{gravNum}) remains valid even when 
$m_{\chi} > \trh$.\footnote{Strictly speaking Eq.~(\ref{nhbigm}) is valid when $m_\chi > \tcross$, when all of the DM is produced solely from 
the gravitational bath}.

The relic density from the gravitational bath when $m_\chi > \trh$ and $n< n_c^h$ is given in general in Eq.~(\ref{ohnltnc})
and for $k=2$ is
 \beq
    \frac{\Omega_{\chi}^h h^2}{ 0.12} \simeq \frac{8.4 \times 10^{-13}}{(-3-2n) \Lambda^{n+2}} \trh m_\chi^{n+\frac52} \, .
    \label{ohnltnck2}
    \eeq
 Once again, units for dimensionful parameters are all in GeV.

Concerning the decay bath, the corresponding general results for the DM number density at $\trh$ when $m_{\chi}>\trh$
is given by \cite{GKMO1}
\begin{align}
 n_{\chi}^{y}(a_{RH}) &=  \left[\frac{ \sqrt{3} M_{P}\left(T^y_{\rm max}\right)^{n+6}}{ \sqrt{\rhorh}\Lambda^{2+n}}\right] \left(\frac{2k+4}{\!3n\!-\!3nk\!+\!30\!-\!6k}\right) \nonumber \\
     & \times \left(\frac{a_{\rm max}}{\arh} \right)^{\frac{6(k+1)}{k+2}} \left(\frac{a_{\rm max}}{a_m}\right)^{\frac{3nk-3n+6k-30}{2k+4}} \,, 
\label{nlt2}
\end{align}
for $n< n_c^y$, or $nk-n+2k-10 < 0$,
where the first (IR) term of Eq.~(\ref{nychiarh})
dominates the production process. 
On the other hand, for $nk-n+2k-10 > 0$, the result in Eq.~(\ref{nychiarh}) is unchanged and there is no dependence on $m_\chi$
because the production is dominated by its UV component, 
{\it i.e}, all of the DM is produced around the end of inflation, at 
$\sim \aend$. 
We can conclude that for all masses $m_\chi>\tcross$, the production 
of DM from thermal bath scattering should be affected by the presence 
of the gravitational bath which adds a new {\it indirect} gravitational
contribution to the relic abundance.

The relic density in this case for $n< n_c^y$ and $k=2$
is
    \beq
    \frac{\Omega_{\chi}^y h^2}{ 0.12} \simeq \frac{9.3 \times 10^{25}}{(6-n) \Lambda^{n+2}} \trh^7 m_\chi^{n-5} \, .
    \label{oh2ymgttrhk2}
\eeq
and the more general expression is found in Eq.~(\ref{oh2ymgttrh}). The expression for $n=n_c^y$ is also found in the Appendix in Eq.~(\ref{oh2ymgttrhnc}).

\subsection{Freeze-in Production when $\trh < m_\chi < \Lambda < \tmax^h$}
\label{sec:FIscat3}

In the above analysis, we considered freeze-in production with a rate 
$R^i_\chi$ 
parameterized by Eq.~(\ref{R(T)}), with $\Lambda > \tmax$. In this 
subsection, we consider the interesting cases where $\Lambda < \tmax$. 
This situation is slightly more complex, because the value of $n$
can change during the production process from $\aend$ to $a_m$.
Indeed, if $\Lambda> m_\chi$, 
there is a point where the temperature drops to $T \lesssim \Lambda$, while the baths are still able to produce DM.  The value of $n$ then increases by 4.\footnote{ From a microscopic perspective, this can be understood by considering the propagator, $\frac{1}{(T^2-M^2)}$, of a mediator of mass, $M$, 
 changing from $\frac{1}{T^4}\rightarrow \frac{1}{M^4}$, inducing a gain of 4 powers of $T$ in the 
amplitude squared.} For example, when $n=-2$, the rate changes from $R^i_\chi:~T^4 \rightarrow \frac{T^{8}}{\Lambda^{4}}$. 
As 
a concrete example, we consider an interaction producing fermionic DM from the 
scattering of fermions in the radiation bath mediated by a heavy gauge boson, $Z'$, with $\tmax > M_{Z'} > m_\chi > \trh$. In the notation introduced earlier this corresponds to FVF. 
If $m_\chi>M_{Z'}$, the UV rate never changes before production stops when $T\sim m_\chi$ and $n = -2$ is fixed. 
For $m_\chi < M_{Z'}$, the form of 
the scattering cross section $\sigma(\overline{f}f \rightarrow \overline{\chi}\chi)$ changes during reheating. In particular, the cross section takes the 
approximate form $\sigma(T) \propto \frac{1}{T^2}$ when the temperature of the radiation bath is much greater than the $Z'$ mass, and takes the 
approximate form $\sigma(T) \propto \frac{T^2}{M_{Z'}^4}$ when the gravitational bath temperature drops significantly below the $Z'$ mass. This corresponds to a change in the rate from $n=-2$ to $n=2$ as the temperature drops below $M_{Z'}$
as discussed earlier.  The same behavior is found for SVF, FVF, and FSF type interactions.

We 
can estimate the DM number density at $\trh$ resulting from this interaction by first integrating the Boltzmann equation from $\aend$ 
to $a_{M_{Z'}}$ using $R_{\chi} \propto T^{4}$ ($n = -2$) and then integrating again from $a_{M_{Z'}}$ to $\arh$ using $R \propto \frac{T^{8}}{M_{Z'}^{4}}$ ($n=2$). 
This neglects resonant production near the $Z'$ pole, and thus provides a 
conservative, minimal estimate of $\overline{\chi}\chi$ production during reheating. If $a_{m} < \arh$, we will simply cut off the second integral at $a_{m}$ rather than $\arh$, and compute the remaining dilution.

    We first compute the resulting number density from freeze-in production sourced by the gravitational radiation bath for 
    $T^{h}_{\rm max} >  M_{Z'} > \trh $ and $k=2$. The result after performing both integrals of the Boltzmann equation, and keeping only the relevant terms is   
    \bea 
   &&
    n^{h}_{\chi}(\trh)= \alpha \frac{16 \sqrt{3}}{7} \left(\frac98\right)^9 \frac{(T^{h}_{\rm max})^{\frac92}\trh^{4}M_{P}}{\sqrt{M_{Z'}}\rhoend^{\frac32}}
    \eea
or
\beq
\frac{\Omega^h_\chi h^2}{0.12}\simeq \sqrt{\frac{10^{11}~{\rm GeV}}{M_{Z'}}}
\left(\frac{\trh}{10^{8}~{\rm GeV}}\right)
\left(\frac{m_\chi}{1.3\times 10^9~{\rm GeV}}\right)
\eeq
where we used the value of $\tmax^h$ given by Eq.~(\ref{Eq:tmax}).
    This is an interesting result. We start with a process mediated by a massive vector and therefore we expect this to be an $n=2$ type process. Since $n_c^h = -\frac32$ for $k=2$, we might expect DM production to be UV dominated.
    But in the UV, for $\Lambda < \tmax^h$, 
    the process is actually characterized by $n=-2$ as we can ignore the mediator mass. But for $n=-2$, DM production is IR dominated and thus the dominant contribution to the DM density occurs at $T\sim M_{Z^\prime}$ rather than either, $\tmax$, $\trh$, or $m_\chi$. This is unique to the gravitational bath, since for the decay bath $n_c^y =6$ for $k=2$ and we are IR dominated. Furthermore, 
    without taking into account the change in $n$, for $m_\chi > \trh$ and $n=2$,
    we could be tempted to use Eq.~(\ref{Ohngtnck2}) for the relic density,
    but this would lead to an over-abundance 
    by a factor of $\mathcal{O}(100)$. 
    Of course, as just remarked, this is a UV dominated expression and in the UV,
    we have $n=-2$ for $M_{Z^\prime} < \tmax$. Using Eq.~(\ref{ohnltnck2}) would also give an overestimate of the density, but by a smaller factor. 
 
Next, we generalize the above results by computing the number density for arbitrary $k$ and $n$. The possible interactions we consider here are where the interaction in the UV corresponds to $n_{1}= -6$, $-4$ or $-2$ and the interaction after $T$ drops below the cutoff scale $\Lambda$ corresponds to $n_{2} = -2$, 0 or 2 respectively. 
 It is indeed easy to see that the change of regime is equivalent to adding 4 units to $n$ at the crossover temperature.
In our analysis below, we have therefore let $n_{1} = n$ and $n_{2} = n+4$.  For freeze-in sourced by the gravitational radiation bath, and $T^{h}_{\rm max} > \Lambda > m_{\chi} > \trh$, we find
\begin{align}
n^{h}_{\chi}(\trh) &= \alpha^{\frac{k+2}{2k}}\sqrt{3}\frac{(T^{h}_{\rm max})^{3}T_{\rm RH}^{\frac{2k+4}{k}}\Lambda M_{P}}{\rho_{\rm end}^{\frac{k+1}{k}}} \left(\frac{T^{h}_{\rm max}}{\Lambda}\right)^{\frac{3k}{k+2}}  \nonumber \\ 
& \times \left(\frac{2k+4}{6k-3}\right)^{\frac{3(k+1)}{7-4k}} \left[\left(\frac{k+2}{-n_1k-2n_1-6}\right)+\nonumber \right. \\ & \left. \left(\frac{k+2}{\!n_1k\!+\!2n_1\!+\!4k\!+\!14}\right) \left(\!1\!-\!\left(\frac{m_{\chi}}{\Lambda}\right)^{\frac{n_1k+2n_1+4k+14}{k+2}}\right)\right] \label{nharhswitch2}
\end{align}

In this case, there are two distinct critical values of $n=n^{h}_{c}$. In 
particular, for the first segment of the integration with $n_1=-6,-4 \text{ or} -2$, we find $n^{h}_{c} = \frac{-6}{k+2}$ which is consistent with 
what we found above. In the second segment with $n_2 = -2, 0$, or 2, we find that $n^{h}_{c} = \frac{-4k-14}{k+2}$. Note that we did not include the case where $\trh > m_{\chi}$, because in this case production sourced by the decay bath will 
always exceed that of the gravitational bath.
The relic density for $k=2$ corresponding to Eq.~(\ref{nharhswitch2}) is
\begin{align}
\frac{\Omega^{h}_{\chi}h^{2}}{0.12} &= 2.1 \times 10^{-12} \frac{\trh m_{\chi}}{\sqrt{\Lambda}}\left[\frac{1}{3-2n_1}+ \right. \nonumber \\ & \left. \frac{1}{2n_1+11}\left(1-\left(\frac{m}{\Lambda}\right)^{\frac{2n_1+11}{2}}\right)\right] \, .
\label{gravchange}
\end{align}
The general result is found in the Appendix in Eq.~(\ref{nharhswitch}).

Finally, we compute the analogous result for the conventional radiation bath produced by inflaton decays. 
 After the second integration, for $T^{y}_{\rm max} > \Lambda > m_{\chi} > \trh$, we find (for $n_1 < \frac{14-6k}{k-1}$)
\begin{align}
&n^{y}_{\chi}(\!\trh\!)\!=\!\sqrt{\frac{3}{\alpha}}M_{P}\!\left[\Lambda^{2}\!\left(\!\frac{\trh}{\Lambda}\!\right)^{\! \frac{2k+6}{k-1}\!}\!\left(\frac{2k+4}{\!3n_1\!-\!3n_1k\!-\!6k\!+\!30}\!\right) \right. \nonumber \\ & \left. +\left(\frac{T_{\rm RH}}{m_{\chi}}\right)^{\frac{4k+4}{k-1}}\left(\frac{m_{\chi}^{n_+10}}{\Lambda^{6+n_1}T_{\rm RH}^{2}}\right)\left(\frac{2k+4}{\!3n_1\!-\!3n_1k\!-\!18k\!+\!42}\right)\right] \label{nyarhswitchk2}
\end{align}
Note that for each of the interactions we consider for regime switching, namely where $n_1 = -6, -4, \text{ or} -2$, the above results hold (that is, $n < \frac{10-2k}{k-1}$ and $n<\frac{14-6k}{k-1}$ in all cases considered). For $T^{y}_{\rm max} > \Lambda > T_{RH} > m_{\chi}$, we can simply replace $m_{\chi}$ with $T_{\rm RH}$ in the above result.
Finally, the relic density for $k=2$ is
\begin{align}
\frac{\Omega^{y}_{\chi}h^{2}}{0.12} &= 9.3 \times 10^{25} \frac{\trh^{7} m_{\chi}}{\Lambda^8}\left[\frac{1}{6\!-\!n_1}+\frac{1}{2\!-\!n_1}\left(\frac{\Lambda}{m_{\chi}}\right)^{2-n_1}\right] \, ,
\label{oh2ynchangek2}
\end{align}
with the general expression in Eq.~(\ref{nyarhswitch}).

The three results for $n_\chi(\arh)$, arising from gravitational direct production from the condensate, scattering in the gravitational bath and from the decay bath, all compete with each other. Direct production is inevitable if the DM couples to gravity. The production from the scattering depends on a coupling of the DM to the Standard model, which has here been parameterized by an effective coupling $\Lambda$. 
It is then the relative values between $\Lambda$, $m_\chi$ and $\trh$
that determine which of the three processes dominates DM production.
We will use these results to compare the different production mechanisms for given values of $m_\chi$ and $\trh$ for several choices of $n$ and $k$.

\section{Results} \label{sec:develop}
    \subsection{Freeze-in for $k=2$}\label{sec:developk2}
    
     Equipped with analytic expressions for the DM number densities produced from 1) direct gravitational production from the inflaton condensate, 2) scattering within the gravitational radiation bath, and 3) scattering within the conventional radiation bath produced by inflaton decay,
     we can now compare the resulting contributions to the relic density from each. Since the production from scattering requires a (non-gravitational) interaction parameterized by an effective scale $\Lambda$, we can also determine the values of $\Lambda$ for which scattering (from either bath)
     is more efficient than the gravitational production from inflaton scattering. 
     We will also determine the set of parameters for which the production 
     from the gravitational bath is dominant. 
     Recall that the gravitational DM production from the condensate is all UV dominated, and DM is effectively produced at $a = \aend$.  
For scattering in the thermal bath, there is a critical value of $n$ which separates UV from IR production.    
For the gravitational bath with $k=2$,  $n^h_c = -\frac{3}{2}$. Production is UV dominated for $n>n^h_c$, and IR 
dominated for $n<n^h_c$. The dependence on $n$ is easy to 
understand. For large $n$, the suppression from the effective scale $\Lambda$ is large, requiring higher energy to produce dark matter.
For the decay bath with $k=2$, $n^y_c=6$.

The UV contribution occurs near $a=\aend$, when $T = \tmax^{h,y}$. The IR contribution generally occurs at
$a = \arh$ when $T = \trh$. However, as
 discussed above, when $m_{\chi} < T_{\rm RH}$, there are some cases (notably $n=-2$) for which production after reheating is significant. In these cases, we instead use 
the number density when the radiation temperature $T=m_{\chi}$ before 
isentropically evolving the number density to determine the contribution to the present relic density. For $m_\chi > \trh$, the IR contribution occurs at $T = m_\chi$.

The DM relic density contribution from freeze-in sourced by the radiation baths is dependent upon $n$, $k$, $\Lambda$, $m_{\chi}$ and $T_{\rm RH}$. 
We provide the general $n$- and $k$-dependent results in the Appendix, where  the corresponding expressions for the relic abundance contributions from the gravitational production from the condensate can also be found.

\subsubsection{$n=-2$}
We first consider the $k=2$ and $n=-2$ case, which corresponds to a 
contact interaction producing scalars, or a light mediator exchange 
producing fermions. In both cases, we will use a dimensionless coupling $\sigma$ (see the parametrization discussion in section \ref{sec:FIscat}). The 
relic density of dark matter produced by the decay bath for $k=2$ and $n=-2$ with $m_{\chi} < \trh$ takes the form
\beq 
\frac{\Omega^{y}_{\chi} h^{2}}{0.12} \simeq 
4.9 \times 10^{7} \sigma^{2} \sqrt{\frac{3}{\alpha}}\frac{M_{P}}{\rm GeV}  \simeq \left(\frac{\sigma}{1.7 \times 10^{-13}}\right)^2
\label{osig}
\eeq
Note that in this case, the relic density is independent of $\trh$ and $m_{\chi}$. This can be seen from Eq.~(\ref{pastrhk2}) which is dominated by the last term for which the dependence on $\trh$ drops out and the dependence on the mass is $m_\chi^{n+2}$ which drops out for $n=-2$.  This characteristic is sometimes called {\it FIMP miracle} in the literature, but is a very specific case of freeze-in 
processes. This corresponds to $n=-2$ within a Universe where the IR contribution dominates.
For $k=6$, this would not be the case anymore because $n^y_c=-2.4$ in this case, and the relic abundance is then proportional to $m_\chi$.
As a result, for $n=-2$ and $k=2$, there is simply a critical value of $\sigma$, $\sigma_c$, that is consistent with the observed relic density, irrespective of $\trh$ and $m_{\chi}$ (provided that $m_{\chi} < \trh$) which is
\beq 
\sigma_{c} \simeq 1.7\times 10^{-13}\,.
\label{sigmack2}
\eeq
When $m_\chi > \trh$, the relic density does depend on 
$m_\chi$ and $\trh$ and is given by

\beq 
\frac{\Omega^{y}_{\chi} h^{2}}{0.12} \simeq \left(\frac{\sigma}{0.01}\right)^{2} \left(\frac{\trh}{10^{6} \text{ GeV}}\right)^{7} \left(\frac{10^{9} \text{ GeV}}{m_{\chi}}\right)^{7}\,.
\label{osig2}
\eeq

On the other hand, the relic density of DM produced by freeze-in from the gravitational bath (for $n=-2$, $k=2$) is very inefficient compared to the decay bath when $m_{\chi}<T_{RH}$, and is given by 
\beq
\frac{\Omega^{h}_{\chi}h^{2}}{10^{-26}} \simeq\! \left(\frac{\sigma}{1.7\times10^{-13}}\right)^{2} \! \left(\frac{\trh}{10^{10} \text{ GeV}}\right)^{1/3} \! \left(\frac{m_{\chi}}{10^{7} \text{ GeV}}\right)\!.
\eeq
For $m_{\chi} > \trh$, however, freeze-in production sourced by the gravitational bath can be significant. We find 
\beq
\frac{\Omega^{h}_{\chi}h^{2}}{0.12} =\left(\frac{\sigma}{0.1}\right)^{2} \left(\frac{\trh}{10^{9} \text{ GeV}}\right) \left(\frac{m_{\chi}}{2 \times 10^{10} \text{ GeV}}\right)^{1/2} \,.
\label{Eq:omegahk2nm2}
\eeq

\begin{figure}[ht!]
        \centering
        \includegraphics[width=3.in]{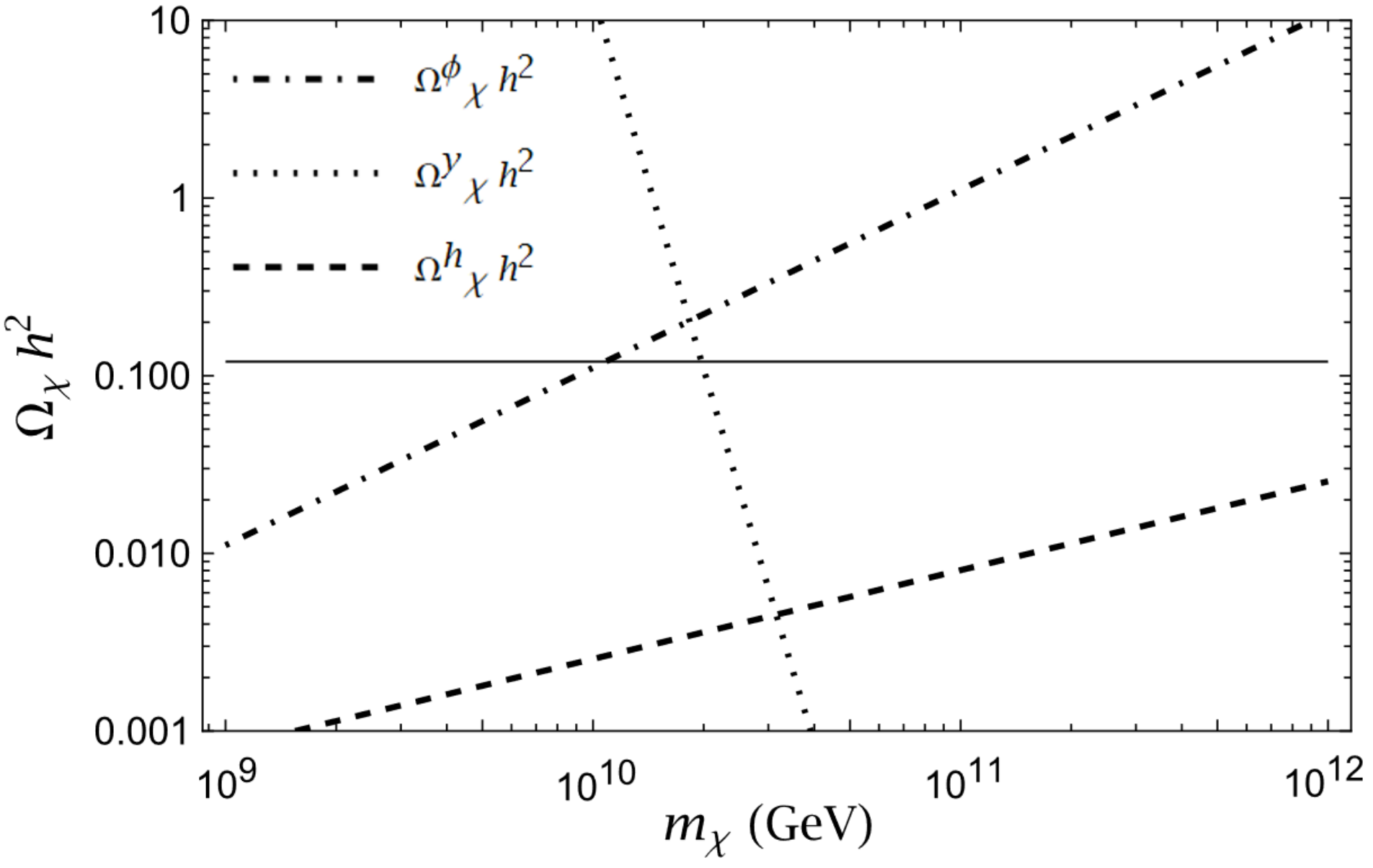}
        \caption{
        Contributions of each of the three freeze-in  sources of scalar dark matter to the final relic density for  $k=2$, $n=-2$, $\trh = 10^{7}$ GeV, and $\sigma = 0.1$. The production from inflaton scattering, from the decay bath and the gravitational bath are shown by the dot-dashed, dotted, and dashed lines respectively. The horizontal line is placed at $\Omega_\chi h^2 = 0.12$ to guide the eye.} 
        \label{nm2k2relicComparisons}
    \end{figure}

To illustrate the contributions of each source to the relic density for scalar DM, we plot $\Omega^i_{\chi}h^2$ as a function of $m_{\chi}$ for fixed values of $\trh = 10^{7} \text{ GeV}$ and $\sigma = 0.1$ for $k=2$ 
and $n=-2$ in Fig. \ref{nm2k2relicComparisons}. The direct gravitational 
production from the inflaton condensate from Eq.~(\ref{Eq:omega0k2}) is 
shown by the dot-dashed line and is the dominant source for the production 
of DM when $m_\chi \gtrsim 2 \times 10^{10}$~GeV.  The production from the decay bath is shown by the dotted line from Eq.~(\ref{osig2}) and 
dominates when $m_\chi \lesssim 2 \times 10^{10}$~GeV. Note the steep dependence on the mass $\propto m_\chi^{-7}$ and that when $m_\chi < \trh$, this line would saturate at a very high value of $\Omega_\chi h^2$ given by Eq.~(\ref{osig}).
As one can see the production from the gravitational bath (shown by the dashed line) never dominates for scalar DM
and this choice of parameters. 

\begin{figure}[ht!]
        \centering
        \includegraphics[width=3.in]{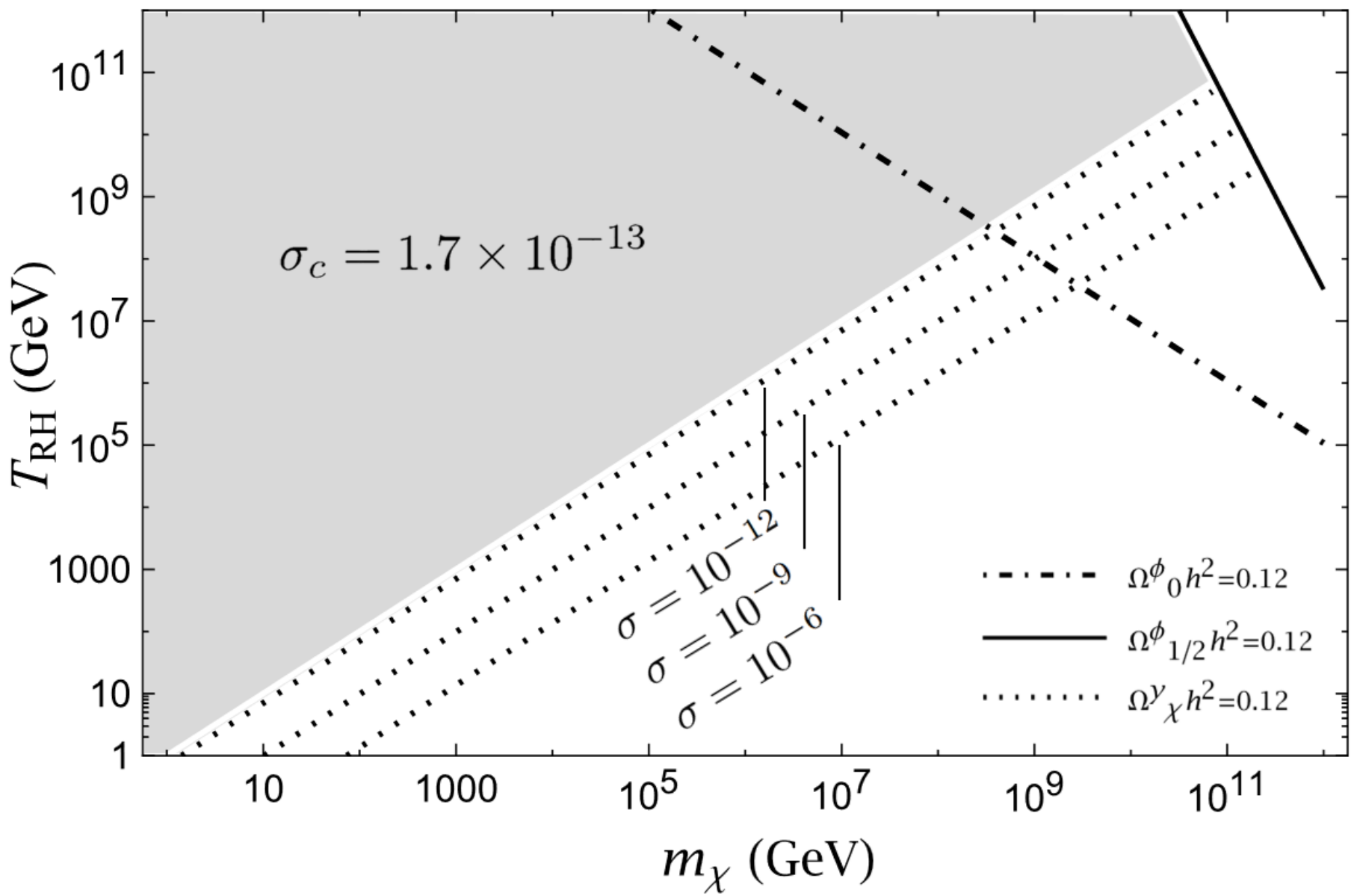}
        \caption{
       The $m_\chi,\trh$ plane for scalar and fermionic dark matter production for $k=2$ and $n=-2$. The relic density is dominated either by the decay bath ($\Omega^{y}h^{2}$) (dotted lines and shaded region), gravitational bath (dashed lines) ($\Omega^{h}h^2$), or by gravitational scattering of the inflaton ($\Omega^{\phi}h^2$) (dot-dashed line for scalar DM, and solid line for fermionic DM). The relic density in the shaded region is determined exclusively by a critical coupling ($\sigma_{c} = 1.7 \times 10^{-13}$), independent of $\trh$ or $m_{\chi}$. In this region if $\sigma > \sigma_{c}$, the decay bath will overproduce DM. Whereas if $\sigma<\sigma_{c}$ in this region, there is no viable combination of $m_{\chi}$ and $\trh$ that will produce sufficient DM to account for the observed relic density.
       }
        \label{nm2k2relicCombined}
    \end{figure}

We show regions of the $(m_\chi, \trh)$ plane for $k=2$ and $n=-2$ for which $\Omega_\chi h^2 = 0.12$ in Fig.~\ref{nm2k2relicCombined}. The dot-dashed line 
shows the relation between $m_\chi$ and $\trh$ needed to obtain the correct relic density,
sourced by the gravitational inflaton scattering
in the case of scalar dark matter, see Eq.~(\ref{Eq:omega0k2}). 
Above this line, the 
production 
is always excessive and is excluded if $\chi$ is stable 
as we are assuming. Below this line, the relic density 
from inflaton scattering is low but freeze-in from the 
decay bath may be sufficient to obtain $\Omega_\chi h^2 = 0.12$, and for $m_\chi > \trh$, depends on the value of
$\sigma$ as shown by the dotted lines. 
As expected from Eq.~(\ref{osig2}), these lines follow $\trh \propto m_\chi$.
When $m_\chi = \trh$, $\sigma \simeq 1.7 \times 10^{-13}$ is required 
and then for all points with $m_\chi < \trh$ the same 
value of $\sigma$ is needed allowing 
all of the shaded region as viable (FIMP miracle scenario from Eq.~(\ref{osig})). 

For values of $\sigma$ larger than those shown in Fig.~\ref{nm2k2relicCombined}, the DM may begin to come into equilibrium and therefore the calculation of the relic abundance is determined from freeze-out rather than freeze-in. We do not treat those cases here. Freeze-in is justifiable for the decay bath, so long as $\sigma^2 \lesssim m_\chi^3/\trh^2M_P$ when $m_\chi > \trh$. 
If in addition we use Eq.~(\ref{osig2}) and fix the relic density, we can eliminate $\trh$ and $\sigma \lesssim 10^5 (m_\chi/M_P)^{\frac{7}{10}}$. For example, for $\sigma = 0.1$ equilibrium is not achieved if $m_\chi \gtrsim 10^{10}$~GeV and freeze-in remains valid. For the gravitational bath, because the dependence of $H$ on temperature is different, the limit becomes $\sigma \lesssim (\rhoend m_\chi/M_P^2 \tmax^3)^\frac14$ and equilibrium is not attained for $\sigma = 0.1$ even for masses as low as $m_\chi \simeq 10^6$~GeV.

It is important to note
that for certain values of ($\sigma$, $m_\chi$, $\trh$), the
dominant process which populates the dark sector 
{\it does not} originate from the thermal bath, but from
the gravitational inflaton scattering. This process is often neglected in many current studies of freeze-in. These analyses in the FIMP framework must be treated with some caution. This is one of the important conclusions of our work. We see that by combining equations (\ref{Eq:omega0k2}) and (\ref{osig2}) and (\ref{Eq:omega0k2}) and (\ref{osig}), the parameter space of concern is 

\beq
\frac{10^{17}~{\rm GeV}}{\trh}< m_\chi<\trh\,,
\eeq
for $m_\chi <\trh$, and

\beq
\sigma \lesssim 10
\left(\frac{m_\chi}{10^9~\rm GeV}\right)^4
\left(\frac{10^6~\rm GeV}{\trh}\right)^3\,,
\eeq
for $m_\chi>\trh$.

\begin{figure}[ht!]
        \centering
        \includegraphics[width=3.in]{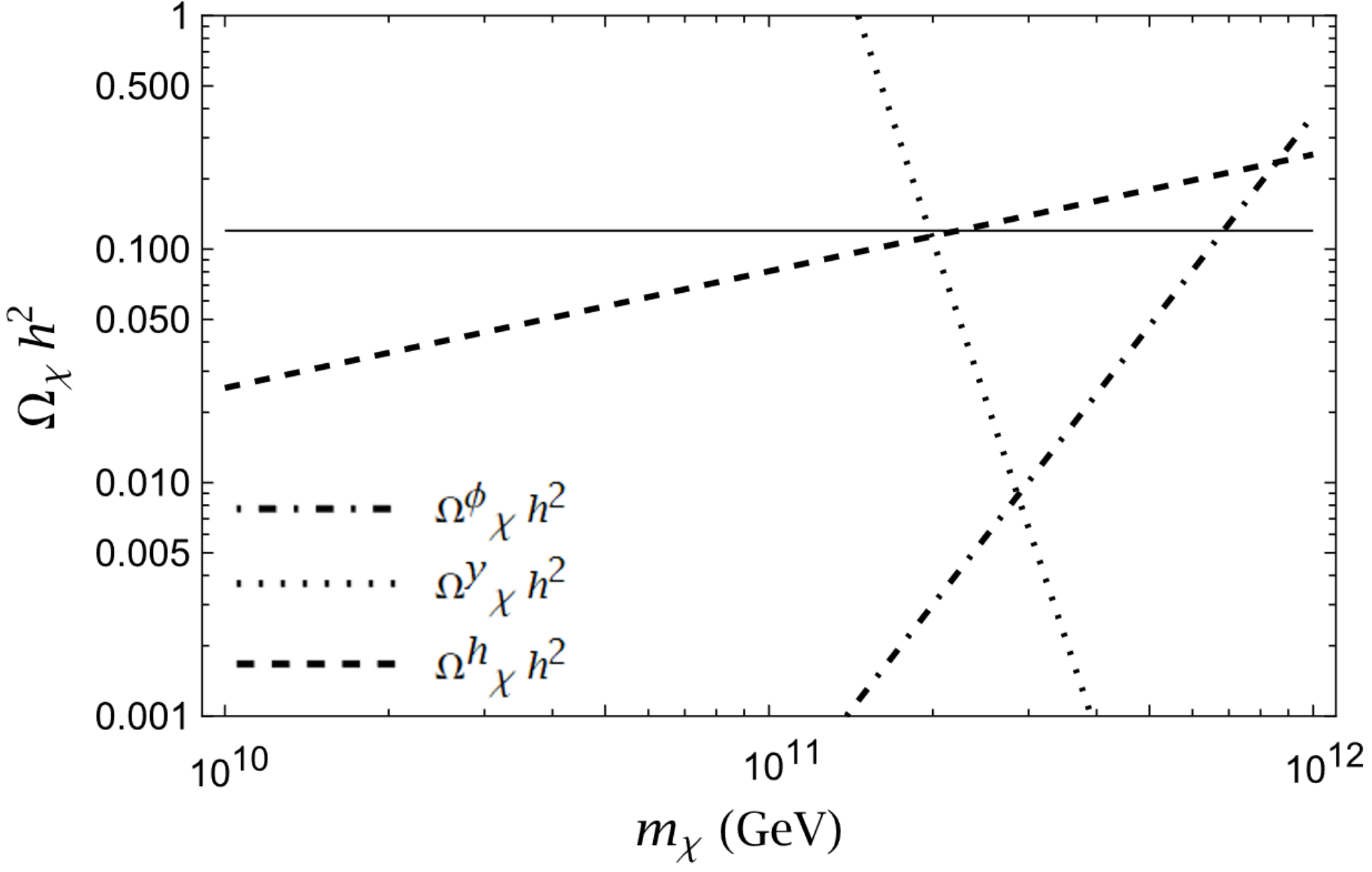}
        \caption{As in Fig.~\ref{nm2k2relicComparisons}, for fermionic DM with $k=2$, $n=-2$, $\trh = 10^{8}$ GeV, and $\sigma = 0.1$.}
        \label{nm2k2relicComparisonsFermions}
    \end{figure}

For a fermionic dark matter candidate, we show
the contribution from each of the three sources in Fig.~\ref{nm2k2relicComparisonsFermions}.  
Because of the additional suppression by a factor of $(m_\chi/m_\phi)^2$ (recall that $m_\phi^2 = 2 \lambda M_P^2$), the resulting production from 
inflaton scattering is significantly lower. We can see that this mode of production begins to dominate only for $m_\chi \gtrsim 8\times 10^{11}$ GeV.
For masses $m_\chi \simeq 2- 8 \times 10^{11}$~GeV, the production from the gravitational bath is the dominant source of production of DM for this particular choice of parameters and for $m_\chi \simeq 2 \times 10^{11}$~GeV, production from the two thermal baths are about equal and for that mass, the desired value of $\Omega_\chi h^2 = 0.12$ is achieved. Note that as per our discussion above, for a mass of order $m_\chi \sim 10^{11}$~GeV, equilibrium is not achieved even for $\sigma  = 0.1$ for either the gravitational or decay bath and the freeze-in calculation provides the correct relic density.  As for scalar DM, for lower masses, the relic density is dominated  by freeze-in sourced by the decay bath though it overproduces DM for this value of $\sigma$.  

The $(m_\chi, \trh)$ plane for fermionic DM is unchanged for the freeze-in production from either of the two thermal baths.  However the inflaton scattering relation is weaker and the region respecting the right fermionic relic abundance for this case is shown by the solid line in Fig.~\ref{nm2k2relicCombined}. 
We notice that a region is still excluded because of an overabundance due
to gravitational scattering of the inflaton, above the solid line,
for larger masses $m_\chi$ than found in the scalar case.
The limit follows $\trh \propto m_\chi^{-3}$, as expected from Eq.~(\ref{Eq:omegaphihalfk2}).
All points below this line are viable if there is a freeze-in source from the thermal bath.

\begin{figure}[ht!]
        \centering
        \includegraphics[width=3.in]{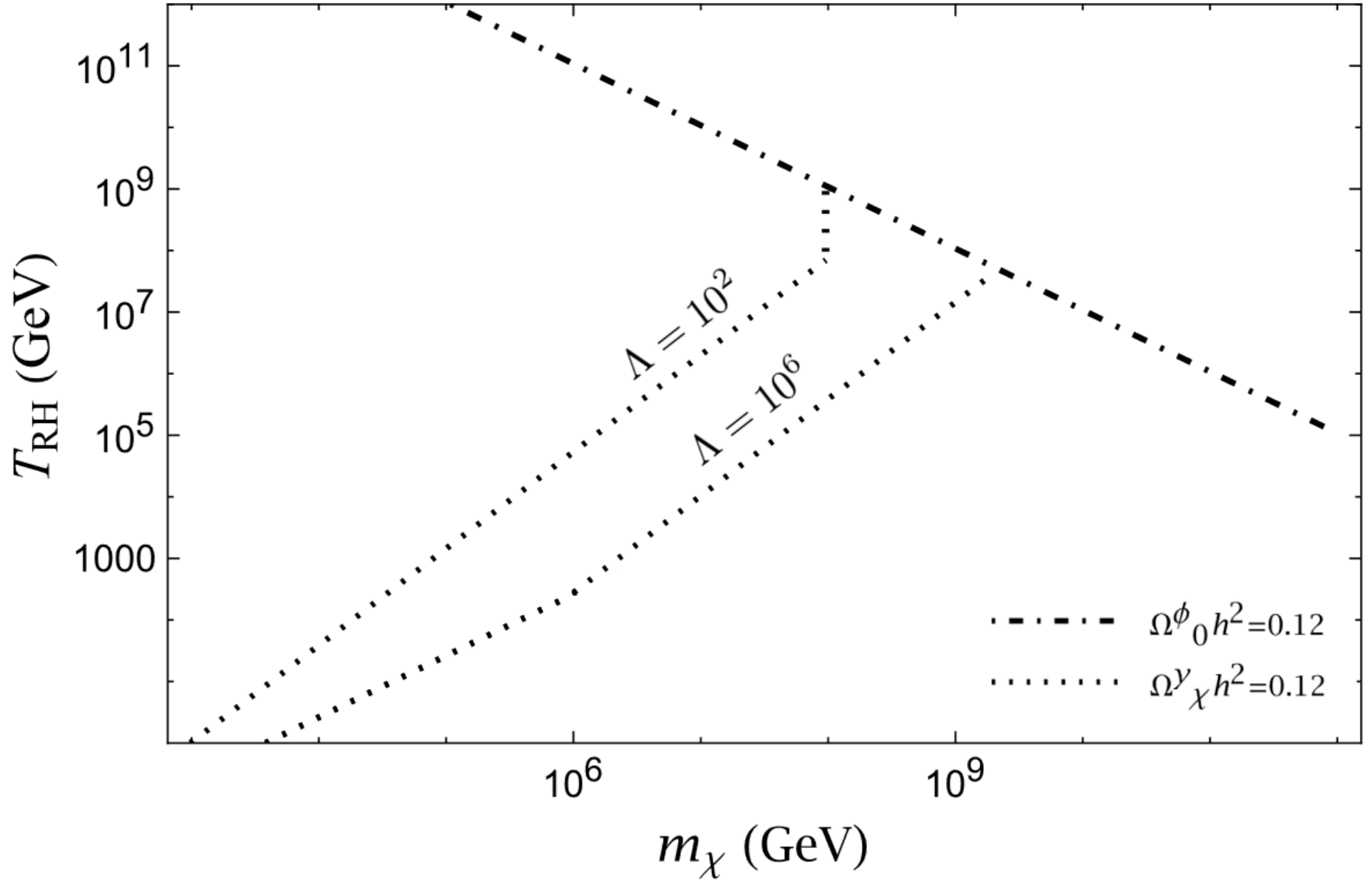}
        \caption{
       The $m_\chi,\trh$ plane for scalar dark matter production for $k=2$, $n_{1}=-6$ and two values of $\Lambda$ with units in GeV. The relic density is dominated either by the decay bath ($\Omega^{y}h^{2}$) (dotted lines) or by gravitational scattering of the inflaton ($\Omega^{\phi}h^2$) (dot-dashed line). When $\Lambda > m_{\chi}$, the interaction corresponds to $n_{2}=-2$.  }
        \label{nm6k2relicCombined}
    \end{figure}

The $n=-2$ interactions for the production of scalar DM can originate from scalar interactions in the thermal bath mediated by a scalar (SSS). In this case, when the temperature is above the mediator scale, $\Lambda$, the production rate is given by $n_1=-6$, and changing to $n_2=-2$ at low energies. For these values of $n$, the dark matter production from the decay bath is IR dominated because $n_i < n^y_c$. 
However, when $\Lambda$ is relatively low the production is more 
characteristic of $n=-6$. A $(m_\chi, \trh)$ plane for this example is shown in Fig.~\ref{nm6k2relicCombined}. 
The behavior of the lines can be understood as follows:
For $\Lambda = 10^{6}$~GeV, the lower part of the line is produced where
$\trh < m_\chi< \Lambda$ so that $n=-2$ and from Eq.~(\ref{oh2ymgttrhk2}), we expect $\trh \propto m_\chi$. 
This is true for smaller values of $\Lambda$ for sufficiently low $m_\chi$. At $\Lambda \simeq m_\chi$, the relic density is still given by Eq.~(\ref{oh2ymgttrhk2}) but now with $n=-6$ in which case, $\trh^7 \propto m_\chi^{11}/\Lambda^4$ corresponding to the change in slope seen in the figure. Finally for $\Lambda = 10^2$~GeV, over the range shown, $n=-6$ all along the curve, but for $m_\chi < \trh$,
we have to integrate the Boltzmann equation past $\arh$ to $a_{\rm m}$ and the relic density is given by Eq.~(\ref{pastrhk2})
but in this case, the dependence on $\trh$ drops out and the line becomes vertical. Note that the $n=-2$ portion of the line for $\Lambda = 10^6$~GeV is not strictly valid as this line is equivalent to $\sigma = 1$ and from our previous discussion equilibrium is achieved for $T > m_\chi$.

   To show the dependence of the results on the type of interaction, we next consider  $n=2$ and in the following subsection, $n=0$, and $n=6$.

\subsubsection{$n=2$}

  The relic density contribution from freeze-in sourced by the decay bath for $-1 < n < 6$, $k=2$ and $m_{\chi}<\trh$ 
   was given in Eq.~(\ref{ohnltncyk2}). 
    The relic density contribution from freeze-in sourced by the gravitational bath for $k=2$ and $n > -\frac{3}{2}$ was given in Eq.~(\ref{Ohngtnck2}).
Both of these sources scale as $\Lambda^{-(n+2)}$ and compete with the gravitational production from inflaton scattering. 
It is then obvious that for a large effective scale, 
the gravitational production of DM
will always dominate, particularly for scalar DM where there is no DM mass suppression due to a helicity flip.

\begin{figure}[ht!]
        \centering
        \includegraphics[width=3.in]{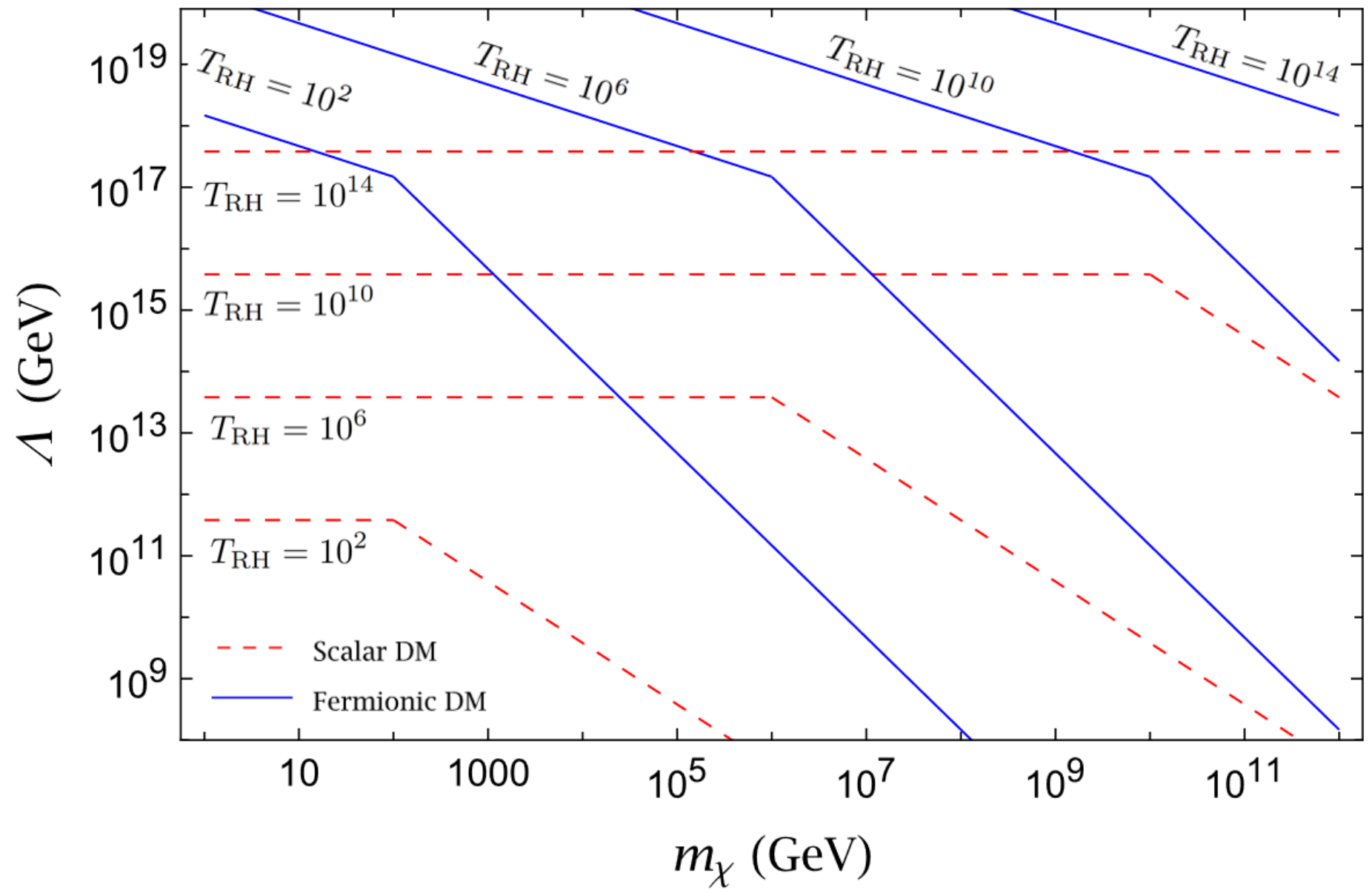}
        \caption{$\Lambda$ vs. $m_{\chi}$ plane displaying where DM production via gravitational scattering of the inflaton equals freeze-in production sourced by the decay bath, for $k=2$ and $n=2$. Both scalar (red-dashed) and fermionic (blue-solid) DM are considered for various choices of $\trh$ with units in GeV. In all cases, regions above the lines correspond to direct gravitational DM production from the inflaton exceeding decay bath production, which constrains the choice of $\Lambda$ for freeze-in models.}
        \label{lambdavsmk2n2}
    \end{figure}

Naively, one might expect gravitational interactions to dominate only when $\Lambda \simeq M_P$. However, even for smaller values of $\Lambda$, gravitational interactions may dominate over thermal scattering. This can be understood since
the scattering rates are proportional to $T^{n+6}/\Lambda^{n+2}$ and the gravitational rates are proportional to $\rho_\phi^2/M_P^4$. But since $\rhoend^\frac14 \gg T$ gravity can dominate even if the gravitational interactions are Planck-suppressed. 
To illustrate this competition,
we show in Fig.~\ref{lambdavsmk2n2}, the regions in the $(m_\chi,\Lambda)$ plane where the decay bath production equals the gravitational production from inflaton scattering at $a \simeq \aend$ for $n=2$. We
show lines of equality for scalar (red-dashed) and fermionic (blue-solid) DM for several choices of the reheating temperature.
For scalars, from Eqs.~(\ref{Eq:omega0k2}) and (\ref{ohnltncyk2}), we expect that $\Lambda^2 \propto T$ for $m_\chi < \trh$ (seen as horizontal lines in the Figure) and from Eq.~(\ref{oh2ymgttrhk2})
we have $\Lambda^2 \propto T/m^2$ for $m_\chi > \trh$. 
For fermions, we use Eq.~(\ref{Eq:omegaphihalfk2}), and find
$\Lambda^2 \propto T/m$ and $\Lambda^2 \propto T/m^3$ for $m_\chi < \trh$ and $m_\chi > \trh$, respectively. This is the behavior seen in the Figure.

As one can see, for any pair ($m_\chi$, $\trh$), 
there exists a beyond the Standard Model (BSM) scale $\Lambda$ above which one must take into account the gravitational scattering of the inflaton for the relic abundance calculation, because 
its net production of DM exceeds that produced from the scatterings in the thermal bath.
This is also a very important conclusion of our work. 
Indeed, this means that within a given microscopic framework, 
where $\Lambda$ represents, for example, a mediator
mass, such as a massive $Z'$ or a Higgs from a grand-unified theory, there is an upper limit to these 
masses beyond which gravitational production of dark 
matter sourced by the inflaton takes over from thermal scattering.
We consider this result as important, since all studies (including some by the authors of this paper) have 
so far neglected this source of dark matter when studying
UV-motivated scenarios of  FIMP DM. We recall however, that the production via gravitational scattering of the inflaton is highly suppressed for fermionic DM which weakens the upper limit on $\Lambda$ as seen in Fig.~\ref{lambdavsmk2n2}, where for a given value of $m_\chi$ and $\trh$, the fermionic lines are significantly higher than the corresponding bosonic lines, though this difference is reduced for large DM masses.

\begin{figure}[ht!]
        \centering
        \includegraphics[width=3.in]{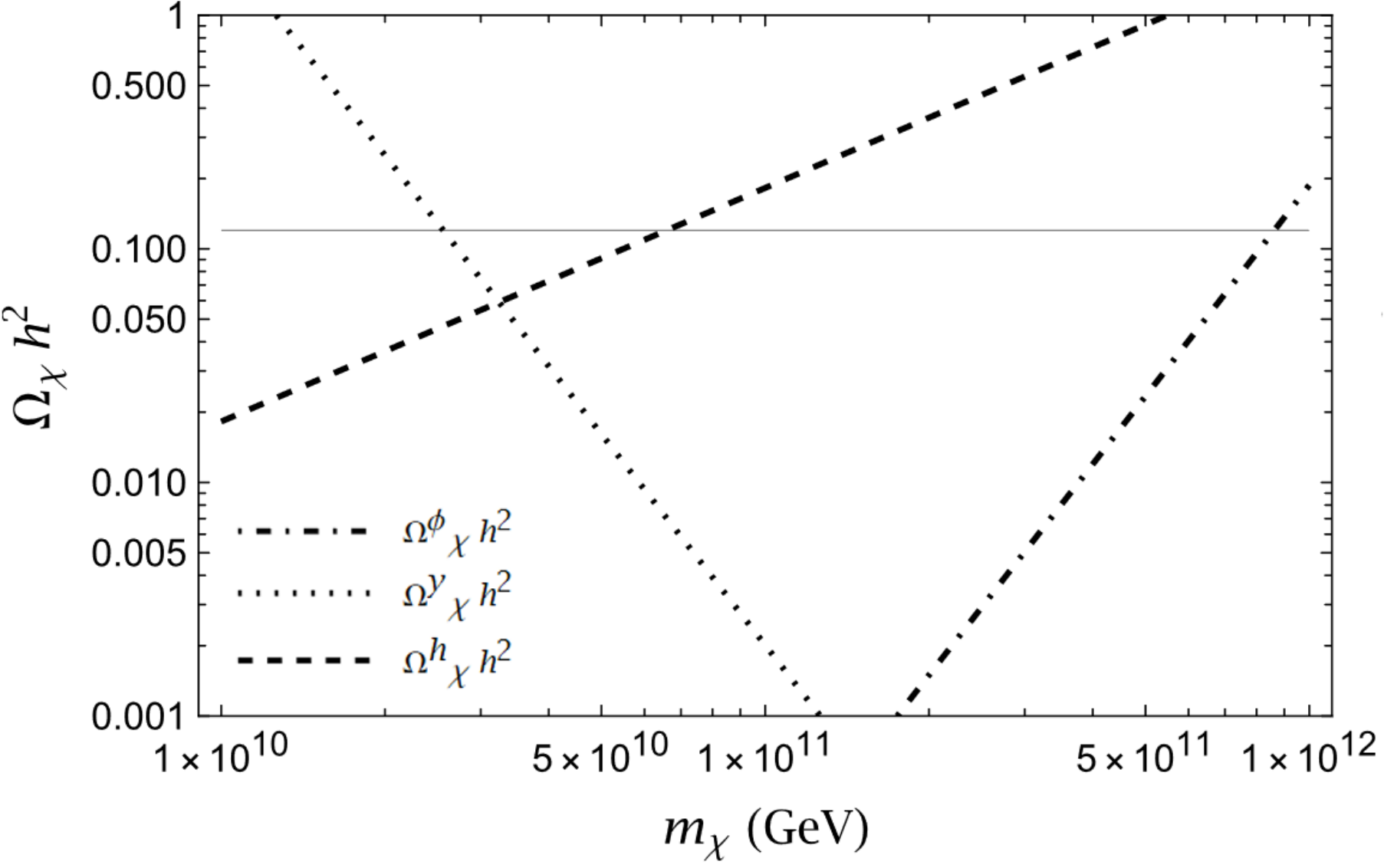}
        \caption{Contributions to the (fermionic) dark matter relic density for each of the three freeze-in sources as a function of $m_{\chi}$ for $k=2$ and $n=2$, $\Lambda =  10^{12}$ GeV and $T_{\rm RH} = 5 \times 10^{7}$ GeV.}
        \label{n2k2relicComparison}
    \end{figure}

We can look in more detail at the case $n=2$, where the effective 
 mass scale, $\Lambda$, appears explicitly in the thermal production rate.
We show each contribution for a fermionic DM candidate in Fig.~\ref{n2k2relicComparison} for $k=2$, 
$\trh = 5 \times 10^7$~GeV and $\Lambda = 10^{12}$~GeV. 
This corresponds to SVF, FVF and FSF interactions,
with $\Lambda$ related to either the vector or scalar mediator mass
which is larger than the maximum temperature, $\tmax^h$.
We observe that relative to the case for $n=-2$ shown in Fig.~\ref{nm2k2relicComparisonsFermions}, there is a larger range of DM 
masses where freeze-in sourced by the gravitational bath 
dominates over the decay bath.
We see that the dominant freeze-in mechanism originates in the gravitational bath when 
$m_\chi \gtrsim 3 \times 10^{10}$~GeV. In this case,
we are able to saturate the relic density from the gravitational bath with $m_\chi \simeq 7 \times 10^{10}$~GeV.

\begin{figure}[ht!]
        \centering
        \includegraphics[width=3.in]{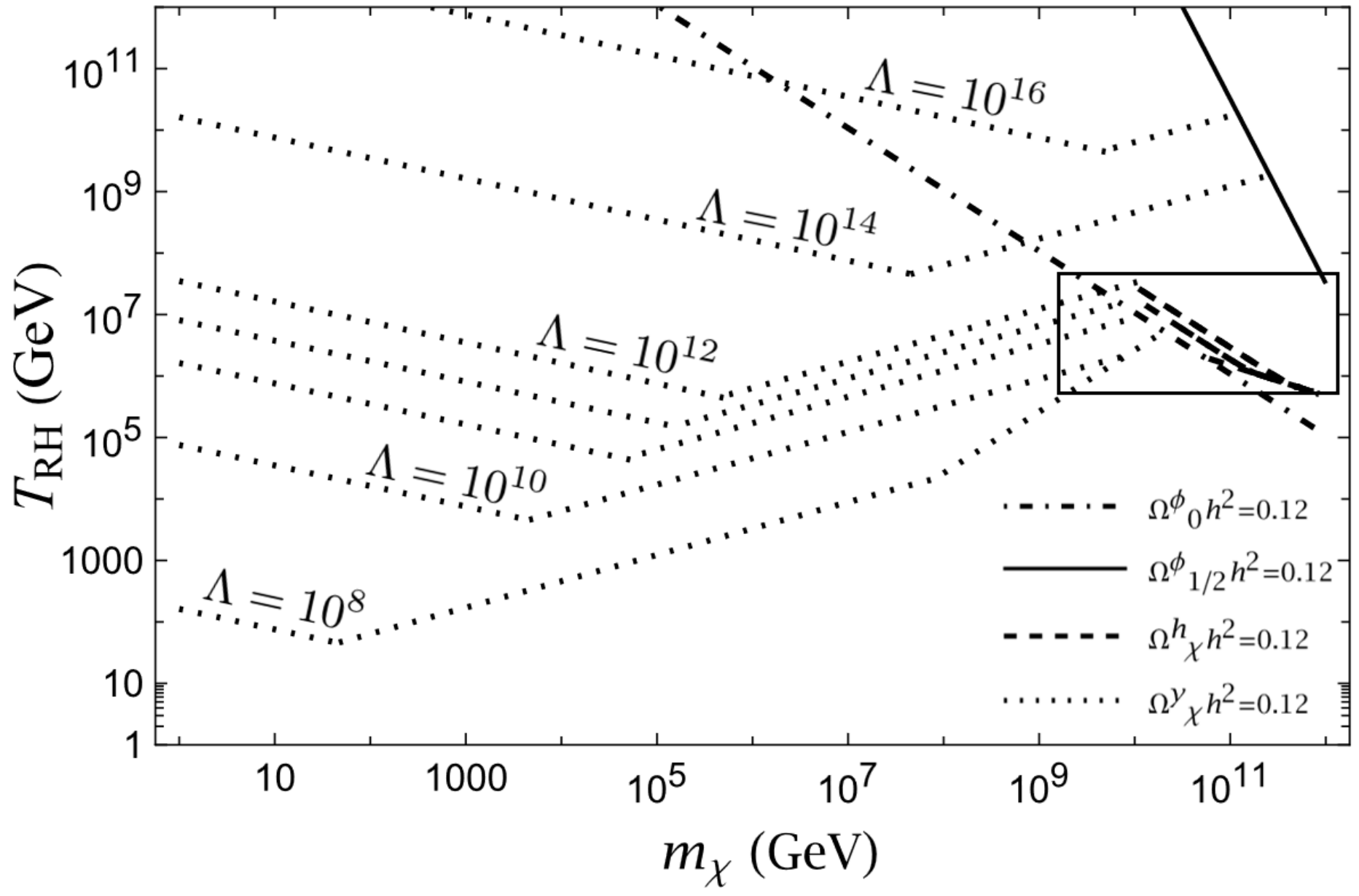}
        \caption{Relic density summary plot for dark matter production for $k=2$ and $n_2=2$. The relic density is dominated either by freeze-in sourced by the decay bath ($\Omega^{y}_{\chi}h^{2}$), freeze-in sourced by the gravitational bath ($\Omega^{h}_{\chi}h^2$), or direct gravitational production via inflaton scattering ($\Omega^{\phi}_{\chi}h^2$). The relic densities are displayed for various values of the suppression scale $\Lambda$. The boxed region is displayed in greater detail in Fig.~\ref{nm2ton2k2relic}.}
        \label{n2k2relic}
    \end{figure}

The corresponding $(m_\chi,\trh)$ plane is shown in Fig.~\ref{n2k2relic}. Once again the dotted lines correspond to values in the plane where $\Omega_\chi h^2 = 0.12$. For values of $\Lambda \ge 10^{12}$~GeV, the 
dependence of the thermal rate follows $R^{h,y}(T)\propto T^8/\Lambda^4$ (i.e., $n=2$) over the entire production process, since in this case $\Lambda >\tmax^h$. 
We observe however a change in slope in these lines occurring 
when $\trh = m_\chi$. This comes from the fact that the integration of 
the Boltzmann equations is cut off before $\trh$ when $m_\chi > \trh$.
 The lines corresponding to DM production from inflaton 
scattering are shown again by the dot-dashed line for bosons and solid line for fermions. As this production is independent of $n$, these lines are the same in all of the $(m_\chi,\trh)$
planes. 
On this plot, the dominant contribution to the relic density from the gravitational bath is limited to the short dashed lines at high masses.
We see that the gravitational bath never plays a role for scalar DM (as the dashed lines lie above the scalar dot-dashed line).

The dotted lines in Fig.~\ref{n2k2relic} make use of many of the derivations performed in the previous section, and so it is worthwhile taking a closer look at the behavior of these.
For each value of $\Lambda$, $\trh < \Lambda$ and at sufficiently low $m_\chi < \trh$, the relic density is given by Eq.~(\ref{ohnltncyk2}). Although $n_1 = -2$ in this case, the relic density is determined in the IR since $n < n_c^y$ and $\Omega_\chi h^2$ is determined by $n_2 = 2$. From Eq.~(\ref{ohnltncyk2}) a constant relic density implies then that 
$\trh^3 \propto \Lambda^4/m_\chi$, which is precisely the behavior we see at low masses. The reheating temperature drops with decreasing $\Lambda$ and with increasing $m_\chi$.  When $m_\chi$ is increased so that $m_\chi > \trh$, the integration of the Boltzmann equation is cut off at $m_\chi$ (rather than $\trh$)
and the relic density is now given by Eq.~(\ref{oh2ymgttrhk2}) and is again determined in the IR so that we take $n = n_2 = 2$.
In this case, we expect $\trh^7 \propto \Lambda^4 m_\chi^3$.
The reheating temperature drops less with respect to $\Lambda$
and now increases with $m_\chi$ as is clearly seen in the Figure.

\begin{figure}[ht!]
        \centering
        \includegraphics[width=3.in]{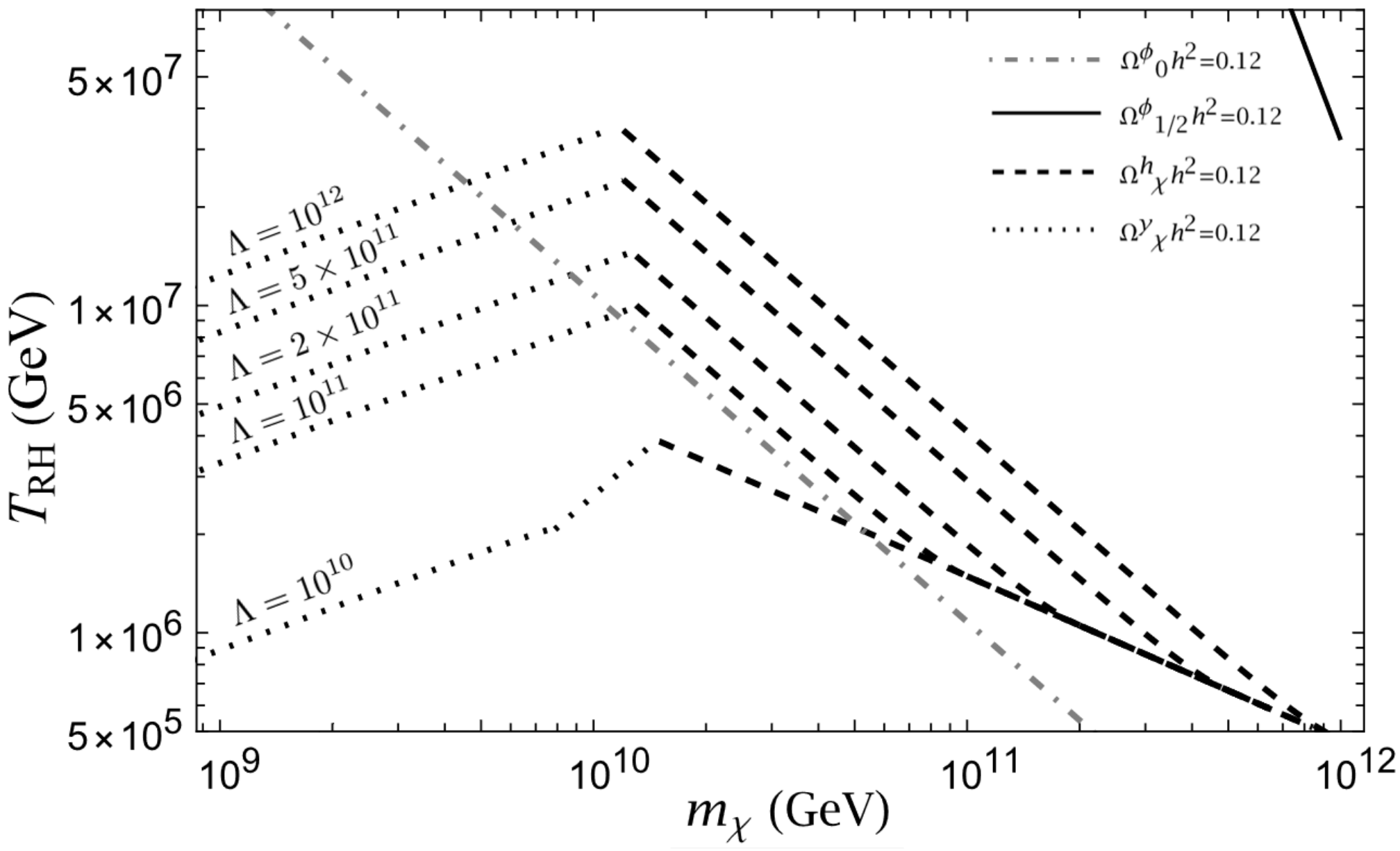}
        \caption{A zoom of the boxed region in Fig.~\ref{n2k2relic} where $\Lambda < T^{h}_{\rm max}$ and $n_{1} = -2$ and $n_{2} = 2$. The relic density is dominated either by freeze-in sourced by the decay bath ($\Omega^{y}_{\chi}h^{2}$), freeze-in sourced by the gravitational bath ($\Omega^{h}_{\chi}h^2$), or direct gravitational production via inflaton scattering ($\Omega^{\phi}_{\chi}h^2$). The relic densities are displayed for various values of the suppression scale $\Lambda$ with units in GeV. }
        \label{nm2ton2k2relic}
    \end{figure}

For $\Lambda < T^h_{\rm max}$, the situation is a little bit more complex.
Indeed, in this case the value of $n$ changes from -2 to 2 in the evaluation of $\Omega^y_\chi h^2$ when $T$ reaches $\Lambda$. As for the higher values of $\Lambda$, the $\Omega^y_\chi h^2$ contours change slope in the $(m_\chi,\trh)$ plane
    when $m_\chi = \trh$. Another change in slope occurs when 
    $m_\chi = \Lambda$. When $\Lambda < m_\chi$, integration of the 
    Boltzmann equations stops at $a_{\rm m}$ and therefore the value of $\Lambda$ no longer enters in to the expression for $\Omega^y_\chi h^2$. 
    This can be better understood by comparing the second term in 
    Eq.~(\ref{oh2ynchangek2}) with Eq.~(\ref{oh2ymgttrhk2}) using $n_1$ in the former equation and $n_2$ in the latter. Recalling that $n_2 = n_1 + 4$, the second term in (\ref{oh2ynchangek2}) is identical to (\ref{oh2ymgttrhk2}) with $n = n_2$. When $m_\chi < \Lambda$,  
    Eq.~(\ref{oh2ynchangek2}) is no longer valid and we must use
    Eq.~(\ref{oh2ymgttrhk2}) but now with $n = n_1$ (appropriate for small $\Lambda$). Then for $n_1 = -2$, we see that $\Omega_\chi^2$ becomes 
    independent of $\Lambda$ (hence a common line \footnote{This line is 
    equivalent to the line corresponding to $\sigma = 1$ in 
    Fig.~\ref{nm2k2relicCombined}. Again as previously discussed points along this line should be computed using freeze-out. } for all values of $\Lambda$) and $\trh \propto m_\chi$, ie. steeper than the behavior at larger masses,
    where $\trh \propto m_\chi^{3/7}$ as we already discussed. 
To the right of this line the relic density drops
    as $\Omega_\chi h^2$ is inversely proportional to a (positive) power of $m_\chi$ as seen in Eq.~(\ref{oh2ymgttrhk2}) with $n=-2$. 
    At still higher masses, and sufficiently low $\Lambda$, the gravitational bath eventually becomes important.
    The boxed region at high masses is zoomed in on in Fig.~\ref{nm2ton2k2relic}. Here, we see more clearly 
    the importance of the gravitational bath at higher masses for 
    a range of $\Lambda$ between $10^{10} - 10^{12}$~GeV. For $\Lambda = 10^{10}$ GeV, the gravitational bath may be dominant even for scalar DM.
    The relic density sourced by the gravitational bath is determined by either from Eq.~(\ref{gravchange}) for $m_\chi < \Lambda$ or Eq.~(\ref{ohnltnck2}) for $m_\chi > \Lambda$. Still using $n_1=-2$, we then expect $\trh \propto \sqrt{\Lambda}/m_\chi$ for low masses and 
     $\trh \propto m_\chi^{-\frac12}$ for higher masses and this is borne out by the dashed lines in Fig.~\ref{nm2ton2k2relic}. 
    As in the case of the decay bath,
    when $m_\chi > \Lambda$, the lines of constant density merge 
    to a single line when $\trh$ is independent of $\Lambda$.

\subsubsection{n=0 and 6}

\begin{figure}[ht!]
        \centering
        \includegraphics[width=3.in]{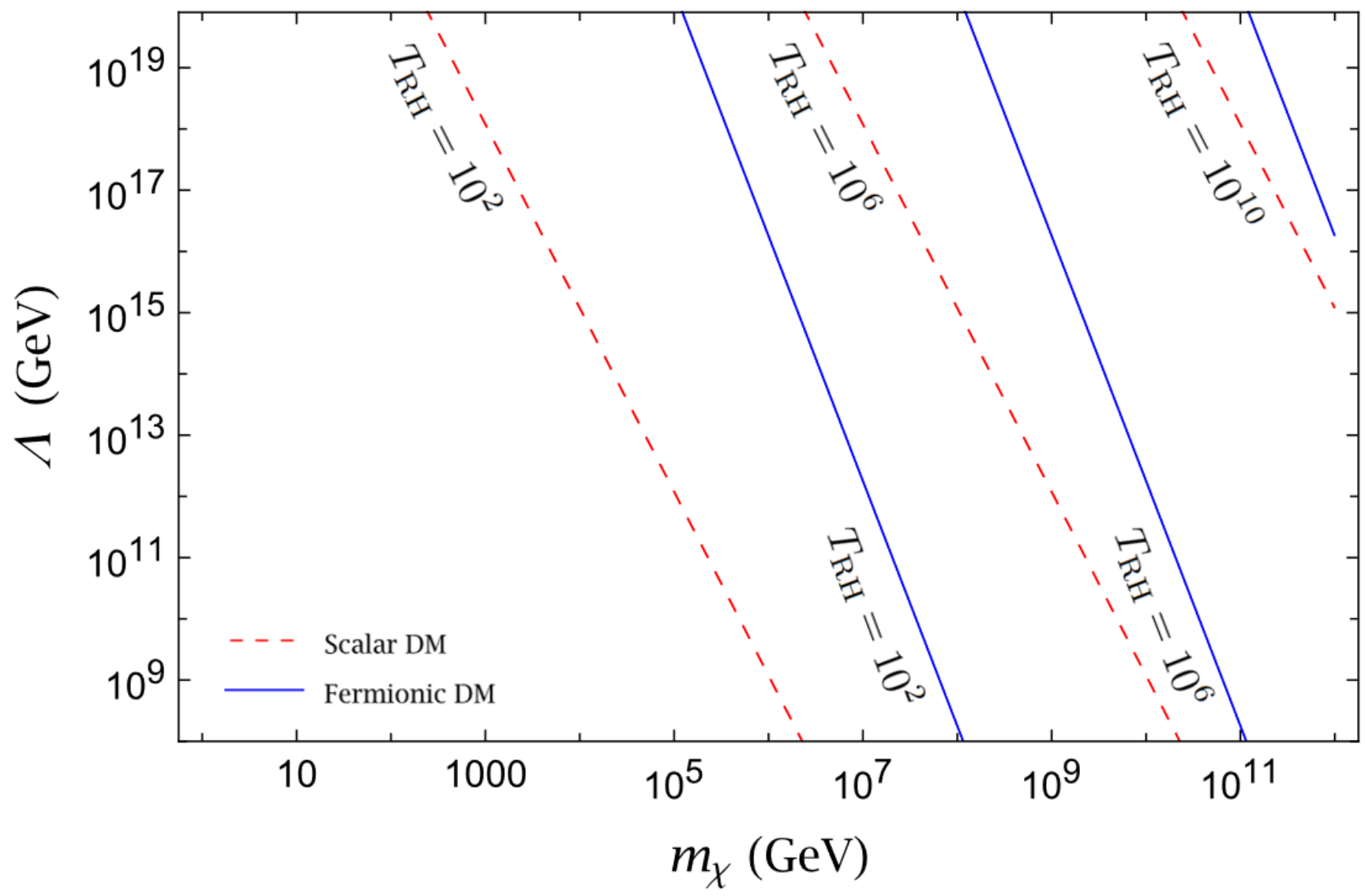}
        \caption{As in Fig. \ref{lambdavsmk2n2} but with $k=2$ and $n=0$.}
        \label{lambdavsmk2n0}
    \end{figure}

\begin{figure}[ht!]
        \centering
        \includegraphics[width=3.in]{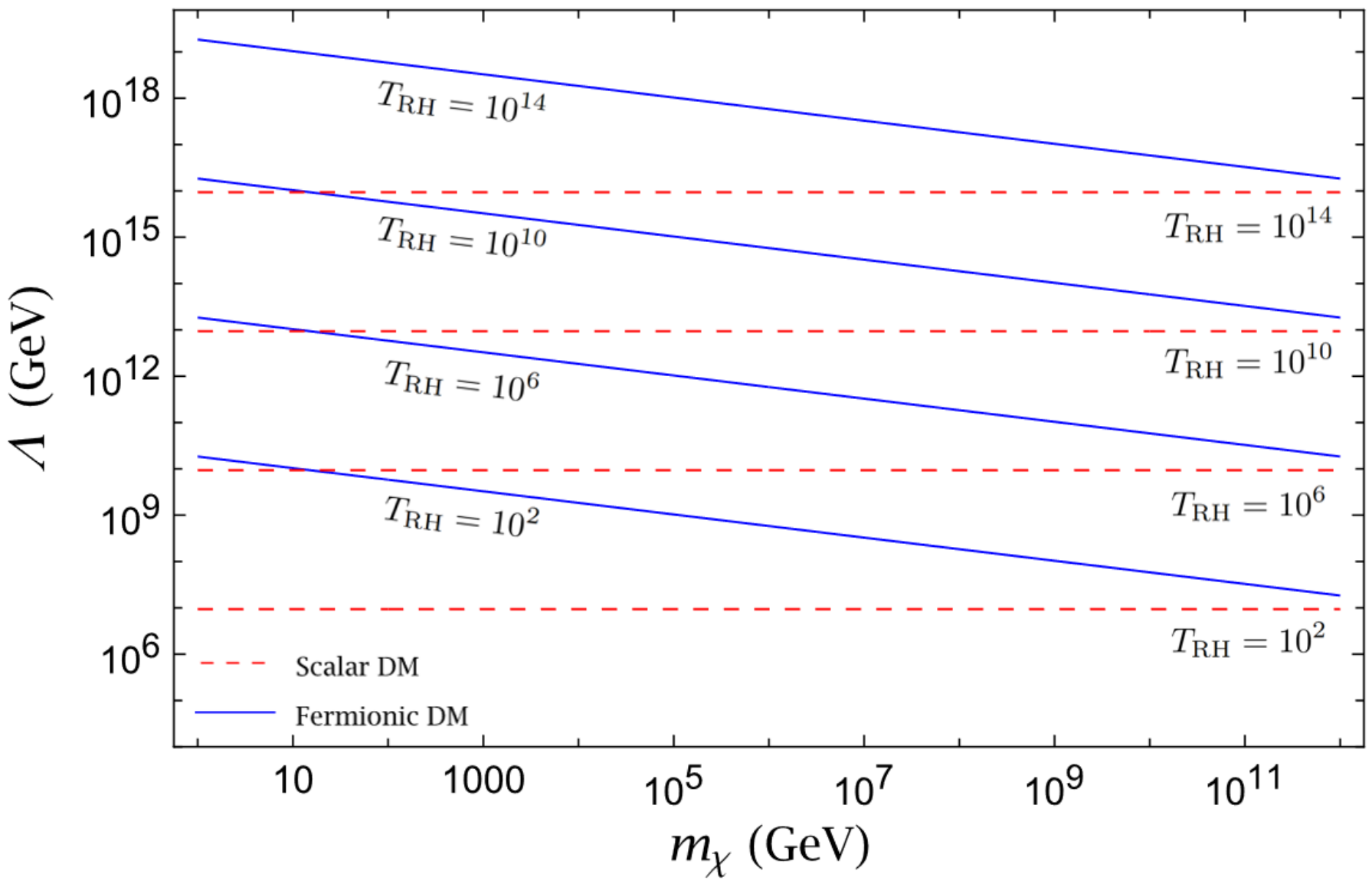}
        \caption{As in Fig. \ref{lambdavsmk2n2} but with $k=2$ and $n=6$.}
        \label{lambdavsmk2n6}
    \end{figure}

As in the case of $n=2$, one can derive the value of $\Lambda$ such that gravitational production from inflaton scattering
is critical in determining the DM density when $n=0$ or $n=6$.  
The analogue of Fig.~\ref{lambdavsmk2n2} for $m_\chi$ vs. $\Lambda$ is shown in Figs.~\ref{lambdavsmk2n0} and \ref{lambdavsmk2n6} for $n=0$ and $n=6$, respectively. 
We again use Eqs.~(\ref{Eq:omega0k2}) and (\ref{Eq:omegaphihalfk2}) for scalars and fermions. These are compared with Eqs.~(\ref{ohnltncyk2}) and (\ref{oh2ymgttrhk2}) for $m_\chi < \trh$  and $m_\chi > \trh$.  For $n=0$, we need only the case of $m_\chi > \trh$ so that we expect
$\Lambda \propto T^3/m_\chi^3$ for scalars and $\Lambda \propto T^3/m_\chi^4$ for fermions. For $n=6$, we have $n=n_c^y$ (for $k=2$)
and we must use either Eq.~(\ref{ohneqncy}) or Eq.~(\ref{oh2ymgttrhnc}) which up to a logarithmic difference both 
give $\Lambda^8 \propto T^6$ (i.e., independent of $m_\chi$) for scalars, and  $\Lambda^8 \propto T^6/m_\chi^2$ for fermions. 

\begin{figure}[ht!]
        \centering
        \includegraphics[width=3.in]{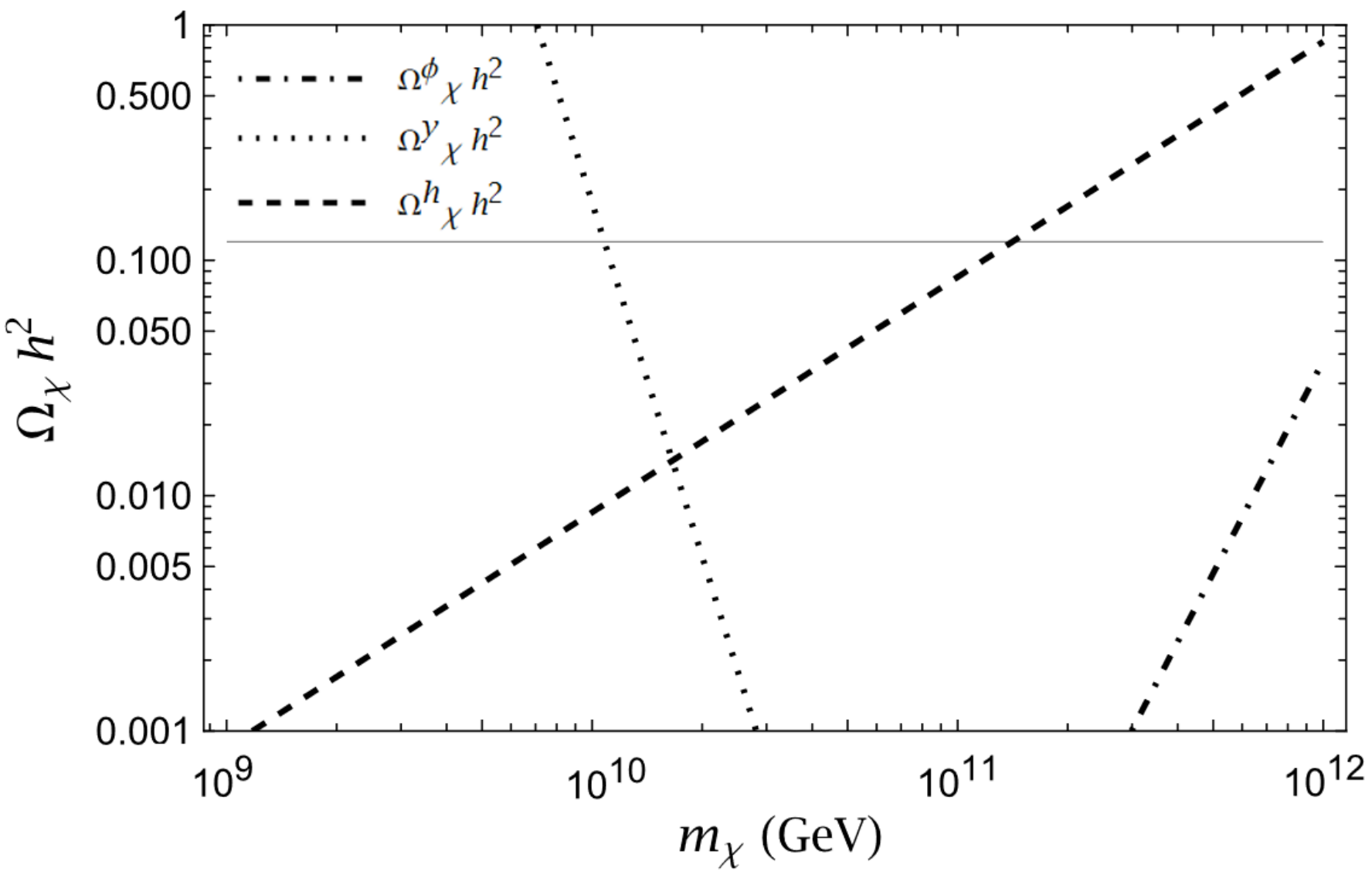}
        \caption{Contributions to the (fermionic) dark matter relic density for each of the three freeze-in sources as a function of $m_{\chi}$ for $k=2$ and $n=0$, $\Lambda = 10^{12}$ GeV and $T_{\rm RH} = 10^{7}$ GeV.}
        \label{n0k2relicComparison}
    \end{figure}

\begin{figure}[ht!]
        \centering
        \includegraphics[width=3.in]{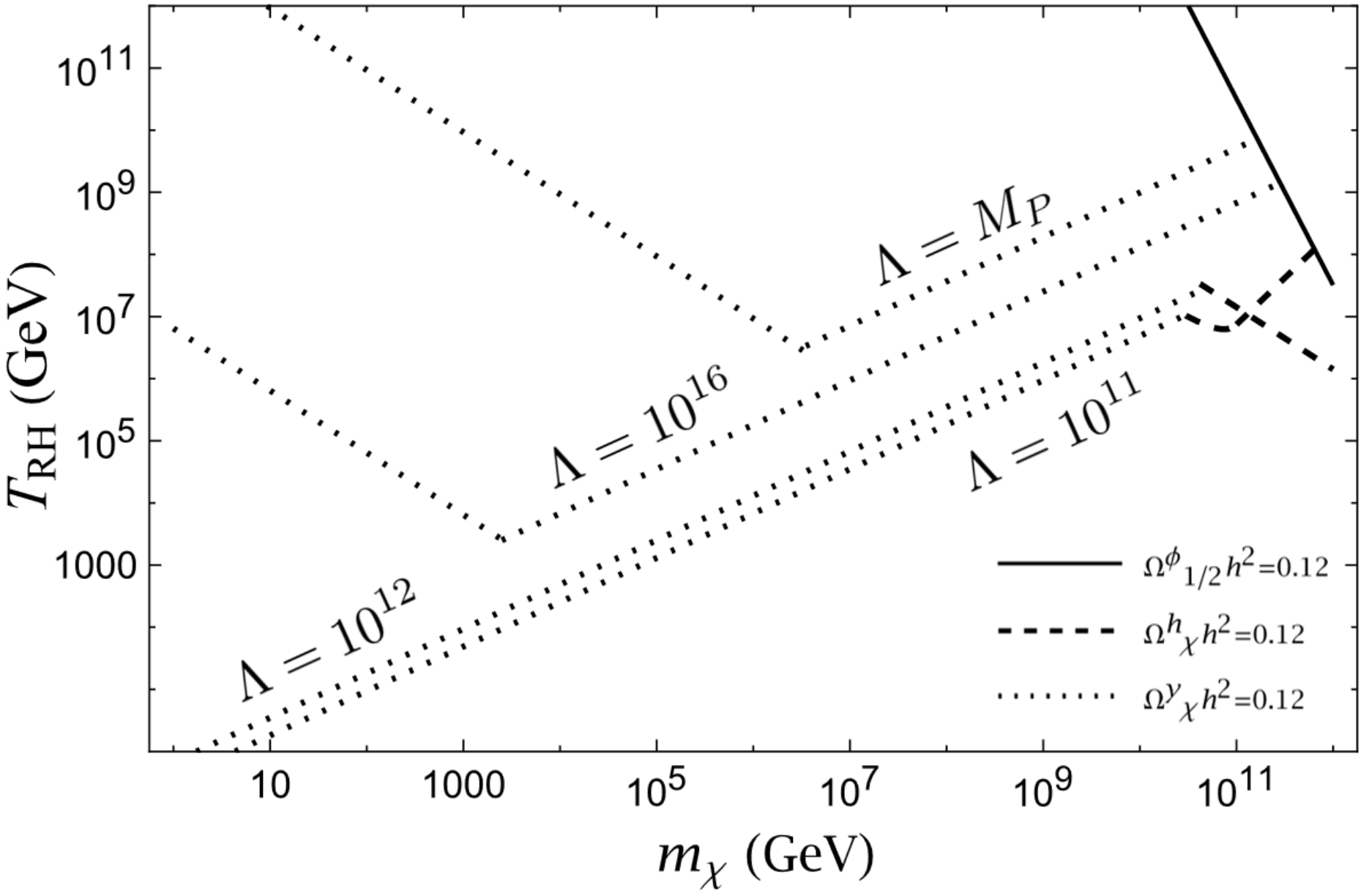}
        \caption{Relic density summary plot for fermionic dark matter production for $k=2$ and $n_2=0$. The relic density is dominated either by freeze-in sourced by the decay bath ($\Omega^{y}h^{2}$-dotted), by the gravitational bath ($\Omega^{h}h^2$-dashed), or from inflaton scattering ($\Omega^\phi_{1/2}h^2$-solid). The relic densities are displayed for various values of the suppression scale $\Lambda$ in units of GeV.}
        \label{n0k2relic}
    \end{figure}

\begin{figure}[ht!]
        \centering
        \includegraphics[width=3.in]{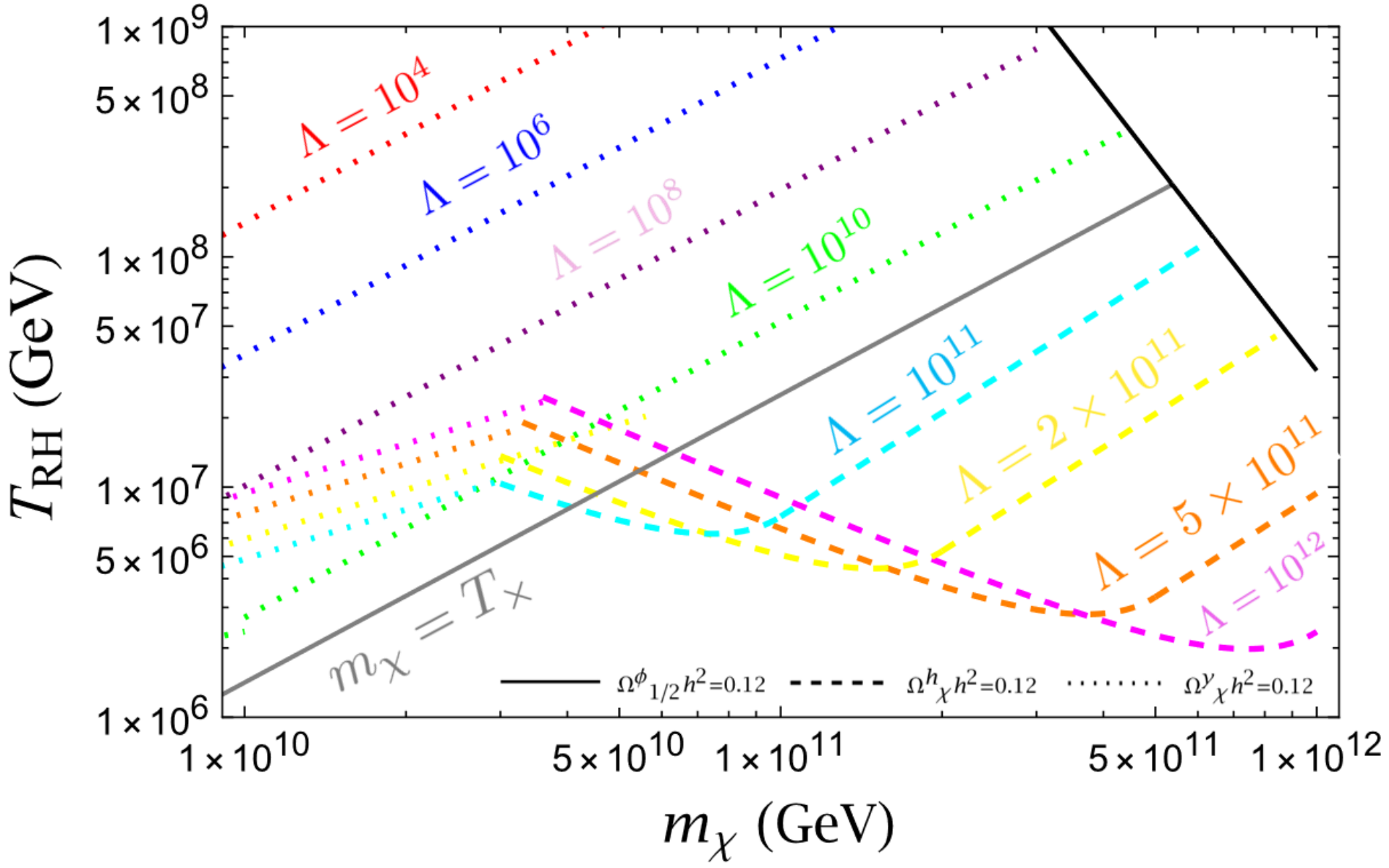}
        \caption{Relic density summary plot for fermionic dark matter production for $k=2$, $n_{1} = -4$, and $n_2=0$. The relic density is dominated either by freeze-in sourced by the decay bath ($\Omega^{y}h^{2}$), the gravitational bath ($\Omega^{h}h^{2}$), or by gravitational scattering of the inflaton ($\Omega^{\phi}h^2$). The gray line corresponding to $T_{\cross} = m_{\chi}$ approximately marks the separation between the regions where freeze-in is dominated by either the decay or gravitational radiation bath.}
        \label{n0k2relicZoomedIn}
    \end{figure}

\begin{figure}[ht!]
        \centering
        \includegraphics[width=3.in]{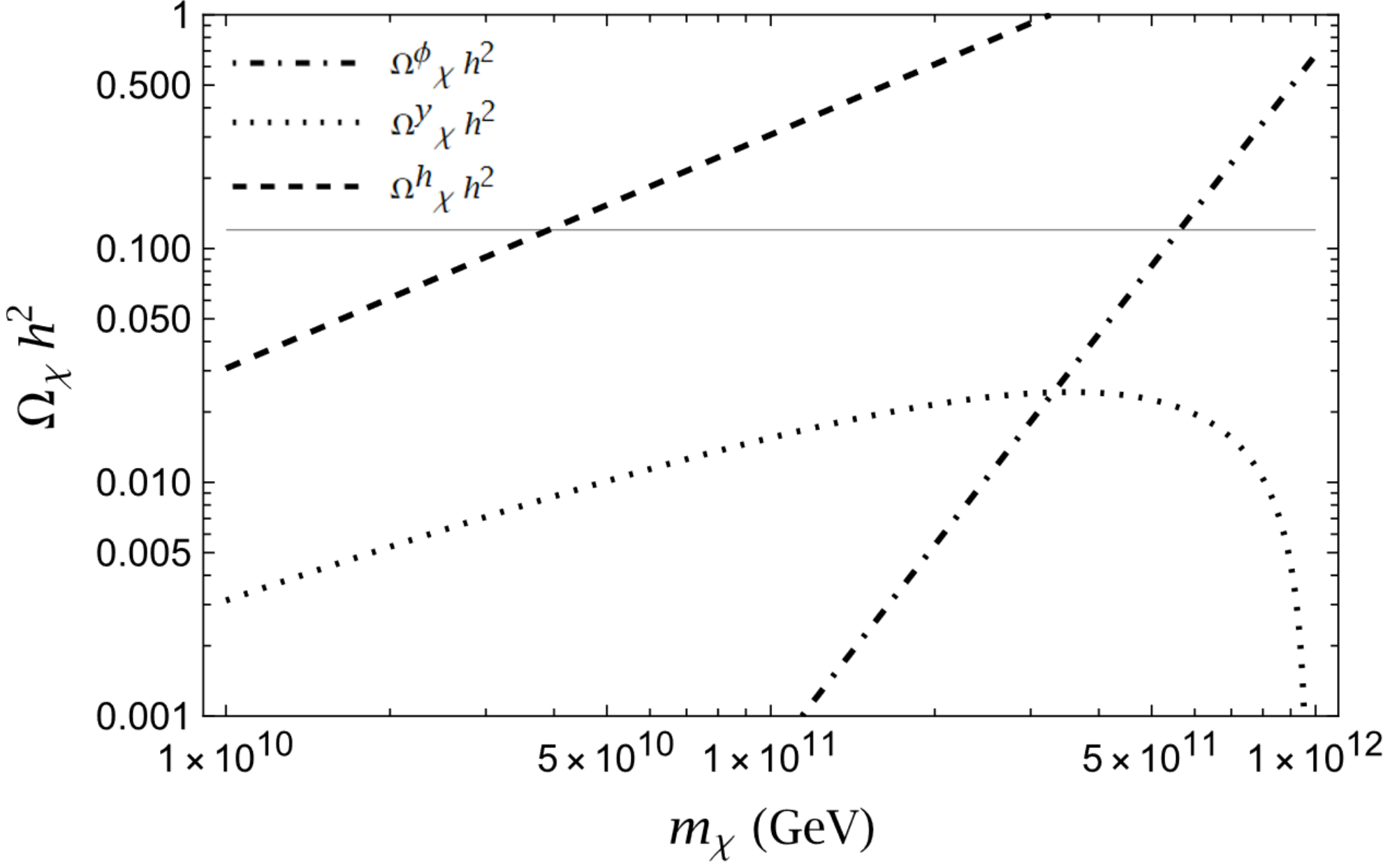}
        \caption{Contributions to the (fermionic) dark matter relic density for each of the three freeze-in sources as a function of $m_{\chi}$ for $k=2$ and $n=6$, $\Lambda =  10^{12}$ GeV and $T_{\rm RH} = 2 \times 10^{8}$ GeV.}
        \label{n6k2relicComparison}
    \end{figure}

\begin{figure}[ht!]
        \centering
        \includegraphics[width=3.in]{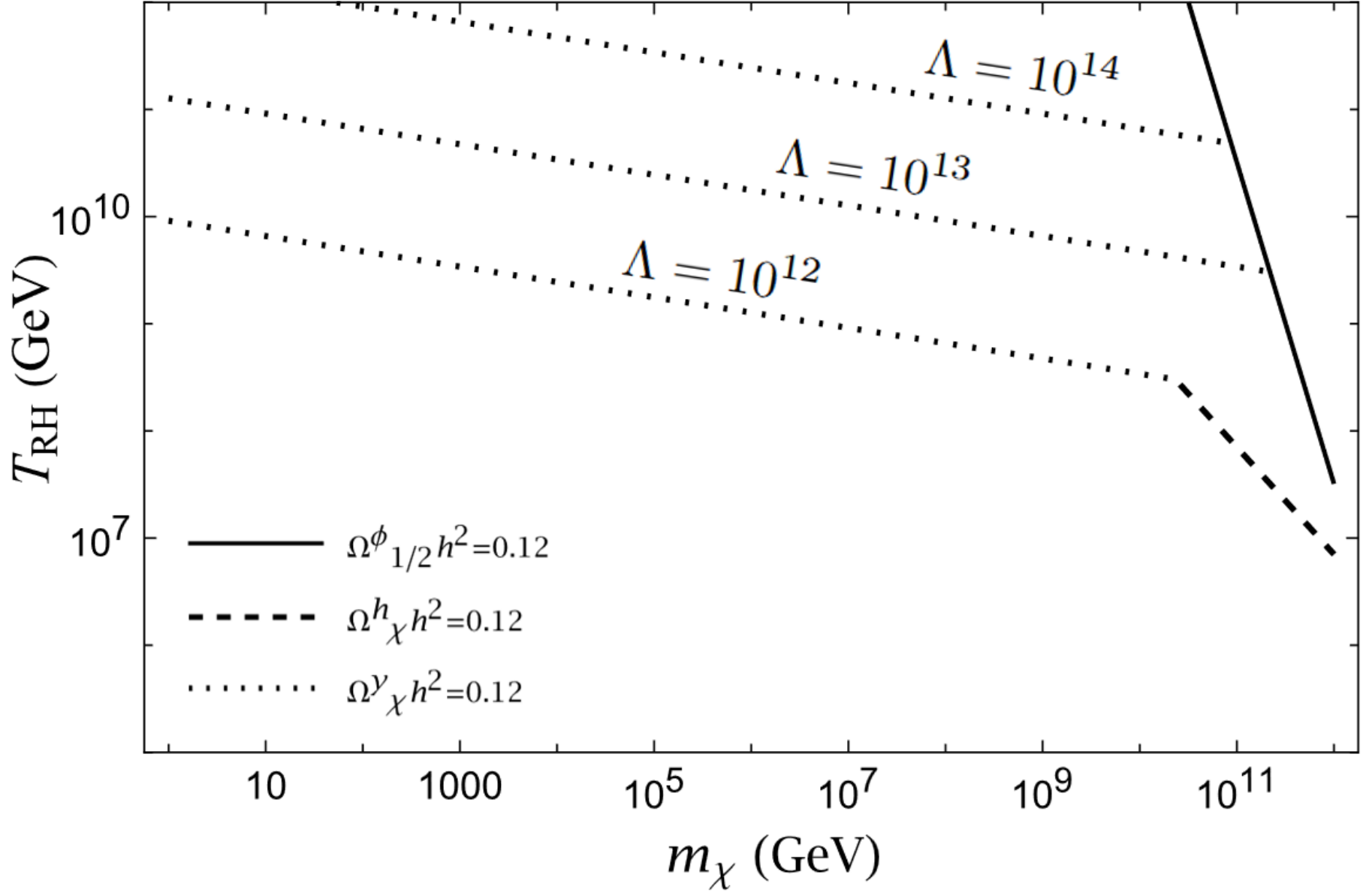}
        \caption{Relic density summary plot for fermionic dark matter production for $k=2$ and $n=6$. The relic density is dominated either by freeze-in sourced by the decay bath ($\Omega^{y}h^{2}$), freeze-in sourced by the gravitational bath ($\Omega^{h}h^2$), or by gravitational scattering of the inflaton ($\Omega^{\phi}h^2$). The relic densities are displayed for various values of the suppression scale $\Lambda$ in units of GeV. }
        \label{n6k2relic}
    \end{figure}

  The analogous results for $n=0$ and $n=6$ are presented in Figs.~\ref{n0k2relicComparison} - \ref{n6k2relic}. For $n=0$, corresponding to an SSF type interaction, the gravitational bath is dominant when $m_\chi \gtrsim  10^{10}$~GeV and yields $\Omega_\chi h^2 = 0.12$ when $m_\chi \approx 10^{11}$~GeV as seen in Fig.~\ref{n0k2relicComparison} for $\trh = 10^7$~GeV and $\Lambda = 10^{12}$~GeV. The corresponding $(m_\chi,\trh)$ plane is shown in Fig.~\ref{n0k2relic}
  and is qualitatively similar to the case with $n=2$ in Fig.~\ref{n2k2relic}.
  The negative sloped dotted lines come from Eq.~(\ref{ohnltncyk2}) with $n_2 = 0$ so that $\trh \propto \Lambda^2/m_\chi$. When $m_\chi > \trh$, the relic density is determined by Eq.~(\ref{oh2ymgttrhk2}) and $\trh^7 \propto \Lambda^2 m_\chi^5$. 
  In this case, when $\Lambda < 10^{12}$~GeV, $n_1=-4$ when $T>\Lambda$. At high masses the gravitational bath becomes important again.

For lower values of $\Lambda$, we again expect a change in $n$ from $n=0$ for $\trh < m_\chi < \Lambda$, to $n=-4$ when $m_\chi > \Lambda > \trh$. For $n=n_2=0$ (since this case is IR dominated)
we would use Eq.~(\ref{oh2ymgttrhk2}) with so that $\trh^7 \propto m_\chi^5 \Lambda^2$. However, unless $m_\chi \lesssim 10^{17} \Lambda^{10/3}/M_P^{7/3}$, equilibrium is established and the freeze-in calculation is not valid (see \cite{Bernal:2022wck} for a discussion of freeze-out during reheating). For larger masses, 
with $n=n_1 = -4$, $\Omega_\chi h^2 = 0.12$ implies $\trh^7 \propto m_\chi^9/\Lambda^2$. In this case there is a lower limit on $m_\chi \gtrsim 10^{-3} M_P^{7/17} \Lambda^{10/17}$, which is relaxed at low $\Lambda$. 
The high mass region for low $\Lambda$ is shown in Fig.~\ref{n0k2relicZoomedIn}. Notice that the dependence of $\trh$ with respect to $\Lambda$ is counter-intuitive, as higher values of $\Lambda$ require lower values of $\trh$.

The high mass region is zoomed in on in Fig.~\ref{n0k2relicZoomedIn}. For values of $\Lambda$ between $10^{11}-10^{12}$~GeV production from the gravitational bath dominates over the decay bath when $m_\chi \gtrsim 3 \times 10^{10}$~GeV. The line where $m_{\chi} = T_{\cross}$ is displayed in Fig.~\ref{n0k2relicZoomedIn}, since this is approximately the line of demarcation which distinguishes whether the gravitational or decay bath is providing the dominant contribution to the relic abundance. The DM density produced by the gravitational bath is given by Eq.~(\ref{gravchange}) for $m_\chi < \Lambda$ with
$n= n_2 = 0$ and $\trh \propto \Lambda^\frac12/m$. Whereas for $m_\chi > \Lambda$, we use Eq.~(\ref{ohnltnck2}) with $n = n_1 = -4$, and $\trh \propto m^\frac32/\Lambda^2$. This change in slope and reversal in the dependence with respect to $\Lambda$ is seen in the Figure.

Finally, for $n=6$, $\trh = 2 \times 10^8$~GeV and $\Lambda = 10^{12}$~GeV, the gravitational bath dominates the freeze-in production for the entire mass range shown in Fig.~\ref{n6k2relicComparison} and we obtain $\Omega_\chi h^2 = 0.12$ when $m_\chi \simeq 4 \times 10^{10}$~GeV. The $(m_\chi,\trh)$ plane for this case is shown in Fig.~\ref{n6k2relic}. In this case, we only consider $\Lambda > 10^{12}$~GeV, since $n=6$ is already an effective interaction and is not described by a (tree-level) massive mediator exchange. The slope of the dotted line is not affected at $\trh = m_\chi$ as can be seen from
Eqs.~(\ref{ohneqncy}) and (\ref{oh2ymgttrhnc}) where in both cases $\trh^7 \propto \Lambda^8/m_\chi$ for $n=6$.

\subsection{Freeze-In with $k \neq 2$} \label{sec:developkneq2}

\begin{figure}[ht!]
        \centering
        \includegraphics[width=3.in]{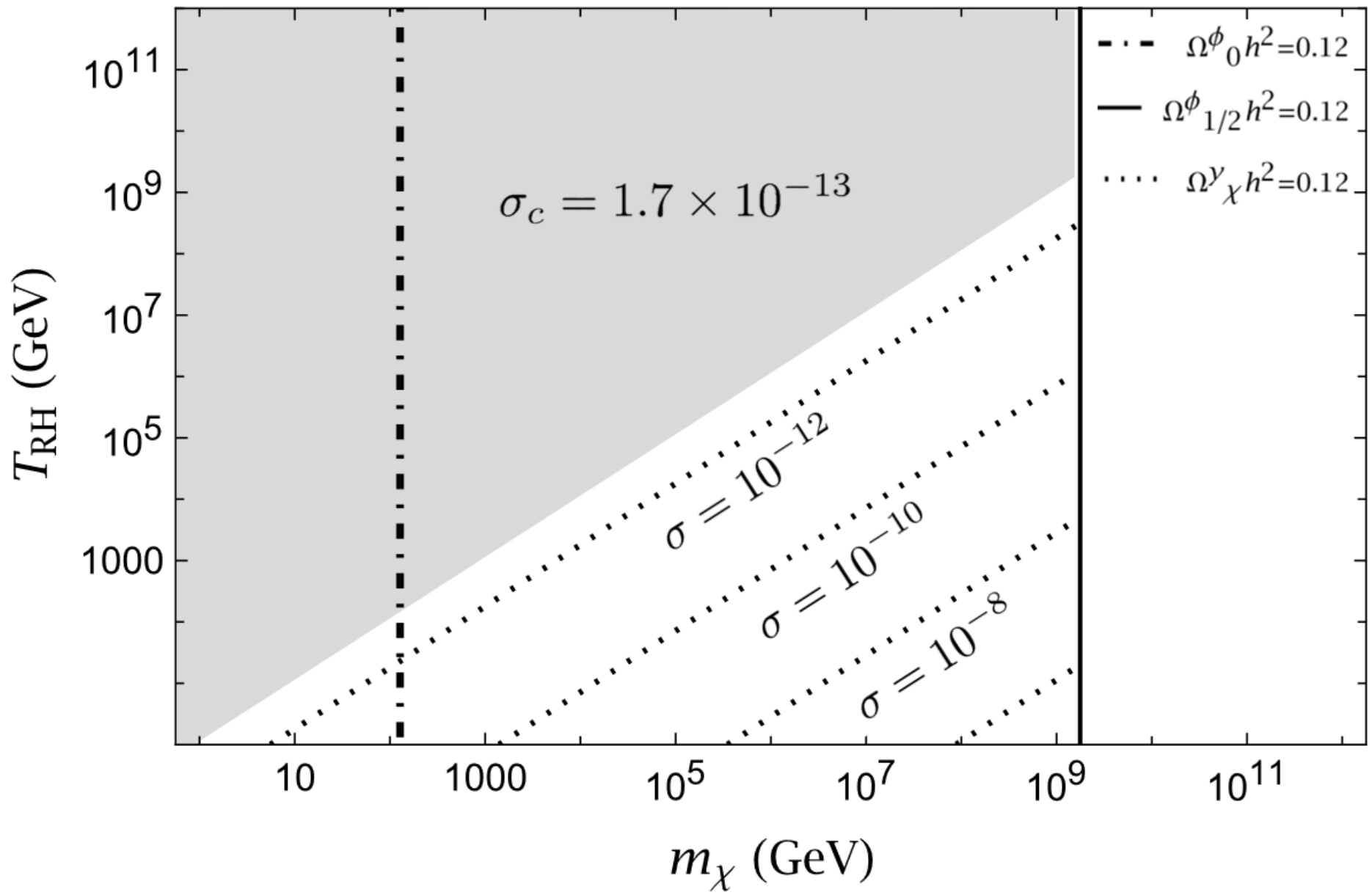}
        \caption{DM relic density summary plot for $k=4$ and $n=-2$ (scalars and fermions). The relic density is dominated by gravitational scattering of the inflaton ($\Omega^{\phi}h^{2}$) or by freeze-in sourced by the decay bath ($\Omega^{y}h^{2}$). Scalar DM masses above 128 GeV and fermionic DM masses above $1.8 \times 10^9 \text{ GeV}$ are not allowed since direct gravitational production from the inflaton will overproduce DM. In the low mass and high $\trh$ region, there is a critical value of $\sigma$ for which freeze-in sourced by the decay bath can account for the observed relic density. }
        \label{nm2k4relic}
    \end{figure}

Next, we investigate the case for $k=4$, which leads to strikingly different results. We first consider scalar 
DM production via a contact term, namely $n=-2$. For $k=2$, we saw that for large masses and reheating 
temperatures, scalar DM production via gravitational scattering of the inflaton exceeded production by 
freeze-in sourced by either of the two thermal baths.  For $k=4$, gravitational scattering of the inflaton 
dominates even at much lower DM masses particularly when the reheating temperature is relatively low \cite{cmov}. 
Indeed, there is a critical mass above which 
gravitational scattering saturates the relic density 
constraint independent of the reheating temperature.
This can be seen as a consequence of the conformal invariance in the case of a quartic potential for $\phi$: 
there is no scale explicitly introduced in the Lagrangian.
The critical mass for $\chi$, which is independent of $n$ as well, is found 
from Eq.~(\ref{n0phi}) for scalars using (\ref{Omega}) to convert $n_\chi(\trh)$ to $\Omega_\chi h^2$ \cite{cmov},
\begin{align} 
{m_{\chi}|}_{\rm crit}&=\frac{24 \pi (0.12)}{\sqrt{3} \alpha^{3/4}} \frac{M_{P}^{3}}{\rho_{\rm end}^{3/4} \Sigma^{4}_{0}} \cdot \frac{1}{5.88 \times 10^{6} \text{ GeV}^{-1}} \nonumber  \\ &= 128 \text{ GeV} \, .
\label{scalarul}
\end{align} 
Scalar dark matter masses greater than ${m_{\chi}|}_{\rm crit}$ are excluded since the gravitational scattering of the inflaton would overproduce DM. 
For $m_\chi < {m_{\chi}}|_{\rm crit}$, gravitational scattering is insufficient in producing $\Omega_\chi h^2 = 0.12$, but freeze-in from the decay bath can provide the necessary density of DM depending on the coupling. For $m_{\chi} < T_{\rm RH}$ and $m_{\chi} < {m_{\chi}|}_{\rm crit}$, we can use Eq.~(\ref{pastrh}) to obtain the relic density. When the last term dominates (when $m_\chi < \trh$ and $n < -1$), it is $k-independent$ and we have the same critical coupling, $\sigma_c$. The $(m_\chi,\trh)$ plane is shown in Fig. \ref{nm2k4relic}. For scalars, only masses to the left of the dot-dashed line are allowed.

    For fermionic DM, because of the helicity suppression,
    much larger masses are allowed \cite{cmov}. We can now use
    Eq.~(\ref{Eq:nhalf}) to obtain a critical mass for fermions
    \begin{align}
{m_{\chi}|}_{\rm crit}&= \left(\frac{24 \pi (0.12)}{\sqrt{3} \alpha^{3/4}} \frac{\sqrt{\lambda} M_{P}^{3}}{\rho_{\rm end}^{1/4} \Sigma^{4}_{12}} \times \frac{1}{5.88 \times 10^{6} \text{ GeV}^{-1}} \right)^\frac13
\nonumber  \\ &= 1.8 \times 10^9 \text{ GeV} \, .
\label{fermiul}
  \end{align}
In this case, all points to the left of the vertical solid line in Fig.~\ref{nm2k4relic} are allowed. In the area below the shaded region, couplings greater than $\sigma_c$ are required. 
Note that this region is exactly the same as in Fig.~\ref{nm2k2relicCombined}. This is understandable. For $m_\chi<\trh$,  DM production from the decay bath is IR-dominated, and the physics above $\trh$ (and thus the dependence on $k$) does not influence the DM production. On 
the other hand, for $m_\chi>\trh$, 
from Eq.~(\ref{nlt2}) we see that while $n_\chi^y\propto (a_m/a_{\rm max})^3$ for $k=2$, it is proportional to $(a_m/a_{\rm max})^2$ for $k=4$. This means that a much smaller value of $\sigma$ is necessary to fulfill the relic abundance constraint, which is clear when comparing Figs.~\ref{nm2k2relicCombined} and \ref{nm2k4relic}. This is also an important conclusion of our work. The larger equation of state parameter for the inflaton field ($w = 1/3$ relative to $w=0$) implies a  much larger density of DM and requires a much smaller coupling to avoid overproduction.

\begin{figure}[ht!]
        \centering
        \includegraphics[width=3.in]{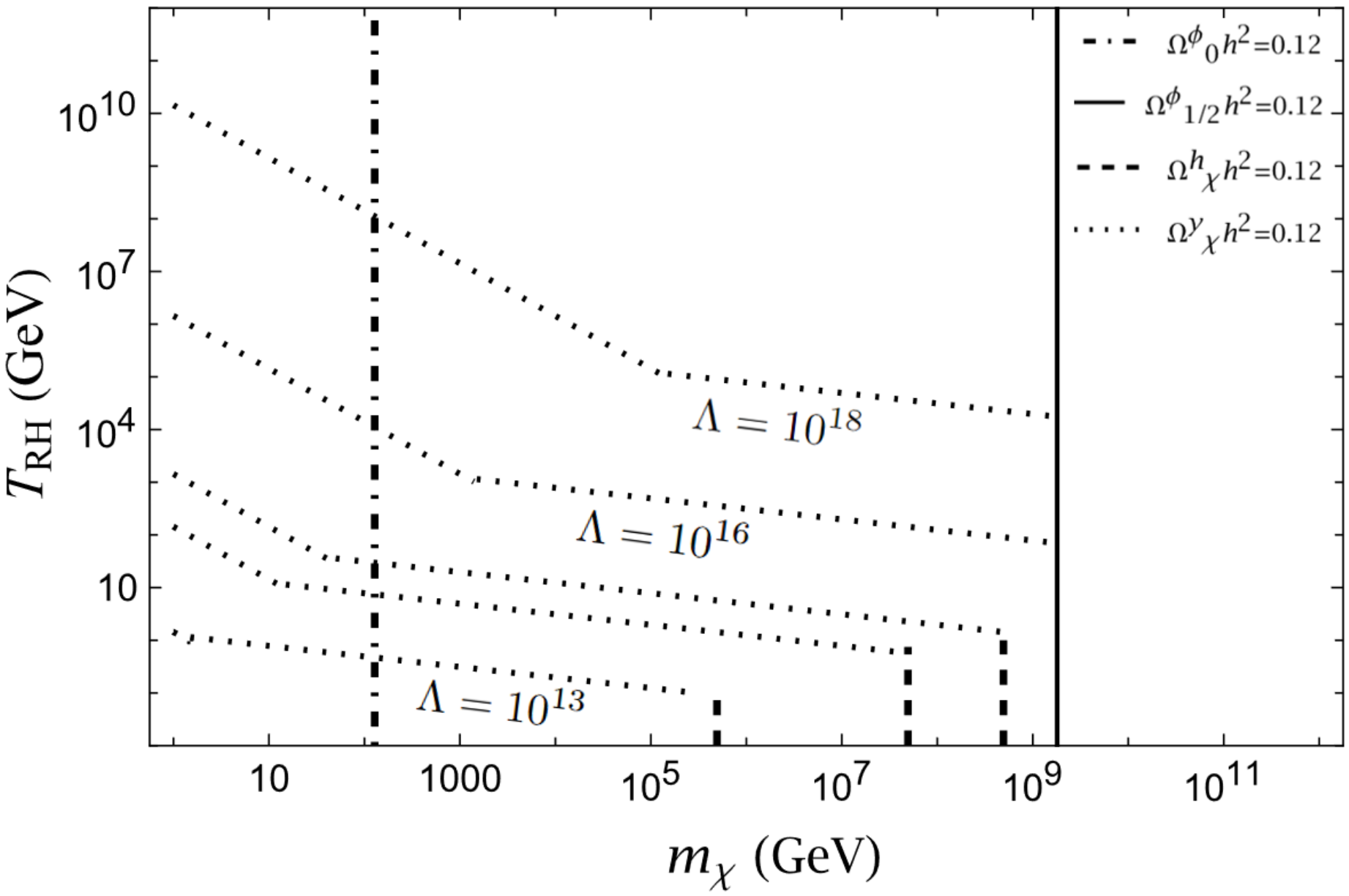}
        \caption{DM Relic density summary plot for $k=4$, $n=0$ and several values of $\Lambda$ with units in GeV. The relic density is dominated by freeze-in sourced by the decay bath ($\Omega^{y}h^{2}$), the gravitational bath ($\Omega^{h}h^{2}$), or by gravitational scattering of the inflaton ($\Omega^{\phi}h^{2}$). The vertical lines correspond to the upper limits given in Eqs.~(\ref{scalarul}) and (\ref{fermiul}). }
        \label{n0k4relic}
    \end{figure}

The analogous results for $k=4$ and $n=0$ are plotted in Fig. \ref{n0k4relic}. As in Fig.~\ref{nm2k4relic}, $m_\chi$ must be less than the mass given by the vertical lines depending on whether the DM is bosonic or fermionic. This figure shows the values of $\Lambda$ needed to obtain $\Omega_\chi h^2 = 0.12$ across the $(m_\chi, \trh)$ plane. Dotted lines correspond to regions where the DM production is dominated by scattering in the thermal decay bath.
The slopes of these lines can be determined from Eqs.~(\ref{ohnltncy}) and (\ref{oh2ymgttrh}) which give
$\trh \propto 1/m_\chi$ for $m_\chi < \trh$ and $\trh \propto 1/m_\chi^\frac35$ for $m_\chi > \trh$. Comparing our previous result for $n=0$ with $k=2$ shown in  Fig.~\ref{n0k2relic},
we observe that, for $k=4$, much larger values of $\Lambda$ are necessary to avoid overclosing the Universe than that for $k=2$.
For large masses and low reheating temperatures, the gravitational bath is relevant as seen by the vertical dashed lines. This can be seen from Eq.~(\ref{Ohngtnc}) which is independent of $\trh$ for $k=4$.

\begin{figure}[ht!]
        \centering
        \includegraphics[width=3.in]{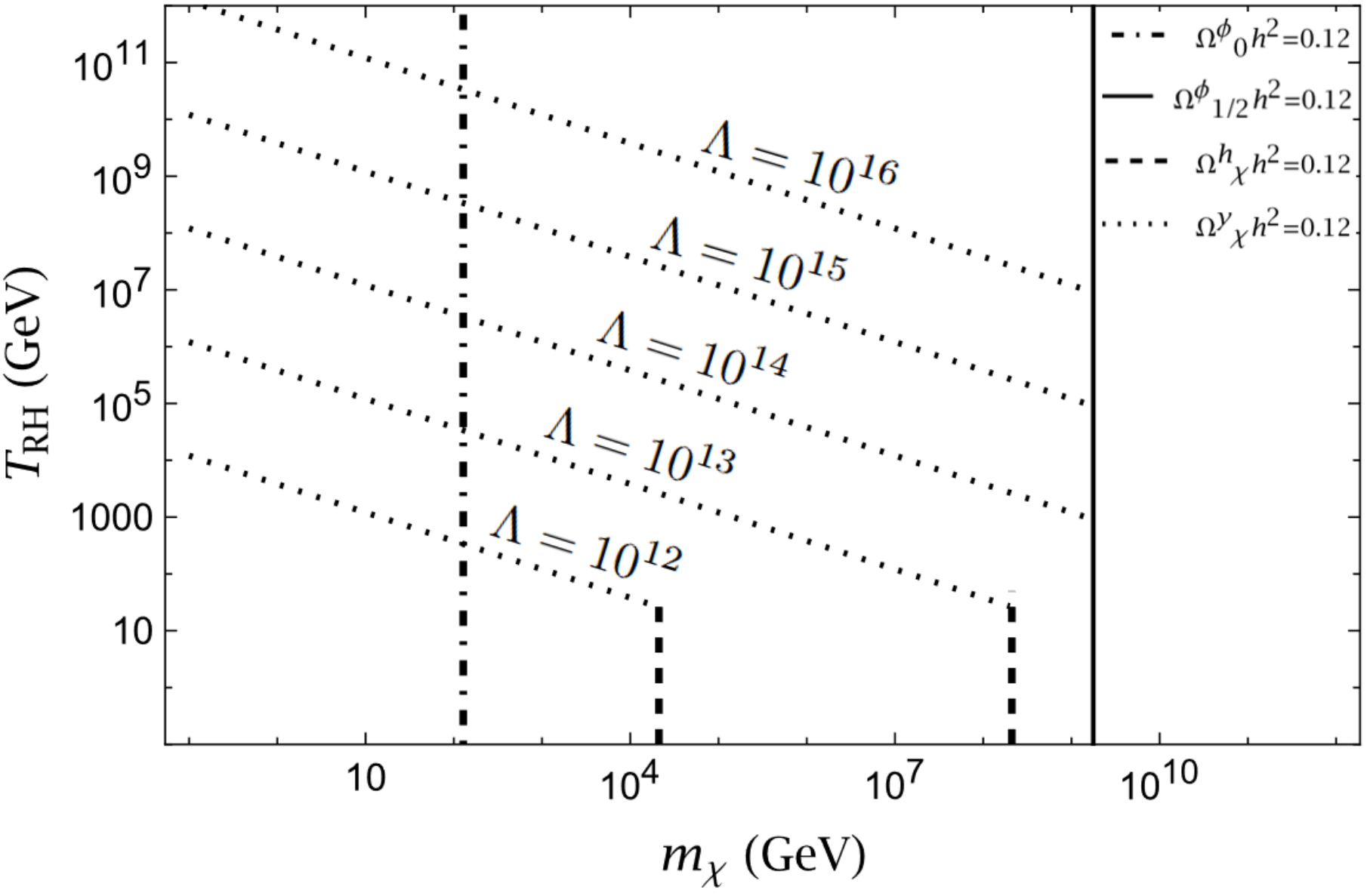}
        \caption{As in Fig.~\ref{n0k4relic} for $k=4$ and $n=2$. }
        \label{n2k4relic}
    \end{figure}

 The case of $k=4$ and $n=2$ is shown in Fig.~\ref{n2k4relic}. Note that while $n=2$ corresponded to IR production for the decay bath when $k=2$, this interaction type now corresponds to UV production when $k=4$ since $n^{y}_{c} = 2/3$.
 The dotted lines for $\Omega_\chi h^2 = 0.12$ from the decay bath are found from Eq.~(\ref{ohngtncy}) and give $\trh \propto (\Lambda^4/m_\chi)^\frac35$. Freeze-in production sourced by the gravitational bath will be the dominant source of DM production consistent with the relic density for a large range of masses at low reheating temperatures, $\trh \lesssim 60$~GeV. 

  In the case of $k=6$, we find that for $\trh \gtrsim 10^{-3} \text{ GeV}$, $T^{y}_{\rm max}$ will always exceed $T^{h}_{\rm max}$ (see Fig. \ref{Fig:TRHmax}). As a result, there are no viable circumstances in which freeze-in DM production sourced by the gravitational bath will exceed production from the decay bath due to the BBN limit of $\trh > 4 \text{ MeV}$. Furthermore, this is compounded by the need for a large reheating temperature ($\gtrsim 10^{10}$~GeV for $k=6$) to avoid the fragmentation of the inflaton condensate \cite{Garcia:2023dyf} which would revert the $k=6$ condensate to a gas of massless inflatons characterized by $k=4$. This is also true for $k > 6$\footnote{The lower bound from fragmentation drops rapidly for larger $k$ \cite{Garcia:2023dyf}.}. For these reasons, we do not present a full set of results for $k=6$. However, it is worth commenting on the behavior of the direct gravitational production from the inflaton for $k=6$. We find that for $k=6$, scalar DM particles with $m_{\chi} \gtrsim 10 \text{ GeV}$ and fermionic DM with $m_{\chi} \gtrsim 10^{8} \text{ GeV}$ are excluded due to overproduction from gravitational scattering of the inflaton. These aren't exact equalities, however, because there is a slight dependence upon $\trh$ in each case, which is in contrast to the $k=4$ results shown above (namely, the vertical lines for $\Omega^{\phi}_{0}h^2 = 0.12$ and $\Omega^{\phi}_{1/2}h^2 = 0.12$ in the $k=4$ relic summary plots). In particular, the $\trh$-dependence of the scalar and fermionic DM relic abundances from gravitational scattering of the inflaton for $k=6$ are given by $\Omega^{\phi}_{0}h^{2} \propto \trh^{-1/3}m_{\chi}$ and $\Omega^{\phi}_{1/2}h^{2} \propto \trh^{-1/3}m_{\chi}^{3}$ respectively. In short, this means that the regions of the ($m_{\chi},\trh$) plane in which direct gravitational production from the inflaton over-produces DM are even larger for $k=6$ than $k=4$ or $k=2$. These results are consistent with those presented in \cite{cmov}.

\section{Conclusions} \label{sec:conclusions}

    In this work, we have made a detailed study and comparison of the sources of DM production stemming from 1) gravitational inflaton scattering, 2) freeze-in from scattering within the thermal bath produced by inflaton decay, and 3) freeze-in from scattering within the thermal bath which is itself produced by gravitational scattering of the inflaton. As inflation ends, the universe enters into a period in which the energy density is dominated by the oscillations of an inflaton condensate. Dark matter production can be highly sensitive to the equation of state of the condensate which in turn depends on the shape of the 
    inflaton potential near its minimum. As an example, 
    we started with a so-called T-model attractor potential \cite{Kallosh:2013hoa} given in Eq.~(\ref{Vatt}) which very simply parameterizes the equation of state, $w = (k-2)/(k+2)$, when expanded about the minimum of the potential,  $V(\phi)\propto \phi^k$. Any inflaton potential which admits a similar parameterization
    would lead to equivalent results. This class of potentials provides a tilt to the CMB anisotropy spectrum in excellent agreement with observations.  The height of the potential is fixed by the normalization of the CMB anisotropy spectrum. 
    
    Direct gravitational production of DM from the inflaton depends only on the equation of state of the inflaton condensate and the spin of the DM \cite{MO,cmov,kkmov,Garcia:2023obw,Bernal:2018qlk}
    if only minimal (universal) gravitational interactions are considered. If the dark matter candidate is coupled to the SM and it comes into thermal equilibrium with the SM,
    then its relic density is found from the extremely well studied thermal freeze-out scenario \cite{Arcadi:2017kky,Arcadi:2024ukq}. However, dark matter can also be produced when it never comes into equilibrium
    through freeze-in \cite{fimp}. One of the earliest examples of freeze-in is the production of the gravitino as dark matter \cite{nos,ehnos,kl}. As a particle with essentially only gravitational interactions, it never achieves thermal equilibrium, yet it remains a good dark matter candidate.

    Gravitational interactions of the inflaton also produce
    SM particles which form a thermal bath (called here the gravitational bath) after thermalization. While this bath 
    can not be responsible for reheating (in minimally coupled gravity), it may dominate the radiation density for
    some period of time after inflation ends. Furthermore, the gravitational bath quickly reaches a maximum temperature of $\simeq 10^{12}$~GeV initially.  Reheating (defined when the radiation density becomes greater than the inflaton density) is most easily accomplished through inflaton decay.
    The thermal bath produced by these decays (the decay bath) eventually comes to dominate the radiation density (for $k< 7$) when $T = T_{\cross}$ (see Fig.~\ref{Fig:rhoR_num}).

    Of course DM may couple to the SM model through a wide variety of interactions, stronger than gravity, 
    yet too weak to achieve equilibrium. We parameterize these interactions with a freeze-in production rate $R(T)=\frac{T^{n+6}}{\Lambda^{n+2}}$, where $\Lambda$ is an effective high energy scale. Depending on the value of $n$,
    the final density of DM particles may be sensitive to the UV scale (near $\tmax$) or the IR scale (near $\trh$ or $m_{\chi}$, depending on the interaction). 
    For $n$ greater than a critical value, $n_c$, the UV scale determines the relic density. The two thermal baths each have a different $n_c$ which depends on $k$.  Most freeze-in analyses have focused on DM production from the decay bath. However, freeze-in processes occurring at $T>T_{\cross}$ originate from the gravitational bath and may be important. 

    In this work, we have undertaken a comprehensive study of the interplay between three sources of DM production. We compare
    the resulting DM densities from inflaton scattering with
    freeze-in from both the gravitational and decay bath. 
    The production rates for the two thermal baths are the same as a function of temperature (not as a function of $a$) and their strength is fixed by $\Lambda$. Among our key results is the determination of lower limits to the scale $\Lambda$ for which the gravitational production from inflaton scattering dominates. These results are well summarized by Figs.~\ref{lambdavsmk2n2}, (\ref{lambdavsmk2n0} and \ref{lambdavsmk2n6}), where we show the value of the BSM 
    scale $\Lambda$ above which 
    the relic density of scalar (red-dashed) and fermionic (blue-full) dark 
    matter is saturated by the gravitational production for $k=2$ and $n=2$ ($n=0$ and $n=6$ respectively). We conclude that for BSM scales above $\sim 10^{10}$ GeV, for a very large range of DM mass, one must consider gravitational effects when computing the relic abundance from the freeze-in mechanism. These effects increase in importance for larger values of $k$.

In the case of a contact interaction or the high temperature limit of certain interactions mediated by scalars or vectors, the value of $n$ is effectively -2. This leads to a DM production rate with
$R(T) = \sigma^2 T^4$, and our result in this case for $k=2$ is depicted in Fig.~\ref{nm2k2relicCombined}.
For $m_\chi<\trh$, the dark matter relic density is solely  determined by $\sigma$ for scattering in the decay bath and  $\sigma\simeq 1.7\times 10^{-13}$ is needed to obtain $\Omega_\chi h^2 = 0.12$.  We have shown that larger values of $\sigma$ are necessary to populate the dark sector for $m_\chi > \trh$. For even larger mass, the production from gravitational inflaton scattering dominates the freeze-in. For $n=-2$, the gravitational bath was only important at very large DM masses and high values of $\sigma$. The importance of the gravitational bath is enhanced
for fermionic DM as the production from inflaton scattering is suppressed. It is also enhanced for large DM masses.

Some scalar or vector mediated interactions with a high BSM scale
($\Lambda > \tmax$) are characterized by $n=2$. In this case, there is a larger range of parameters for which the gravitational bath is important. 
Moreover, for smaller BSM scales ($< \tmax$), we found that the interaction type switches during reheating with $n_{1}=-2$ at the onset of reheating changing to $n_{2} = 2$ when $T\lesssim \Lambda$. This analysis (found in Sections \ref{sec:FIscat3}, \ref{nharhswitch}, \ref{nyarhswitch}) is new even for the decay bath, and we have shown that it can significantly impact the resulting DM relic density. 
The change in production rate also occurs for other interactions 
such as SSS (scalars producing scalar DM mediated by scalars) where $n$ changes from -6 at high temperature to -2 at low temperature, or SSF where $n$ changes from -4 to 0. 

One gravitational portal interaction during reheating that we did not discuss in detail in this work is SM SM $\rightarrow h_{\mu \nu} \rightarrow$ DM DM. However, this interaction is easily accounted for (up to phase space factors) by setting $\Lambda = M_{P}$ in our relic density expressions in the Appendix, and extrapolating our relic summary plots to include $\Lambda = M_{P}$. This interaction was also considered in detail in \cite{cmov}.

Here, we have also stressed the importance of the temperature $T_{\cross}$. For $T > T_{\cross}$, the radiation density originates from gravitational scattering. Then for $m_\chi > T_{\cross}$, the relic density from freeze-in is always produced from the gravitational bath, even for interaction types and values of $n$ which give IR freeze-in. Additionally, $T_{\cross}$ is quite sensitive to $k$ for a fixed value of $\trh$. As a result, the DM mass range over which the gravitational bath can be the dominant source of freeze-in dark matter is in turn $k$-dependent, which we explicitly demonstrate for $k=4$.

We have not examined here in detail the parameter space where 
freeze-out is important. That is when the dark matter comes into thermal equilibrium with the thermal bath. 
For $n=-2$, we have remarked that for the large values of $\sigma \gtrsim 0.1$, freeze-in only takes place for very large masses ($\gtrsim 10^8$~GeV) as equilibrium occurs with the decay bath
 for lower masses. We 
expect that it is difficult to attain equilibrium with the gravitational bath.  We have also not examined other possible 
reheating profiles. For example, a model of reheating where the production of the decay bath is accomplished through an 
intermediate decay was proposed in \cite{oleg1,oleg2}. We expect 
the gravitational bath to be significant in this case (for large DM masses) and we plan to explore these possibilities in future work. 

\section*{Acknowledgements} \label{sec:acknowledgements}
  We would like to thank W. Ke, Jong-Hyun Yoon, Simon Clery and Mathieu Gross for helpful conversations. 
  This project has received support from the European Union's Horizon 2020 research and innovation program under the Marie Sklodowska-Curie grant agreement No 860881-HIDDeN and the CNRS-IRP project UCMN. 
  The work of K.A.O.~was supported in part by DOE grant DE-SC0011842  at the University of
Minnesota. Y.M. acknowledges support by Institut Pascal at Université Paris-Saclay during the Paris-Saclay Astroparticle Symposium 2024, with the support of the P2IO Laboratory of Excellence (program “Investissements d’avenir” ANR-11-IDEX-0003-01 Paris-Saclay and ANR- 10-LABX-0038), the P2I axis of the Graduate School of Physics of Université Paris-Saclay, as well as the CNRS IRP UCMN.

\appendix*
\section{} 
\label{sec:appendix}

    In this section, we have collected expressions for the relic abundance contributions from freeze in sourced by the gravitational scattering of the inflaton as well as from scattering within the radiation baths as a function of $n$ and $k$. 

For the production from the gravitational scattering of the inflaton, the general expression for the relic density is
\bea
\frac{\Omega_0^\phi h^2}{0.12} &\simeq& 3.5 \times 10^{13-\frac{24}{k}} \left(\frac{k+2}{6k-6}\right) \alpha^{\frac{4-k}{4k}} \left(\frac{\rhoend}{10^{64} ~{\rm GeV}^4}\right)^{\frac{k-1}{k}}
\nonumber
\\
&
\times& \left(\frac{\trh}{10^{10}~{\rm GeV}}\right)^{\frac{4-k}{k}} 
\left( \frac{m_\chi}{10^7 ~{\rm GeV}}\right) \Sigma_0^k\,,
\label{Eq:omega0}
\eea
which reduces to Eq.~(\ref{Eq:omega0k2}) for $k=2$.
The corresponding result for the production of fermions is 
\bea
\frac{\Omega_{1/2}^\phi h^2}{ 0.12} &=&
\frac{0.14 \times 10^{-\frac{18}{k}} \alpha^{\frac{4-k}{4k}} }{(2.4)^{\frac{8}{k}}}\frac{k+2}{k(k-1)}
\left(\frac{10^{-11}}{\lambda}\right)^{\frac{2}{k}} 
\nonumber
\\
& \times &
\left(\frac{\rhoend}{10^{64} ~{\rm GeV}^4}\right)^{\frac{1}{k}} \left(\frac{\trh}{10^{10}~{\rm GeV}}\right)^{\frac{4-k}{k}} \nonumber \\
&\times &
\left(\frac{m_\chi}{ 10^{7} ~{\rm GeV}}\right)^3  \Sigma_{1/2}^k \, ,
\label{Eq:omegaphihalf}
\eea
which reduces to Eq.~(\ref{Eq:omegaphihalfk2}) for $k=2$.
    
Next we list the results for the relic density from scattering within the gravitational bath.   
    For $\Lambda > \tmax^h$, we find

    (i) For $n > n_c^h = \frac{-6}{k+2}$,
\begin{align} 
\Omega^{h}_{\chi}h^{2} &= \alpha^{\frac{k+2}{2k}}\sqrt{3}\left(\frac{k+2}{nk+2n+6}\right) \left(\frac{2k+4}{6k-3}\right)^{\frac{3(k+1)}{7-4k}} \nonumber \\ & \times \frac{\tmax^{h \hspace{1mm} n+6}T_{RH}^{\frac{4-k}{k}}m_{\chi}M_{P}}{\Lambda^{2+n}\rhoend^{\frac{k+1}{k}}} (5.88 \times 10^{6} \text{ GeV}^{-1})
\label{Ohngtnc}
\end{align}

(ii) For $n < n_c^h = \frac{-6}{k+2}$, as noted in the text, the relic density from the gravitational bath is never dominant if $m_\chi < \trh$ and $n < n_c^y$, as we are IR dominated and the decay bath always wins if we require that $\trh$ is above the BBN limit. Therefore
we provide only the case with $m_\chi > \trh$ for $n<n_c^h$:
\begin{align} 
\Omega^{h}_{\chi}h^{2} &= \alpha^{\frac{k+2}{2k}}\sqrt{3}\left( \frac{k+2}{-nk-2n-6}\right) \left(\frac{2k+4}{6k-3}\right)^{\frac{3(k+1)}{7-4k}} \times \nonumber \\ & \frac{T^{h \hspace{1mm} \frac{6(k+1)}{k+2}}_{\rm max}\trh^{\frac{4-k}{k}}m_{\chi}^{\frac{nk+2n+k+8}{k+2}}M_{P}}{\Lambda^{2+n}\rhoend^{\frac{k+1}{k}}}(5.88 \times 10^{6} \text{ GeV}^{-1})
\label{ohnltnc}
\end{align}
Furthermore, if $m_\chi > T_{\cross}$, where $T_{\cross}$ is the temperature when the energy density of the two radiation baths are equal (see Eq.~(\ref{Eq:rhocross}), then the contribution to the relic density from the gravitational bath again can dominate over the decay bath. Furthermore, for fermionic dark matter the from the gravitational bath can dominate over the direct production of DM from inflaton scattering. 
Simplified expressions for the relic density when $k=2$
are given in Eqs.~(\ref{Ohngtnck2}) and (\ref{ohnltnck2}).

In both cases, a non-negligible contribution from the gravitational bath to the total relic density
 requires $\tmax^y < \tmax^h$. When $n>n^h_c$, as in Eq.~(\ref{Ohngtnc}), and $n>n_c^y$, the relic density is dominated by the gravitational bath. 
 In addition, for this case, the hierarchy between $m_\chi$ 
 and $\trh$ is unimportant for the gravitational bath, though we still require $m_\chi < \tmax$.
 For $n_c^h<n<n_c^y$ (for $k\le 6$, $n_c^h < n_c^y$) we must sum the contributions from the two baths.

The decay bath relic abundance contributions are as follows:

    (i) For $n < n_c^y = \frac{10-2k}{k-1}$ and $m_{\chi} < \trh$
\begin{align} 
\Omega^{y}_{\chi}h^{2}&=  \sqrt{\frac{3}{\alpha}} \left(\frac{2k+4}{3n-3nk+30-6k}\right) \frac{\trh^{n+1}m_{\chi}M_{P}}{\Lambda^{2+n}} \nonumber \\ & \times 
(5.88 \times 10^{6} \text{ GeV}^{-1})
\label{ohnltncy}
\end{align}
To convert between $\tmax^y$ and $\trh$, we use
$\tmax^y = \trh (\arh/\amax^y)^{(3k-3)/(2k+4)}$. 
The simplified expression when $k=2$ for this case is given in Eq.~(\ref{ohnltncyk2}). When $m_\chi<\trh$, $n<n_c^y$ and $n<-1$, we need to integrate past $\arh$ to $a_{\rm m}$ in which case
\begin{align}
\Omega^{y}&_{\chi}h^{2}=
 \frac{\sqrt{3} M_{P}\trh^{n+1} m_{\chi}}{\sqrt{\alpha} \Lambda^{n+2}} 
\left[\left(\frac{2k+4}{3n-3nk+30-6k}\right)  \right. \nonumber \\ 
&+ \! \left. \left(\!\frac{1}{1+n}\!\right)\!\left(1\!-\!\left(\frac{m_{\chi}}{\trh}\right)^{n+1}\right)\right] 
(5.88 \times 10^{6} \text{ GeV}^{-1}) \, ,
\label{pastrh}
\end{align}
and the simplified expression for $k=2$ is given in Eq.~(\ref{pastrhk2}).

(ii) For $n = n_c^y$ and $m_{\chi} < T_{RH}$
\begin{align} 
\Omega^{y}_{\chi}h^{2}&= \sqrt{\frac{3}{\alpha}}  \left(\frac{2k+4}{3k-3}\right) \frac{\trh^{n+1}m_{\chi}M_{P}}{\Lambda^{2+n}} \ln \left(\frac{T^{y}_{\rm max}}{\trh}\right) \nonumber \\ & \times  (5.88 \times 10^{6} \text{ GeV}^{-1}) 
\label{ohneqncy}
\end{align}

(iii) For $n >n_c^y$ 
\begin{align} 
\Omega^{y}_{\chi}h^{2}&=  \sqrt{\frac{3}{\alpha}} \left(\frac{2k+4}{3nk-3n-30+6k}\right) \frac{m_{\chi}M_{P}}{\Lambda^{2+n}} \nonumber \\ & \times \trh^{\frac{9-k}{k-1}}{T^{y}_{\rm max}}^{\frac{nk+2k-n-10}{k-1}}(5.88 \times 10^{6} \text{ GeV}^{-1})\, .
\label{ohngtncy}
\end{align}
Because $n > n_c^y$, production is UV dominated and the relative
values of $m_\chi$ and $\trh$ are unimportant.

Finally we provide the relic densities for the decay bath when $m_\chi > \trh$.

(iv) For $n < n_c^y$ and $m_{\chi} >\trh$
\begin{align} 
\Omega^{y}_{\chi}h^{2}&= \sqrt{\frac{3}{\alpha}}  \left(\frac{2k+4}{3n-3nk+30-6k}\right) \frac{M_{P}}{\Lambda^{2+n}}\nonumber \\ & \times \trh^{\frac{9-k}{k-1}}m_{\chi}^{\frac{nk+3k-n-11}{k-1}}(5.88 \times 10^{6} \text{ GeV}^{-1}) \, .
\label{oh2ymgttrh}
\end{align}
This reduces to Eq.~(\ref{oh2ymgttrhk2}) for $k=2$.

(v) For $n = n_c^y$ and $m_{\chi} > \trh$
\begin{align} 
\Omega^{y}_{\chi}h^{2}&= \sqrt{\frac{3}{\alpha}}  \left(\frac{2k+4}{3k-3}\right) \frac{\trh^{n+1}m_{\chi}M_{P}}{\Lambda^{2+n}} \ln \left(\frac{T^{y}_{\rm max}}{m_{\chi}}\right) \nonumber \\ & \times (5.88 \times 10^{6} \text{ GeV}^{-1}) 
\label{oh2ymgttrhnc}
\end{align}

Once again for $n > n_c^y$, we are UV dominated and the hierarchy between $m_\chi$ 
 and $\trh$ is unimportant.

As discussed above, there are several motivated examples
which not always fall into one of the above possibilities. 
In particular, when $\Lambda < \tmax$, the value of $n$ 
changes during the reheating process and the integration must be broken up as described in the text. The solutions for the relic density are given by:

(i) For the gravitational bath with $T^{h}_{\rm max} > \Lambda > m_{\chi} > \trh$: 
\begin{align}
\Omega^{h}_{\chi}h^{2} &= \alpha^{\frac{k+2}{2k}}\sqrt{3}\frac{(T^{h}_{\rm max})^{3}T_{\rm RH}^{\frac{4-k}{k}}\Lambda M_{P}m_{\chi}}{\rho_{\rm end}^{\frac{k+1}{k}}} \left(\frac{T^{h}_{\rm max}}{\Lambda}\right)^{\frac{3k}{k+2}}  \nonumber \\ 
& \times \left(\frac{2k+4}{6k-3}\right)^{\frac{3(k+1)}{7-4k}} \left[\left(\frac{k+2}{-nk-2n-6}\right)+\nonumber \right. \\ & \left. \left(\frac{k+2}{\!nk\!+\!2n\!+\!4k\!+\!14}\right) \left(\!1\!-\!\left(\frac{m_{\chi}}{\Lambda}\right)^{\frac{nk+2n+4k+14}{k+2}}\right)\right] \nonumber \\ &\times (5.88 \times 10^{6} \text{ GeV}^{-1}) \, .
\label{nharhswitch}
\end{align}

(ii) For the decay bath with $T^{h}_{\rm max} > \Lambda > m_{\chi} > \trh$:
\begin{align}
&\Omega^{y}_{\chi}h^{2}\!=\!\sqrt{\frac{3}{\alpha}}\frac{M_{P}m_{\chi}}{\trh}\!\left[\!\left(\!\frac{\trh}{\Lambda}\!\right)^{\! \frac{8}{k-1}\!}\!\left(\frac{2k+4}{\!3n\!-\!3nk\!-\!6k\!+\!30}\!\right) \right. \nonumber \\ & \left. +\left(\frac{T_{\rm RH}}{m_{\chi}}\right)^{\frac{8}{k-1}}\left(\frac{m_{\chi}}{\Lambda}\right)^{n+6}\left(\frac{2k+4}{\!3n\!-\!3nk\!-\!18k\!+\!42}\right)\right] \nonumber \\ & \times (5.88 \times 10^{6} \text{ GeV}^{-1}) \, .
\label{nyarhswitch}
\end{align}

\end{document}